\documentclass[reprint,aip,jcp,groupedaddress]{revtex4-2}
\usepackage{graphicx}
\usepackage{placeins}
\usepackage{color}
\usepackage{latexsym}
\usepackage{amsmath}
\usepackage{amsfonts}
\usepackage{amssymb}
\usepackage{bm}
\usepackage{graphicx}
\usepackage{mathtools}
\usepackage{amsbsy}
\usepackage{physics}
\usepackage{enumitem}

\usepackage{soul}

\usepackage[normalem]{ulem}

\graphicspath{{./figures/}}

\newcommand{\lla}{\left\langle}
\newcommand{\rra}{\right\rangle}

\newcommand{\txi}{%
  \mspace{2mu}%
  \tilde{\mspace{-2mu}\rule{0pt}{1.4ex}\smash[t]{\xi}}%
}

\let\Gamma\varGamma
\let\Delta\varDelta
\let\Theta\varTheta
\let\Lambda\varLambda
\let\Xi\varXi
\let\Pi\varPi
\let\Sigma\varSigma
\let\Upsilon\varUpsilon
\let\Phi\varPhi
\let\Psi\varPsi
\let\Omega\varOmega

\begin{document}

\title{Tangentially Driven Active Polar Linear Polymers --- An Analytical Study}

\author{Christian A. Philipps}
\email{c.philipps@fz-juelich.de}
\affiliation{
Theoretical Physics of Living Matter, Institute of Biological Information Processing and Institute for Advanced Simulation, Forschungszentrum J{\"u}lich and JARA, D-52425 J{\"u}lich, Germany}
\affiliation{Department of Physics, RWTH Aachen University, 52056 Aachen, Germany}
\author{Gerhard Gompper}
\email{g.gompper@fz-juelich.de}
\affiliation{
Theoretical Physics of Living Matter, Institute of Biological Information Processing and Institute for Advanced Simulation, Forschungszentrum J{\"u}lich and JARA, D-52425 J{\"u}lich, Germany.}
\author{Roland G. Winkler}
\email{r.winkler@fz-juelich.de}
\affiliation{
Theoretical Physics of Living Matter, Institute of Biological Information Processing and Institute for Advanced Simulation, Forschungszentrum J{\"u}lich and JARA, D-52425 J{\"u}lich, Germany.}

\date{\today}

\begin{abstract}
The conformational and dynamical properties of isolated flexible active polar linear polymers (APLPs) are studied analytically. The APLPs are modeled as Gaussian bead-spring linear chains augmented by tangential active forces, both in a discrete and continuous representations. The polar forces lead to linear non-Hermitian equations of motion, which are solved by an eigenfunction expansion in terms of a biorthogonal basis set. Our calculations show that the polymer conformations are independent of activity. On the contrary, tangential propulsion strongly impacts the polymer dynamics and yields an active ballistic regime as well as activity-enhanced long-time diffusion, regimes which are both absent in passive systems. The polar forces imply a coupling of modes in the eigenfunction representation, in particular with the translational mode, with a respective strong influence on the polymer dynamics. The total polymer mean-square displacement on scales smaller than their radius of gyration is determined by the active internal dynamics rather than the collective center-of-mass motion, in contrast to active Brownian polymers, reflecting the distinct difference in the propulsion mechanism.     

\end{abstract}


\maketitle

\section{Introduction} \label{sec:introduction}

Polymeric and filamentous structures are an integral part of living matter as they are fundamental for diverse molecular processes and the perpetuation of its out-of-equilibrium state. \cite{demi:10,fang:19,wink:20} Fueled by Adenosine Triphosphate (ATP), molecular machines, such as motor proteins and ribosomes,\cite{kapr:13}  imply an enhanced diffusional motion of biological macromolecular structures.\cite{lau:03,bran:08,robe:10,kapr:13,fakh:14,guo:14,parr:14,gole:15,mikh:15,webe:12,kapr:16,gnes:18,wu:19} {\em In  vivo}, kinesin  motors couple microtubular filaments and generate forces, which affect the dynamics of the cytoskeleton network, the transport processes in  the cell plasma, and the organization of the cell interior.\cite{lau:03,mack:08,lu:16,ravi:17,wink:20}  Similarly, nuclear ATPases such as DNA and RNA polymerase cause nonthermal fluctuations,\cite{guth:99,meji:15,beli:19}  contribute  to chromatin motion, \cite{webe:12,jave:13,zido:13} and are involved in the spatial segregation of the eukaryotic genome. \cite{lieb:09,crem:15,solo:16,sain:18.1,gana:14,smre:17} 

{\em In vitro}, motility assays are a paradigm of active systems. Here, actin or microtubule filaments are translocated by molecular motors whose tails are anchored on a planar substrate. \cite{kawa:08,liu:11,keya:20} Experiments reveal microtubular structures such as rotating rings in dilute systems, or persistently moving large-scale swirls and bands at high filament densities. \cite{butt:10,scha:10,scha:11,doos:18} Likewise, in mixtures of microtubules, kinesin motors, and a depletion agent, bundles emerge, which translocate due to the kinesin motors walking along the filaments.\cite{sanc:12,need:17,doos:18} At high enough concentration, the microtubules form a percolating active network characterized by internally driven chaotic flows, hydrodynamic instabilities, enhanced transport, and fluid mixing.\cite{sanc:12,need:17,vliegenthart_2019,aler:20,mart:21}   

Aside from the biological filamentous structures, diverse concepts are applied to obtain synthetic active colloidal polymers. \cite{sasa:14,mart:15,yan:16,dile:16,zhan:16,zhan:16.1,vutu:17,koko:17,bisw:17,nish:18,loew:18,nish:18,wink:20,shaf:20} Here, propulsion is typically achieved by phoretic effects, based on local gradients such as electric fields (electrophoresis), concentration (diffusiophoresis),  and temperature (thermophoresis).\cite{hows:07,jian:10,vala:10,wurg:10,volp:11,thut:11,butt:13,hage:14,bech:16,maas:16} 

Various models have been proposed to elucidate the conformational and dynamical properties of active polymers and filaments, and, most of the time, have been employed in computer simulations. Particular interesting aspects are the coupling between the polymer conformations, their (internal) dynamics, and the nonequilibrium active forces. Linear polymers comprised of active Brownian particles (ABPOs) \cite{elge:15,bech:16} exhibit an activity-induced polymer collapse, typically in two dimensions, \cite{hard:14,kais:14,bianco_globulelike_2018,anan:20,das:21} or swelling, \cite{ghos:14,shin:15,hard:14,eisenstecken_conformational_2016,eisenstecken_internal_2017,mart:18.1,mous:19,wink:20} and a polymer-length-dependent suppression of phase separation.\cite{suma:14,sieb:17} In contrast to passive polymers, hydrodynamic interactions affect the polymer conformations, with a shrinkage for moderate and a reduced swelling for large activities.\cite{mart:20,wink:20} The polymer dynamics is enhanced and new intramolecular time regimes appear. \cite{mart:19,mart:20} Motor-driven filaments are typically modeled by a tangential active force.\cite{isel:15,duma:18,anan:18,bianco_globulelike_2018,prat:18,fogl:19,moor:20,peterson_statistical_2020,nguy:21,loca:21,qiao:22,philipps_ring_2022} Also in this case strong conformational changes appear, typically polymer shrinkage at very high activities. Computer simulations yield an enhanced center-of-mass dynamics, with a ballistic regime for short times and an activity-amplified long-time diffusive regime for isolated filaments. \cite{isel:15, peterson_statistical_2020, bianco_globulelike_2018} Analytical calculations of semiflexible active polar ring polymers (APRPs) predict a tank-treading-like active motion along the polymer contour,\cite{philipps_ring_2022} consistent with the ring rotation observed in experiments.\cite{kawa:08,keya_synchronous_2020}
Aside from these rather generic approaches, particular realizations of the active environment are considered, such as active dipoles in the modeling of coherent chromatin motion.\cite{sain:18.1}  

Computer simulations provide deep and valuable insight into the properties of individual active polymers and their collective behavior. Yet, for a detailed understanding of dynamical aspects, a theoretical model and analytical predictions are desirable. Specifically for tangentially driven active polar linear polymers (APLPs) an elaborate model is lacking, nonetheless basic considerations have been presented.\cite{peterson_statistical_2020}  

In this article, we describe consistent discrete and continuous theoretical approaches to characterize the properties of flexible APLPs. The discrete polymer model is comprised of beads with active forces along the bond vectors. In the continuum limit, these forces act along the local tangent of the polymer contour. The polar interactions break the polymer end-to-end symmetry, which results in non-symmetric/non-Hermitian eigenvalue problems for the linear equations of motion, which are solved by expansions into biorthogonal basis sets. In contrast to equivalent passive polymers,\cite{harnau_dynamic_1995} the discrete model yields complex eigenvalues for activities exceeding a critical value. This results in a single activity-independent relaxation time and a particular dynamics determined by a superposition of time-dependent trigonometric functions with mode-dependent frequencies. 
The biorthogonal basis leads to a coupling of modes in the mode-amplitude correlation functions with an emerging maximum at a mode, which shifts to a higher mode number for increasing activity, whereas passive flexible polymers show an exponential decay and decoupled modes.\cite{doi_theory_1986,harnau_dynamic_1995} As a consequence, an increasing number of modes is required to ensure convergence of sums over modes in a continuous polymer description. Our approach yields polymer conformational properties, e.g., the mean-square end-to-end distance, which are unaffected by the tangential propulsion. On the contrary, the dynamics is strongly activity dependent. The center-of-mass and bead mean-square displacements show a ballistic regime for short times and a long-time activity-enhanced diffusive overall polymer motion. The total polymer mean-square-displacement is dominated by its internal dynamics, in contrast to active Brownian polymers at higher activities, where the center-of-mass motion prevails.\cite{eisenstecken_internal_2017} The long-time diffusion coefficient increases linearly with increasing  propulsion strength, in agreement with simulation results,\cite{bianco_globulelike_2018} while for active Brownian polymers it increases quadratically with activity.  

Furthermore, the end-to-end vector correlation function is dominated by the first relaxation time for long times, whereas multiple relaxation times contribute at short times resulting in a complex time dependence.

\section{Discrete Model --- Active Polar Linear Bead-Spring Polymer} \label{sec:model_dis}

\begin{figure}[t]
	\includegraphics[width=0.8\columnwidth]{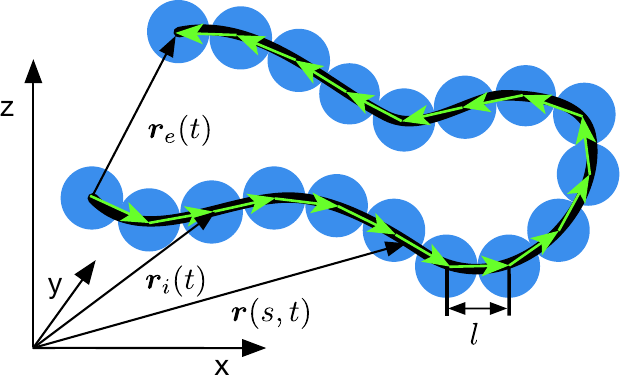}
	\caption{Illustration of an active polar linear polymer. The black line indicates a continuous polymer, the blue beads the discrete one, and the green arrows the active forces along the bonds, which are tangential to the polymer in the continuum limit. Arbitrary position vectors $\bm r_i(t)$, $\bm r(s,t)$, and the end-to-end vector $\bm{r}_e(t)$ are shown, and the bond length $l$ is indicated.}
	\label{fig:sketch}
\end{figure}

\subsection{Equations of Motion}

A flexible discrete linear polymer is composed of $N+1$ beads connected by harmonic bond potentials. Their positions $\bm{r}_{j}(t)$ ($j = 0, \ldots, N$) evolve in time $t$ (Fig.~\ref{fig:sketch}).\cite{harnau_dynamic_1995} An active force $f_a \bm R_j(t)$ is applied along the bond vector $\bm R_j(t) =[\bm r_j(t) - \bm{r}_{j-1}(t)]$ of force density $f_a$ per bond length $l$. The total active force on bead $j$ is then $\bm F_j^a(t) = f_a [\bm R_{j+1}(t) + \bm R_j(t)]/2 = f_a [\bm r_{j+1}(t) - \bm r_{j-1}(t)]/2$. \cite{isel:15,wink:20} The beads' overdamped equations of motion (EOMs) are ($j=1,\ldots, N-1$)
\begin{align}
	\tilde{\gamma} \frac{d}{d t} \bm{r}_{0}(t) = & \ \left(\frac{f_a}{2} + \frac{3k_BT}{l^2} \right) \bm{R}_{1}(t) + \bm{\Gamma}_{0}(t)  ,  
	\label{align:EOM_discrete_start}
	\\
	\begin{split}
    	\tilde{\gamma} \frac{d}{d t} \bm{r}_{j}(t) = & \ \frac{f_a}{2} \big[ \bm{R}_{j+1}(t) + \bm{R}_{j}(t) \big] + \bm{\Gamma}_{j}(t) 
    	\\ 
    	& \ + \frac{3k_BT}{l^2} \big[ \bm{R}_{j+1}(t) - \bm{R}_{j}(t) \big] , 
	\end{split}
	\label{align:EOM_discrete_inner}
	\\
	\tilde{\gamma} \frac{d}{d t} \bm{r}_{N}(t) = & \ \left(\frac{f_a}{2} - \frac{3k_BT}{l^2} \right) \bm{R}_{N}(t) + \bm{\Gamma}_{N}(t)  ,
    \label{align:EOM_discrete_end}
\end{align}	
with the bond force $\bm F_j^b(t) = 3k_BT \big[\bm{R}_{j+1}(t) - \bm{R}_{j}(t) \big]/l^2$ and the stochastic force $\bm \Gamma_j(t)$. The strength of the bond force, $3 k_B T/l^2$, ensures that the local constraint $\langle \bm{R}_j^2 \rangle = l^2$ is satisfied  for all activities, which accounts for the inextensibility of a polymer in a mean-field manner, \cite{winkler_models_1994,winkler_deformation_2003} where $T$ is the temperature, $k_B$ the Boltzmann constant, and $L = Nl$ the polymer contour length. In the matrix formulation of the EOMs, the polar active force leads to a non-symmetric matrix $\mathbf M$ (see Eqs.~(S1) and (S2) of the Supplementary Material (SM)). The thermal fluctuations, $\bm \Gamma_j (t)$, are described as a stationary Gaussian and Markovian stochastic process with zero mean and the second moments\cite{risken_fokker-planck_1984,doi_theory_1986,harnau_dynamic_1995} 
\begin{equation}
	\langle\Gamma_{\alpha, j}(t)\Gamma_{\beta,k}(t') \rangle = 2 \tilde{\gamma} k_BT \delta_{\alpha \beta} \delta_{jk}  \delta(t-t'),
\end{equation}	
with the translational friction coefficient $\tilde{\gamma}$, and the Euclidean coordinates $\alpha, \beta \in \{ x, y, z \}$. To characterize activity, the P\'eclet number  
\begin{equation}
	Pe = \frac{f_a L^2}{k_BT} 
\end{equation}	
is introduced, which is the ratio between the total active energy of a polymer of length $L$ and the thermal energy. \cite{isel:15} 

\subsection{Solution of the Equations of Motion} \label{sec:sol_eom_dis}

The solution of the non-symmetric (non-Hermitian) equations of motion (\ref{align:EOM_discrete_start})-(\ref{align:EOM_discrete_end}) is obtained by the eigenvector expansion  
\begin{equation}
	\tilde {\bm{r}}_j(t) = \sum_{m=0}^{N} \tilde{\bm{\chi}}_m(t) b_m^{(j)}, \  \
	\tilde{\bm{\Gamma}}_j (t) = \sum_{m=0}^{N} \tilde {\bm{\Gamma}}_m(t) b_m^{(j)} ,
	\label{eq:discrete_expansion}
\end{equation}	
with the eigenvectors $\bm b_m = (b_m^{(0)}, \ldots, b_m^{(N)})^T$ of the non-symmetric matrix $\mathbf M$ (Eq.~(S2)). A biorthogonal basis set is formed together with the adjoint eigenvectors $\bm b_m^{\dag} = (b_m^{(0) \dag}, \ldots, b_m^{(N) \dag})^T$ of the transposed matrix ${\mathbf M}^{T}$. The eigenvectors are normalized such that $\bm b_m^{\dag} \cdot \bm b_n = \delta_{mn}$. \cite{risken_fokker-planck_1984} Explicitly, the eigenvectors are given by 
\begin{align} 
	&b_m^{(j)} = \sqrt{\frac{2}{N+1}} \frac{e^{-dj}}{\sqrt{(1-r_d^2) \sin^2 \tilde k_m + (r_d - \lambda_m/2)^2}} 
	\label{align:eigen_dis_b}
	\\		
	& \times \Big[ \sqrt{1-r_d^2} \sin \tilde k_m \cos(\tilde k_m j) + (r_d - \lambda_m/2) \sin(\tilde k_m j) \Big], \nonumber
	\\[0.5em]
	&b_m^{(j)^{\dagger}} = e^{2 d j} b_m^{(j)}, \hspace{1cm} m \in \mathbb{N}_0,
	\label{align:eigen_dis_b_dag}
	\\[0.5em] 
	&b_0^{(j)} = \sqrt{\frac{\sinh(d)}{e^{dN}\sinh(d(N+1))}}, 
	\label{align:eigen_dis_norm}
	\\[0.5em]
	&d = \ln (\sqrt{1+r_d}/\sqrt{1-r_d}), 
	\hspace{1.44cm}
	r_d <1 , \\
	&d = \ln (\sqrt{1+r_d}/\sqrt{r_d-1}) - i \frac{\pi}{2}, 
	\hspace{0.6cm}
	r_d >1 , 
\end{align}	
with the wave numbers $\tilde k_m = m \pi / (N+1)$, the abbreviation 
\begin{equation} \label{eq:abb_d}
    r_d = \frac{Pe}{6 N^2} ,    
\end{equation}
and the eigenvalues  
\begin{equation}
	\txi_m = \frac{N^2 \tilde{\gamma}}{\pi^2 \tilde{\tau}_R} \lambda _m  = \frac{2N^2 \tilde{\gamma}}{\pi^2 \tilde{\tau}_R} \left[1 - \sqrt{1-r_d^2} \cos \tilde k_m \right], 
	\hspace{0.3cm} 
	\txi_0 = 0,
	\label{eq:discrete_eigenvalues}
\end{equation}	
in terms of the relaxation time $\tilde{\tau}_R = \tilde{\gamma} L^2 /(3\pi^2k_BT)$ of a flexible passive polymer.\cite{doi_theory_1986,harnau_dynamic_1995} The eigenvalues $\txi_m$ can be complex for $r _d >1$, however, only conjugate complex pairs, $\txi_m = \txi^{*}_{N+1-m}$, appear in the eigenvector expansion, which implies real position vectors $\tilde{\bm{r}}_j(t)$ and random forces $\tilde{\bm{\Gamma}}_j(t)$ (SM, Sec.~S-II). 

Insertion of the expansion \eqref{eq:discrete_expansion} into the EOMs \eqref{align:EOM_discrete_start}-\eqref{align:EOM_discrete_end} yields the equations for the mode amplitudes
\begin{align}
	\tilde{\gamma} \frac{d}{dt} \tilde{\bm{\chi}}_m(t) & = - \txi_m \tilde{\bm{\chi}}_m(t) + \tilde{\bm{\Gamma}}_m(t) . 
	\label{align:discrete_EOM_AMF} 
\end{align}	
In the stationary state, their solutions are
\begin{align} 
	\tilde{\bm{\chi}}_m(t) & = \frac{1}{\tilde{\gamma}} e^{- \tilde \xi_m t/ \tilde{\gamma}}  \int_{-\infty}^{t}  e^{ \tilde \xi_m t'/ \tilde{\gamma}} \tilde{\bm{\Gamma}}_{m}(t')  dt' , 
	\label{align:discrete_EOM_AMF_mn}
	\\ 
    \tilde{\bm{\chi}}_0(t) & =  \tilde{\bm{\chi}}_0(0) +  \frac{1}{\tilde \gamma} \int_{0}^{t}  \tilde{\bm{\Gamma}}_{0}(t')  dt'  .
    \label{align:discrete_EOM_AMF_translational}
\end{align}	
As long as $r_d < 1$, $\txi_m \in \mathbb{R}$ and Eq.~\eqref{align:discrete_EOM_AMF} describes relaxation processes with the relaxation times 
\begin{align}
	\tilde \tau_m = \frac{\tilde{\gamma}}{\txi_m} = \frac{\pi^2}{2N^2} \frac{\tilde{\tau}_R}{(1 - \sqrt{1-r_d^2} \cos \tilde k_m)}.
\end{align}		
In the case of $r_d > 1$, the $\txi_m$ are complex with the single mode-independent relaxation time $\tau$, and mode-dependent frequencies $\omega_m$, which, via $\txi_m = \xi_m^R + i  \xi_m^I = \tilde{\gamma}/ \tau - i \tilde{\gamma} \omega_m$, are given by
\begin{align} \label{eq:frequency}
	\tau  = \frac{\pi^2}{2N^2} \tilde{\tau}_R ,
	\ \ 
	\omega_m = \frac{2N^2}{\pi^2} \frac{1}{\tilde{\tau}_R} \sqrt{r_d^2-1} \cos \tilde k_m .
\end{align}	
For $N$ even, there is one real eigenvalue, $\txi_0=0$, whereas for $N$ odd, the additional real eigenvalue $\txi_{(N+1)/2} = \tilde{\gamma}/\tau$ is present. 

The spectrum of complex eigenvalues cannot as straightforwardly be interpreted as that for real eigenvalues only, since the solution of the eigenvalue problem is typically a superposition of many eigenfunctions.       

\section{Continuous Model --- Tangentially Propelled Polar Linear Polymer}

\subsection{Equation of Motion}

In the continuum limit of a flexible linear polymer, \cite{winkler_models_1994,winkler_deformation_2003,harnau_dynamic_1995} the equation of motion of the APLP is
\begin{equation} 
    \hspace{-0.18cm}
	\gamma \frac{\partial}{\partial t}{\bm{r}}(s,t) = f_a \frac{\partial}{\partial s} \bm{r}(s,t) + 3pk_{B}T \frac{\partial^{2}}{\partial s^{2}}\bm{r}(s,t)  + \bm{\Gamma}(s,t) 
	\label{eq:continuum}
\end{equation}	
for $pL \gg 1$, where $p$ is related to the polymer persistence length $l_p$ via $p =1/(2l_p)$.\cite{winkler_models_1994,ha_1995,winkler_deformation_2003} Here, $s$ is the contour variable ($s \in [0, L]$), $\gamma$ the friction coefficient per length, and $\bm{\Gamma}(s,t)$ a stationary, Gaussian, and Markovian stochastic process of zero mean and the second moments \cite{harnau_dynamic_1995}
\begin{equation}
    \lla \bm \Gamma (s,t) \cdot \bm \Gamma(s',t') \rra = 6 \gamma k_B T \delta(s-s') \delta(t-t') .    
\end{equation}
Equation~\eqref{eq:continuum} has to be solved with the free-end boundary conditions  $\partial \bm r(s,t)/ \partial s = 0$ at $s=0,L$.

As for the discrete polymer, the strength of the bond potential $3 pk_BT$ ensures that the mean-field constraint 
\begin{equation}
    L = \int_0^L \lla \left( \frac{\partial \bm r(s)}{\partial s} \right)^2 \rra ds     
\end{equation}
of polymer inextensibility is satisfied. 

\subsection{Solution of EOM --- Eigenfunction Expansion}

The solution of the non-Hermitian equation (\ref{eq:continuum}) is obtained by the eigenfunction expansion 
\begin{equation} 	
	\bm{r}(s,t) = \sum_{m=0}^{\infty} \bm{\chi}_m(t) \phi_m(s) , \ 
	\bm{\Gamma}(s,t) = \sum_{m=0}^{\infty} \bm{\Gamma}_m(t) \phi_m(s) ,
	\label{align:continuous_expansion}
\end{equation}	
in terms of a biorthogonal basis $\{ (\phi_m,\phi_m^{\dagger}) ; \ m \in \mathbb{N}_0 \}$, following as solution of the two eigenvalue equations
\begin{align}
    \left(  f_a \frac{\partial}{\partial s} + 3pk_{B}T \frac{\partial^2}{\partial s^2} \right) \phi_m& = - \xi_m \phi_m , 
    \\ 
    \left( -f_a \frac{\partial}{\partial s} + 3pk_{B}T \frac{\partial^2}{\partial s^2} \right) \phi_m^{\dagger}& = - \xi_m \phi_m^{\dagger} .
\end{align}
The eigenfunctions are normalized such that
\begin{equation}
	\int_{0}^{L} \phi_m^{\dagger}(s) \phi_n(s) ds = \delta_{mn} .
\end{equation}	
The free-end boundary conditions for $\bm r(s,t)$ lead to the conditions
\begin{align}
	\left[\frac{d}{ds} \phi_{m}(s) \right]_{s = 0, L} &= 0 ,
	\label{align:BC}
	\\
	\left[-2r_c \frac{\pi}{L} \: \phi_{m}^{\dag}(s) + \frac{d}{ds}\phi_{m}^{\dag}(s)\right]_{s = 0, L} &= 0 ,
	\label{align:BC_adj}
\end{align}
for the eigenfunctions. Note the distinct boundary conditions for the adjoint eigenfunction. Here, the abbreviation 
\begin{equation} \label{eq:abb_c}
    r_c = \frac{Pe}{6 \pi pL}
\end{equation}
is introduced. Explicitly, the eigenfunctions and adjoint ones are obtained as ($m \ge 1$)
\begin{align}	
	\phi_{m}(s) = & \ \sqrt{\frac{2}{L}} \frac{e^{-\pi r_c s/L}}{\sqrt{m^2 + r_c^2}} \big[ m \cos(k_m s) + r_c \: \sin(k_m s) \big],
	\label{align:eig_func}
	\\
	\phi_{m}^{\dagger}(s) = & \ e^{2 \pi r_c s/L} \phi_{m}(s), \hspace{1cm} m \in \mathbb{N}_0,
	\\[1em]
	\phi_0 = & \ \sqrt{\frac{\pi r_c}{L e^{\pi r_c}\sinh(\pi r_c)}}, 
\end{align}	
with the wave numbers $k_m=m \pi/L$. The real eigenvalues are given by
\begin{equation}
	\xi_m = \frac{\gamma}{\tau_R} (m^2 + r_c^2),
	\label{eq:continuous_eigenvalues}
\end{equation}	 
with the relaxation time $\tau_R = \gamma L^2 /(3\pi^2pk_BT)$ of a flexible passive polymer.\cite{harnau_dynamic_1995}
The exponential decay along the polymer contour in Eq.~\eqref{align:eig_func} unveils the head-tail symmetry breaking due to the polarity of the driving force. 

The equations for the mode amplitudes $\bm{\chi}_m(t)$ and their solutions exhibit the same structure as in the discrete model, Eqs.\eqref{align:discrete_EOM_AMF}-\eqref{align:discrete_EOM_AMF_translational}, but with the distinct relaxation times
\begin{equation} 
	\tau_m = \frac{\gamma}{\xi_m} = \frac{\tau_R}{m^2 + r_c^2} .
	\label{eq:relaxation_time_cont}
\end{equation}	

In the following, results of the continuous with those of the discrete polymer model are compared, where the continuous description corresponds to the discrete one for $pL=N \gg 1$, $p=1/l$, and $\gamma =\tilde{\gamma}/l$.






\begin{figure}[t]
	\centering
	\includegraphics[width=\columnwidth]{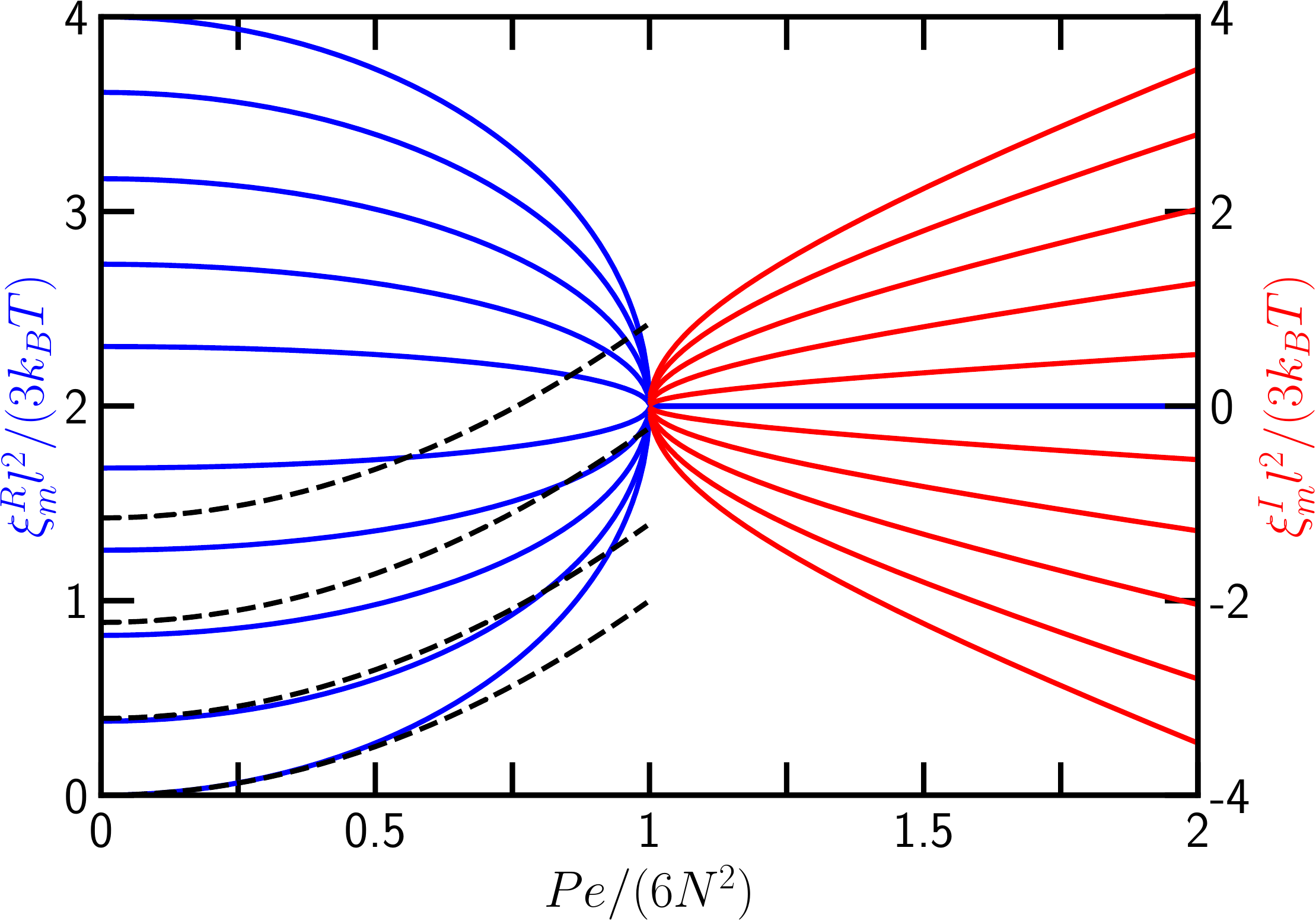}
	\caption{Normalized eigenvalues $\txi_m$ as a function of the scaled activity $Pe/(6N^2)$ for the modes $m \in \{ 1,\, 100,\, 150,\, 190,\, 225,\, 275,\, 310,\, 350,\, 400,\, 500 \}$ (bottom to top) of a discrete polymer model of length $L/l=N=500$.  
	The left axis corresponds to the real part (blue) and the right axis to the imaginary part (red) of the eigenvalues \eqref{eq:discrete_eigenvalues}. The black dashed lines indicate the eigenvalues $\xi_m$ (Eq.~\eqref{eq:continuous_eigenvalues}) of the continuous polymer model for the modes $m \in \{1,\, 100,\, 150,\, 190 \}$ (bottom to top) and $pL=N$.}
	\label{fig:eig_val_dis_con}
\end{figure}

\section{Conformational Properties} \label{sec:conformations}

The conformational properties of the polymers are determined by the relaxation times and the mode amplitudes. 

\subsection{Eigenvalue Spectrum}

Figure~\ref{fig:eig_val_dis_con} illustrates the dependence of the eigenvalues of the discrete polymer model (Eq.~\eqref{eq:discrete_eigenvalues}) on the P\'eclet number for various modes. In the interval $0 \le Pe/(6N^2) \leq 1$, the eigenvalues $\txi_m$ are real, determining relaxation times $\tilde \tau_m$, and they increase/decrease with increasing $Pe$. For $Pe/(6N^2) > 1$, the eigenvalues are given by the real part $\xi^R_m = \tilde\gamma/\tau$ (Eq.~\eqref{eq:frequency}), with the mode-independent relaxation time $\tau$, and the imaginary part  $\xi^I_m = - \tilde\gamma \omega_m$ with the frequencies  $\omega_m$ (Eq.~\eqref{eq:frequency}). These frequencies increase for modes $m \le (N+1)/2$ and decrease for $m \ge (N+1)/2$ with increasing $Pe$. As long as $\tilde k_m = m \pi/(N+1) \ll 1$ and $Pe/(6N^2) \ll 1$, the eigenvalues of the discrete and continuous polymer model agree with each other, as reflected by the modes $m=1$ and $m=100$ in Figure \ref{fig:eig_val_dis_con}. For large $m$ and $Pe/(6N^2)$, the eigenvalue spectra differ due to discretization effects.   

The fundamental differences in the eigenvalue spectra for $Pe/(6N^2)>1$ reflect the distinctiveness of a discrete and a continuous polymer model. The difference equations for the eigenvalues and eigenvectors of the discrete model provide other solutions than the eigenvalue equations of the continuum approach, since the latter requires a smooth and infinitesimal change of the eigenfunctions as the contour variable $s$ varies, whereas the difference equations allow only for finite discrete changes.    

\subsection{Relaxation Times} \label{subsec:relaxation_times}

\begin{figure}[t]
	\centering
	\includegraphics[width=\columnwidth]{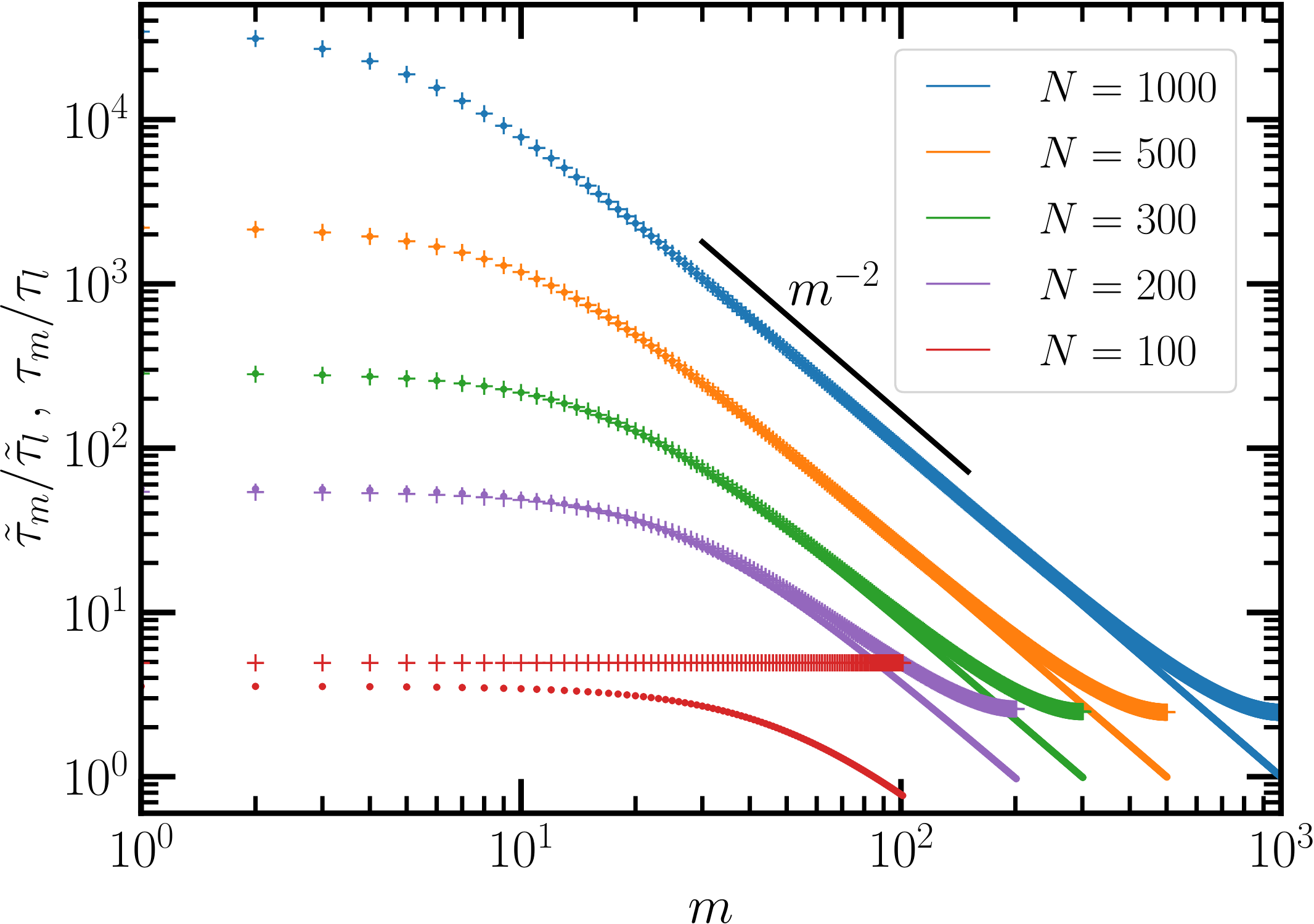}
	\caption{Normalized relaxation times of the discrete, $\tilde{\tau}_m/\tilde{\tau}_l$ (crosses), and the continuous polymer model, $\tau_m/\tau_l$ (dots), as a function of the mode number $m$ for the Péclet number $Pe=10^{5}$ and various number of beads as indicated in the legend. The normalization factor is $\tau_l = \tilde \gamma l^2/(3 \pi^2 k_B T) = \tilde \tau_R/N^2$, also for the continuous polymer with $pL = N$. The black line indicates the quadratic power law decay.}
	\label{fig:tau_m_tau_R_pe_dep_con_dis}
\end{figure}

The relaxation times $\tilde \tau_m= \tilde \gamma/\txi_m$ of discrete APLPs are presented and compared with those of continuous APLPs in Figure~\ref{fig:tau_m_tau_R_pe_dep_con_dis}. 
In a passive polymer, $\tilde \tau_m/ \tilde \tau_R \sim \tau_m/\tau_R  \sim 1/m^2$ for $m > 1$ and $m\pi/(N+1) \ll 1$, i.e., the relaxation times decrease quadratically with increasing mode number.\cite{doi_theory_1986} This is no longer the case for $Pe \gg 1$, where the relaxation times assume a progressively extended plateau at small $m \ge 1$ with increasing $Pe$, as is apparent from Eq.~\eqref{eq:relaxation_time_cont}.   

As long as the polymers are sufficiently long such that $Pe/(6 N^2) \lesssim 0.1$, the relaxation times of the discrete closely agree with those of the continuous polymer model, except of the expected discretization discrepancies for large mode numbers.  Deviations at small $m$ appear for $Pe/(6 N^2) > 0.4$ and are most pronounced for $Pe/(6 N^2)>1$, where there is only one mode-independent relaxation time $\tau = \tilde \gamma l^2/(6 k_BT)$ for the discrete polymer model (Fig.~\ref{fig:tau_m_tau_R_pe_dep_con_dis}). In contrast, the relaxation times of the continuous polymer model $\tau_m $ (Eq.~\eqref{eq:relaxation_time_cont}) are proportional to $Pe^{-2}$ for $Pe/(6 \pi pL) \gg m$ and decrease with increasing P\'eclet number.

\FloatBarrier

\subsection{Correlation Functions of Mode Amplitudes}

For the discrete polymer model, the stationary-state time correlation functions of the mode amplitudes, obtained from Eqs. (\ref{align:discrete_EOM_AMF_mn}) and (\ref{align:discrete_EOM_AMF_translational}), are ($m, n \ge 1$)
\begin{equation}
	\langle \tilde{\bm\chi}_{m}(t)\cdot \tilde{\bm\chi}_{n}(t') \rangle = \frac{6k_{B}T}{\txi_m + \txi_n} e^{- \tilde \xi_x |t-t'| / \tilde{\gamma}} \: 	\bm{b}_m^{\dagger} \cdot \bm{b}_n^{\dagger} ,
	\label{eq:msq_mode_dis}
\end{equation}	
where $\txi_x= \txi_m$ for $t > t'$ and $\txi_x= \txi_n$ for $t < t'$, respectively. Correlation functions comprising the mode $m=0$ are given by 
\begin{align}
	\langle \tilde{\bm\chi}_{m}(t) \cdot \tilde{\bm\chi}_{0}(t') \rangle = & \ \frac{6k_BT \tilde{\tau}_m}{\tilde{\gamma}} \: \bm{b}_m^{\dagger} \cdot \bm{b}_0^{\dagger}
		\begin{dcases}
			e^{-(t-t')/ \tilde{\tau}_m} , & t>t' 
			\\
			1 , & t \le t'
		\end{dcases}, 
		\\
		\langle \tilde{\bm\chi}_{0}(t) \cdot \tilde{\bm\chi}_{0}(t') \rangle = & \ \frac{6k_BT}{\tilde{\gamma}} \: \bm{b}_0^{\dagger} \cdot \bm{b}_0^{\dagger} \: t' + \tilde{\bm\chi}_0^2(0), \ \ \  t \ge  t' \ge 0 .	
\end{align}
For the continuous polymer model, the stationary-state time correlation functions of the mode amplitudes are ($m,n \ge 1$)
\begin{align}
	\begin{split}
	&\langle \bm\chi_{m}(t) \cdot \bm\chi_{n}(t') \rangle = 
	\\
	& \hspace{1cm} \frac{6k_{B}T}{\xi_m + \xi_n} e^{- \xi_m |t-t'| / \gamma} \: \int_{0}^{L} \phi_m^{\dagger}(s)\phi_n^{\dagger}(s) \: ds.
	\end{split}
\end{align}
Spatial integration yields
\begin{widetext}
    \begin{align}
        \langle \bm\chi_{m}(t)\cdot \bm\chi_{n}(t') \rangle = & \ \frac{8 r_c L^3}{\pi^3 pL}  \frac{m n (-1 + (-1)^{m+n}e^{2 \pi r_c})}{[(m-n)^2 + 4 r_c^2][(m+n)^2 + 4 r_c^2] \sqrt{m^2+r_c^2}\sqrt{n^2+r_c^2} } e^{-\xi_x|t-t'| / \gamma}  , 
        \\[7pt]  
        \langle \bm \chi_{m}(t)\cdot \bm\chi_{0}(t') \rangle = & \ \frac{4r_cL^3}{\pi^3 e^{\pi r_c/2} pL} \sqrt{\frac{2 \pi r_c}{\sinh(\pi r_c)}} \frac{m (-1 + (-1)^m e^{3 \pi r_c})}{(m^2+r_c^2)^{3/2}(m^2+9r_c^2)}
        \begin{dcases}
        	e^{-\xi_m(t-t')/ \gamma} , & t>t' 
            \\
        	1 , & t \le t'
        \end{dcases} ,  
        \\[7pt]
        \langle \bm\chi_{0}(t)\cdot \bm\chi_{0}(t') \rangle = &  \ \bm \chi_0^2(0) + \frac{3k_BT}{\gamma}e^{\pi r_c} \frac{\sinh(2 \pi r_c)}{\sinh(\pi r_c)} \ t' , \ \  t \ge t' \ge 0 ,
    \end{align}
\end{widetext}
with the abbreviation $r_c = Pe/(6 \pi pL)$ (Eq.~\eqref{eq:abb_c}).
At equal-times, $t=t'$, and for $m=n \ge 1$ the mean-square mode amplitudes are 
\begin{equation} 
	\langle \bm{\chi}_m^2 \rangle = \frac{L^3}{2 \pi^3 r_c pL}  \frac{m^2}{(m^2+r_c^2)^2} (e^{2\pi r_c}-1).
	\label{eq:msq_mode_con}
\end{equation}	
The mode-amplitude correlation functions reduce to those of a passive system in the limit of $Pe =0$.\cite{harnau_dynamic_1995} In strong contrast to the passive case, the non-Hermitian nature of the equations of motion of the APLPs implies a coupling of the mode amplitudes for $Pe >0$. Not only couple the modes with $m \neq n$, $m, n \ge 1$, but also the mode $m=0$, which describes the center-of-mass translation motion in case of a passive polymer. Such a coupling of modes was assumed in Ref.~[\onlinecite{batt:16}] to describe the broken detailed balance in the internal dynamics of semiflexible polar filaments.

\begin{figure}[b]
	\centering
	\includegraphics[width=\columnwidth]{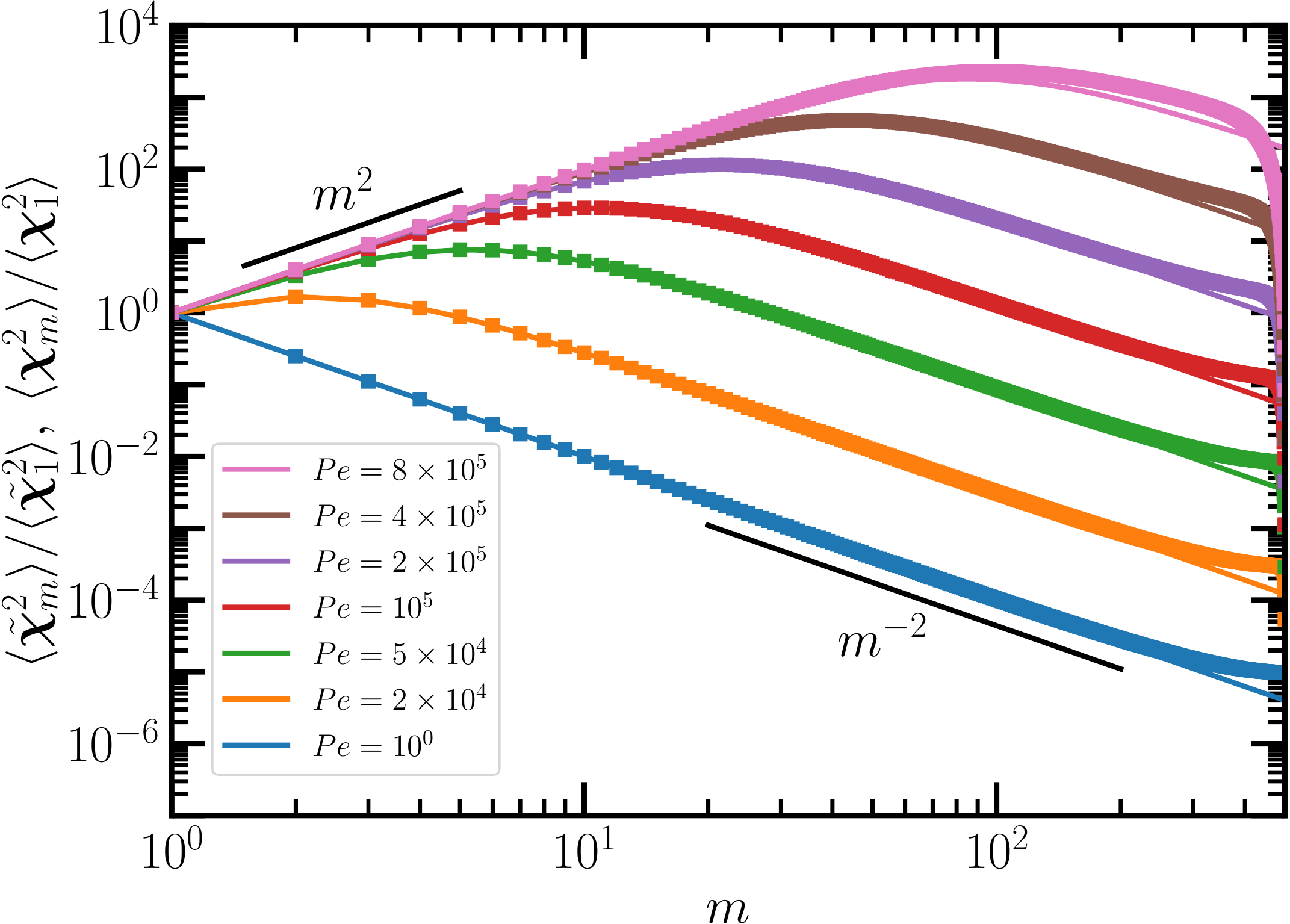}
	\caption{Normalized stationary-state mean-square mode amplitudes $\langle \bm \chi_m^2 \rangle$ for discrete (squares) and  continuous (lines) APLPs as a function of the mode number $m$ for various P\'eclet numbers $Pe$ as indicated in the legend (increasing from bottom to top) and $N= pL = 500$. $\langle \bm \chi_1^2 \rangle$ is the mean-square mode amplitude for the mode $m=1$. The black lines indicate quadratically decreasing and increasing power laws.}
	\label{fig:tau_m_tau_1_mode_omega_dep_con_dis}
\end{figure}

\begin{figure}[t]
	\centering
	\includegraphics[width=\columnwidth]{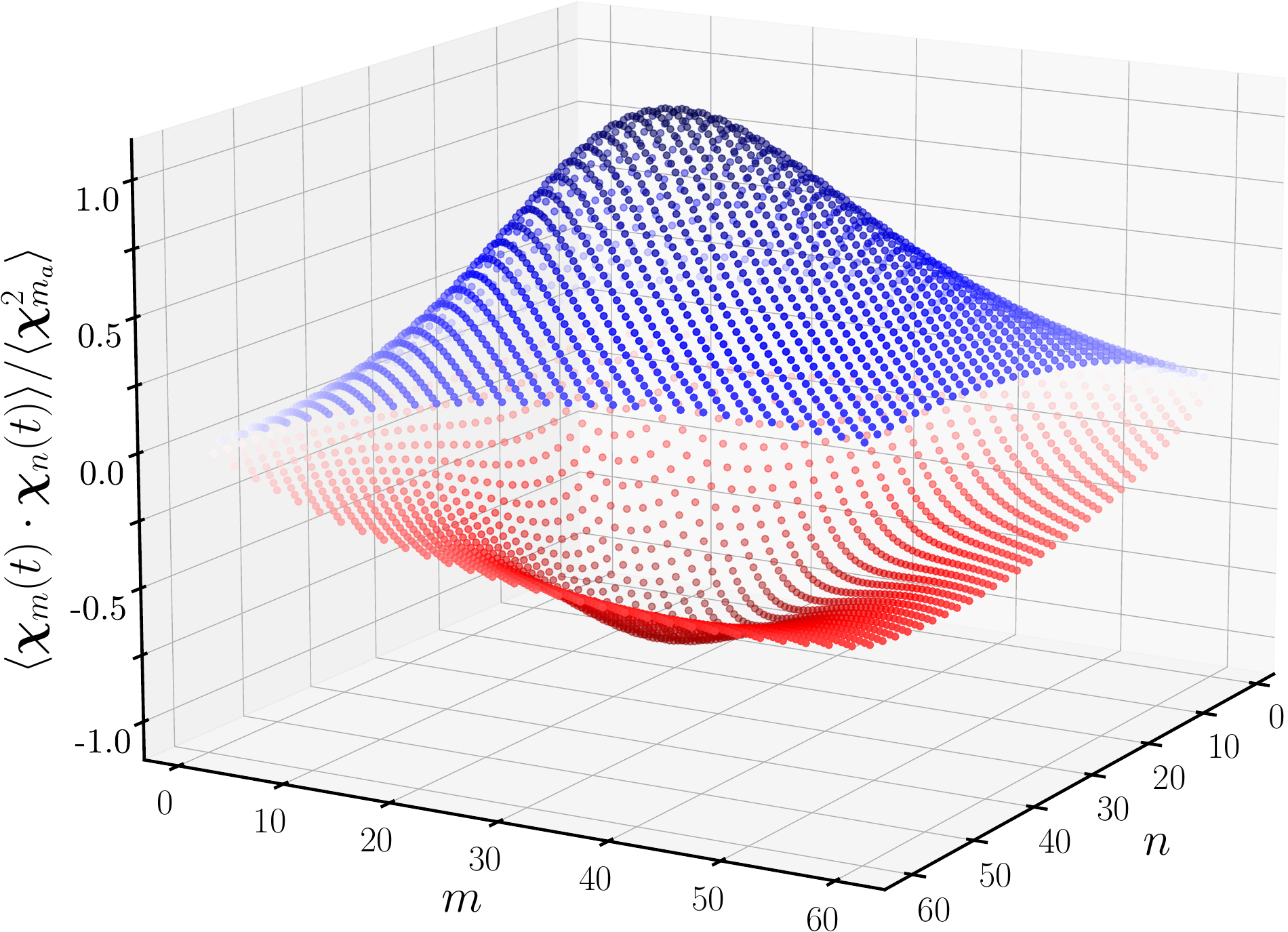}
	\caption{Normalized stationary-state mode-amplitude correlations $\langle \bm \chi_m(t) \cdot \bm \chi_n(t) \rangle$ of a continuous APLP as a function of the mode numbers $m$ and $n$ for $Pe =1.9\times 10^5$ and $pL=500$. The normalization factor $\langle \bm\chi_{m_a}^2 \rangle$ is the maximum of $\langle \bm \chi_m^2 \rangle$ determined by the mode $m_a = 20$. The correlations $\langle \bm \chi_m(t) \cdot \bm \chi_n(t) \rangle$ are arranged on two ``bell-shaped'' surfaces, with positive (blue) and negative (red) values, respectively.}
	\label{fig:AMCF_con_3dim}
\end{figure}

Figure~\ref{fig:tau_m_tau_1_mode_omega_dep_con_dis} depicts mean-square mode amplitudes for discrete and continuous APLPs (Eqs.~\eqref{eq:msq_mode_dis} and \eqref{eq:msq_mode_con}). In the limit of $Pe \to 0$ (bottom curve), the well-known dependence $\langle \bm \chi_m^2 \rangle \sim 1/m^2$ of passive flexible polymers is obtained,\cite{doi_theory_1986,harnau_dynamic_1995} with discretization differences at large $m$. With increasing $Pe$, gradually a maximum appears, $\langle \bm \chi_{m_a}^2 \rangle$, which shifts to larger $m$ with increasing $Pe$. The relaxation times in the vicinity of the maximum, corresponding to the mode number $m_a = \mathrm{Integer}\big[Pe/(6 N^2)\big]$, determine the APLP dynamics, because they yield the largest contribution in the sum over modes in terms of the correlation functions. Below the maximum value $m_a$, the mean-square mode amplitudes increase as $\langle \bm \chi_m^2 \rangle \sim m^2$ for $m \ll m_a$, and for $m>m_a$ they decrease  as $\langle \bm \chi_m^{2} \rangle \sim m^{-2}$. The full dependence of  $\langle \bm \chi_m(t) \cdot \bm \chi_n(t) \rangle$ on the mode numbers $m$ and $n$ is shown in Figure~\ref{fig:AMCF_con_3dim}. The values of the correlations are arranged on two ``bell-shaped'' surfaces, corresponding to even and odd mode numbers. This demonstrates the strong coupling between the modes and, most importantly, the presence of negative correlations. 

The non-Hermitian nature of the equations of motion of the APLPs leads to a tight coupling of the various modes, implying substantial deviations to the mode correlation functions of passive polymers.\cite{doi_theory_1986,harnau_dynamic_1995} Specifically, the maximum at $m_a$ for $Pe/(6N^2)>1$ implies that the relaxation behavior of macroscopic quantities is typically no longer determined by the largest relaxation time, but rather by  $\tau_{m}$ in the vicinity of $m \approx m_a$. Moreover, the presence of negative values of the correlations $\langle \bm \chi_m(t) \cdot \bm \chi_n(t) \rangle$ causes cancellation of positive contributions in sums over modes, and hence, requires summation over many modes to achieve convergence, in particular for continuous APLPs. On the contrary, the number of modes of discrete APLPs is limited by the number of beads.

\subsection{Mean-Square End-to-End Distance, Radius of Gyration} \label{subsec:mse2ed_rog}

With the expansions Eqs.~\eqref{eq:discrete_expansion} and  \eqref{align:continuous_expansion} for the discrete and continuous APLPs, respectively, the numerical evaluation of the mean-square end-to-end distances yields
\begin{align} 
	\langle \tilde {\bm{r}}_e^2 \rangle = & \ \langle ( \tilde{\bm{r}}_N(t) - \tilde{\bm{r}}_0(t))^2 \rangle  = N l^2 , 
	\label{align:msq_end_dis}
	\\ 
    \langle \bm{r}_e^2 \rangle = & \ \langle ( \bm{r}(L,t) - \bm{r}(0,t))^2 \rangle  = \frac{L}{p} ,
    \label{align:msq_end_con}
\end{align}	
over the considered range of P\'eclet numbers, $0 \leq Pe \leq 10^8$, various $N$, and even in the case of complex eigenvalues ($Pe/(6 N^2)>1$) --- identical to those of passive polymers. The double sums over modes have to be performed numerically and, for the continuous polymer, a sufficiently large number of modes has to be taken into account to achieve convergence (SM, Sec.~S-III). Similarly, the radii of gyration, $\langle \bm{r}_g^2 \rangle = \langle \bm{r}_e^2 \rangle/6$, are identical with those of passive polymers, and even the mean-square bond lengths satisfy the constraints $\langle \bm{R}_i^2 \rangle = l^2$ for all Péclet numbers. Moreover, subsequent bond vectors are independent, i.e., $\langle \bm{R}_{i+1} \cdot \bm{R}_{i} \rangle = 0$, as applies for flexible phantom polymers.\cite{doi_theory_1986}  Hence, within the adopted model, the polymer conformational properties of tangentially driven APLPs are independent of propulsion.\cite{peterson_statistical_2020} This is in contrast to some computer simulations of polar polymers, which predict polymer shrinkage, despite accounting for excluded-volume interactions,
however, applying a different bond potential \cite{isel:15,anan:18,bianco_globulelike_2018} and a different tangential active force.\cite{bianco_globulelike_2018}      

The simulations of Ref.~[\onlinecite{bianco_globulelike_2018}] predict a shrinkage of a discrete phantom polymer for $Pe/N^2 \gtrsim 20$ --- note the different definition of the active force in [\onlinecite{bianco_globulelike_2018}] --- and thus, a deviation from the linear $N$ dependence in Eq.~\eqref{align:msq_end_dis} occurs. Although, this differs from the predictions of the presented Gaussian APLP model, these simulation results of the dynamics can be compared to the analytical ones as long as $Pe/N^2 \lesssim 20$, which for a polymer with $N=500$ beads corresponds to $Pe \lesssim 5 \times 10^6$. As will be shown below, the APLP dynamics exhibits generic activity effects for such P\'eclet numbers.  \\

\section{Dynamical Properties} \label{sec:dynamics}

The translational motion of polymers is characterized by the total mean-square displacement (MSD), averaged over the polymer contour, which for a continuous polymer is
\begin{equation} \label{eq:msd}
    \langle \Delta \bm r^2_{tot}(t) \rangle = \frac{1}{L} \int_{0}^{L} \lla \left(\bm r(s,t) - \bm r(s,0) \right)^2 \rra ds ,
\end{equation}
and comprises contributions from the center-of-mass motion, $\langle \Delta \bm r^2_{cm}(t) \rangle$, and the internal dynamics in center-of-mass reference frame, $\langle \Delta \bm r^2(t) \rangle$, such that 
\begin{equation} \label{eq:msd_sum}
    \langle \Delta \bm r^2_{tot} (t) \rangle = \langle \Delta \bm r^2_{cm}(t) \rangle + \langle \Delta \bm r^2(t) \rangle.
\end{equation}
Analogous definitions apply for the discrete polymer model. 

Primarily analytical results for continuous APLPs are presented, because appearing integrals and sums can often be evaluated analytically, whereas the sums in case of discrete APLPs cannot. However, numerically the discrete APLPs can be treated more rigorously than the continuous APLPs, because the latter require summation over a huge number of modes to achieve convergence, specifically at large Péclet numbers.

\subsection{Center-of-Mass Mean-Square Displacement}

Calculation of the center-of-mass mean-square displacement (CM-MSD)
\begin{align}
\langle \Delta \bm{r}_{cm}^2(t) \rangle = \langle (\bm{r}_{cm}(t) - \bm{r}_{cm}(0))^2 \rangle,    
\end{align}
with $\bm{r}_{cm}(t) = \int_{0}^{L} \bm{r}(s,t) ds / L $,  for discrete and continuous APLPs yields
\begin{widetext} 
    \begin{align}  
        \nonumber
        \lla \Delta \tilde{\bm{r}}_{cm}^2(t) \rra  = & \ \frac{2L^2}{\pi^2} \frac{\coth(d(N+1))}{\coth(d)} \frac{t}{\tilde \tau_R}  
        +
        \frac{12 k_BT}{(N+1)^2} \sum_{m=1}^{N} \frac{\bm{b}_m^{\dagger} \cdot \bm{b}_0^{\dagger}}{\txi_m} \sum_{i,j=0}^{N} b_m^{(i)} b_0^{(j)} \left(1 - e^{- \tilde \xi_m t/ \tilde{\gamma}} \right) 
        \nonumber
        \\
        & + \frac{12 k_BT }{(N+1) } \sum_{m,n=1}^{N}  \ \frac{\bm{b}_m^{\dagger} \cdot \bm{b}_n^{\dagger}}{\txi_m + \txi_n} \sum_{i,j=0}^{N} b_m^{(i)} b_n^{(j)} \left(1 - e^{- \tilde \xi_m t/ \tilde{\gamma}} \right) , \label{align:msd_cm_dis} 
        \\  
        \nonumber
        \lla \Delta\bm{r}_{cm}^2(t) \rra  = &  \ 6 D_R \pi r_c \coth(\pi r_c) \, t 
        \\    
         & + \ \frac{32r_c^3L}{\pi^3 e^{\pi r_c}\sinh(\pi r_c)p} \sum_{m=1}^{\infty} 
        \frac{m^2 \left(1-(-1)^me^{-\pi r_c} \right) \left(-1+(-1)^me^{3\pi r_c} \right)}{(m^2+r_c^2)^3(m^2+9r_c^2)}\left(1-e^{-t/\tau_m} \right) 
        \nonumber
        \\ 
         & + \ \frac{128r_c^3 L}{\pi^5 p} \sum_{m,n=1}^{\infty} 
        \frac{m^2n^2 \left(1-(-1)^me^{-\pi r_c} \right) \left(1-(-1)^ne^{-\pi r_c} \right) \left(-1+(-1)^{m+n}e^{2\pi r_c} \right)}{[(m-n)^2+4r_c^2][(m+n)^2+4r_c^2](m^2+r_c^2)^2(n^2+r_c^2)^2}  \left(1-e^{-t/\tau_m} \right) , \label{align:msd_cm_con}
    \end{align}
\end{widetext}
with the abbreviation $d$ of Sec.~\ref{sec:sol_eom_dis}, $r_c$ of Eq.~\eqref{eq:abb_c}, and the diffusion coefficient $D_R =k_BT/(\gamma L)$ of a continuous passive flexible polymer.\cite{doi_theory_1986,harnau_dynamic_1995} Evidently, the non-Hermitian nature of the equations of motion implies a coupling of the translational mode ($m=0$) with higher modes ($m \ge 1$) as well as a coupling between higher modes ($m,n \ge 1$).  
Thus, the CM-MSDs Eqs.~\eqref{align:msd_cm_dis} and \eqref{align:msd_cm_con} differ distinctively from those of passive \cite{doi_theory_1986} and active Brownian polymers, \cite{eisenstecken_conformational_2016} where the CM-MSD is solely determined by the translational mode.

\begin{figure}[t]
	\centering
	\includegraphics[width=\columnwidth]{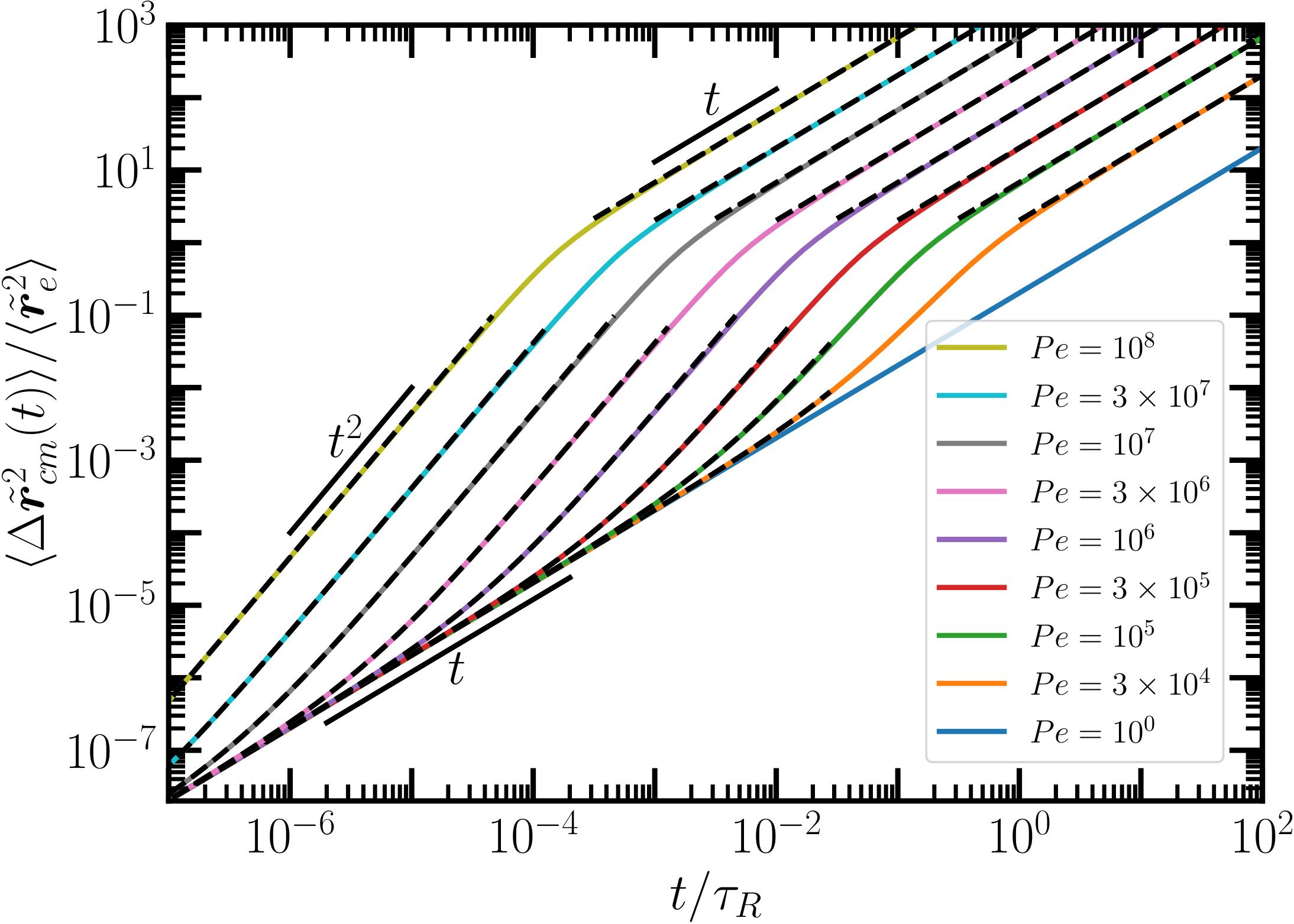}
	\caption{Normalized center-of-mass mean-square displacements $\langle \Delta \tilde{\bm{r}}_{cm}^2(t) \rangle$ as a function of the time $t/\tau_R$ for discrete APLPs of length $L/l =N=500$ and various Péclet numbers $Pe$ as indicated in the legend. The blue curve for $Pe=1$ represents the CM-MSD of a passive polymer. The black dashed lines show the approximations of the short time ballistic regime Eq.~\eqref{eq:MSD_cm_con_approximation} and the long time activity-enhanced diffusion Eq.~\eqref{eq:MSD_cm_con_long_time_approx} for continuous APLPs. The black solid lines indicate power laws.}
	\label{fig:MSD_cm}
\end{figure}

Figure~\ref{fig:MSD_cm} displays the CM-MSD of discrete APLPs and the comparison with the analytical results of the continuous model for various P\'eclet numbers. Evidently, the two representations yield the same results and discretization effects are of minor importance for this polymer length. In the limit $Pe \to 0$, the diffusive behavior of passive flexible polymer is found.\cite{doi_theory_1986,harnau_dynamic_1995} Clearly, three time regimes can be identified for $Pe>1$. Taylor expansion of Eq.~\eqref{align:msd_cm_con} for short times $t/\tau_R \ll 1$ and $Pe>0$ yields (SM, Sec.~S-IV A)
\begin{equation}
	\frac{\langle \Delta\bm{r}_{cm}^2(t)}{\langle \bm{r}_e^2 \rangle} = \frac{2}{\pi^2} \frac{t}{\tau_R} + \frac{Pe^2}{9 \pi^4 (pL)^2} \left( \frac{t}{\tau_R} \right)^2 ,
	\label{eq:MSD_cm_con_approximation}
\end{equation}	 
with the diffusive regime of a passive polymer for $t/\tau_R \ll 18 \pi^2 (pL)^2/Pe^2$, and
an active ballistic regime for $t/\tau_R \gg  18 \pi^2 (pL)^2/Pe^2$ and $Pe \gg 1$. Both terms in Eq.~\eqref{eq:MSD_cm_con_approximation} are determined by the internal polymer dynamics with all modes contributing. In particular, the term linear in $t$ in Eq.~\eqref{align:msd_cm_con} is cancelled by a similar term resulting from the sums in Eq. \eqref{align:msd_cm_con}. 
For $t/\tau_R \gg 3 \pi^2 pL /Pe$, the exponential terms, $e^{-t/\tau_m}$, in Eq.~\eqref{align:msd_cm_con} are negligible, and the time dependence of the CM-MSD is entirely determined by the linear term, i.e., the APLPs exhibit diffusive motion with the activity-enhanced diffusion coefficient 
\begin{equation}
	\frac{D}{D_R} = \frac{Pe}{6 pL} \coth\left(\frac{Pe}{6 pL} \right) = 
	\begin{cases}
		1 + \frac{1}{3} \left(\frac{Pe}{6 pL}\right)^2, & Pe \ll 1
		\\
		\frac{Pe}{6 pL},  & Pe \gg 1
	\end{cases}	
	,
	\label{eq:MSD_cm_con_long_time_approx}
\end{equation}	
with $D_R = k_BT/(\gamma L)$.\cite{peterson_statistical_2020} Similarly, the calculations for the discrete polymer model yield

\begin{equation}
	\frac{D}{D_R} = 
	\begin{cases} \displaystyle
        \frac{N \coth ( d(N+1))}{\coth ( d )} ,  & \displaystyle \frac{Pe}{6 N^2} < 1
		\\[10pt] \displaystyle
		\frac{N \tanh ( \tilde d(N+1))}{\tanh ( \tilde d )}
        ,  & \displaystyle \frac{Pe}{6 N^2} > 1, N \; {\mathrm{even}}
        \\[10pt] \displaystyle
		\frac{N \coth ( \tilde d(N+1))}{\tanh ( \tilde d )}
        ,  & \displaystyle \frac{Pe}{6 N^2} > 1, N \; {\mathrm{odd}}
	\end{cases}	,
	\label{eq:MSD_cm_dis_long_time_approx}
\end{equation}	
with $\tilde d = \ln (\sqrt{Pe+6 N^2}/\sqrt{Pe-6 N^2})$. For the discrete APLP model, there is a pronounced odd-even effect in terms of the bead number $N$ as long as $Pe/(6 N^2)>1$ and $\tilde d (N+1) \ll 1$. In the case $Pe/(6 N^2) \gg 1$, Taylor expansion yields
\begin{equation}
	\frac{D}{D_R} = 
	\begin{cases} \displaystyle
		N (N+1) ,  &  N \; {\mathrm{even}}
        \\[10pt] \displaystyle
		\frac{Pe^2}{36 N^3(N+1)}
        ,  &  N \; {\mathrm{odd}}
	\end{cases}	.
	\label{eq:MSD_cm_dis_diff_expan}
\end{equation}	
Hence, for $N$ even, the long-time diffusion coefficient is independent of activity, whereas for $N$ odd, $D$ shows a strong dependence on $Pe$. This is confirmed by numerical evaluation of Eq.~(S35) of the SM. The manifestation of the odd-even difference requires large activities, because $Pe/(6 N^3) \gg 1$, and is reached for rather large $Pe$ only, even for polymers with a moderate number of beads. Figure~S1 of the SM provides an example for the number of beads $N=50$ and $N=51$.  
In contrast, as long as $\tilde d (N+1) \gg 1$, $D/D_R = 1/\tanh \tilde d $ is independent of the odd-even nature of $N$, and $D/D_R =  Pe/(6 N)$ for $Pe/(6N^2) \gg 1$, with a $Pe$ and $N$ dependence comparable to that of Eq.~\eqref{eq:MSD_cm_con_long_time_approx}. 

The diffusion coefficient in Eq.~\eqref{eq:MSD_cm_con_long_time_approx} increases linearly with $Pe$ for $Pe \gg 1$.\cite{peterson_statistical_2020} This agrees with simulations of active filaments in two  \cite{isel:15} and three \cite{bianco_globulelike_2018} dimensions. Moreover, the long-time diffusion coefficient $D=D_R Pe/(6 pL) = f_a/(6 \gamma p) = f_al^2/(6 \tilde \gamma)$ is independent of the polymer length and depends linearly on the activity, as has also been found in Ref.~[\onlinecite{bianco_globulelike_2018}].
The linear $Pe$ dependence of $D$ (Eq.~\eqref{eq:MSD_cm_con_long_time_approx}) for $Pe \gg 1$ differs from that of $D$ of individual active Brownian particles (ABPs) and active Brownian polymers, which exhibit a quadratic $Pe$ dependence, \cite{howse_abp_2007,elge:15,eisenstecken_internal_2017,wink:20} reflecting the different underlying propulsion mechanisms. For APLPs, the active force on the center-of-mass is $\bm F_a^{cm} = f_a (\bm r_N - \bm r_0)/(N+1) = f_a \bm r_e/(N+1)$, hence, depends on the polymer conformations. In contrast, the propulsion force in ABPs and ABPOs is related to a solid-body rotation of an ABP, and, in the case of ABPOs, is independent of the polymer conformations.      

For shorter discrete polymers, the active diffusion with the diffusion coefficient in Eq.~\eqref{eq:MSD_cm_con_long_time_approx} and $1 \ll Pe \ll 6 N^3$ can be considered as motion with a constant velocity, $v$, along the contour of a flexible polymer. This assumption implies
\begin{equation}
\lla (\bm r_i(t) - \bm r_i(0))^2 \rra = l v t
\end{equation}
for any monomer, with $v= f_al/\tilde \gamma = Pe k_BT/(l \tilde \gamma N^2)$, and yields the diffusion coefficient $D/D_R=Pe/(6N)$ as in Eq.~\eqref{eq:MSD_cm_con_long_time_approx}. This argument is valid as long as the ratio $L/l=N$ between the polymer length and the persistence length ($l_p = 1/(2p) =l/2$) is not too large. It does not apply to the adopted continuum description, because in that case $pL \gg 1$. 

\begin{figure}[t]
	\centering
	\includegraphics[width=\columnwidth]{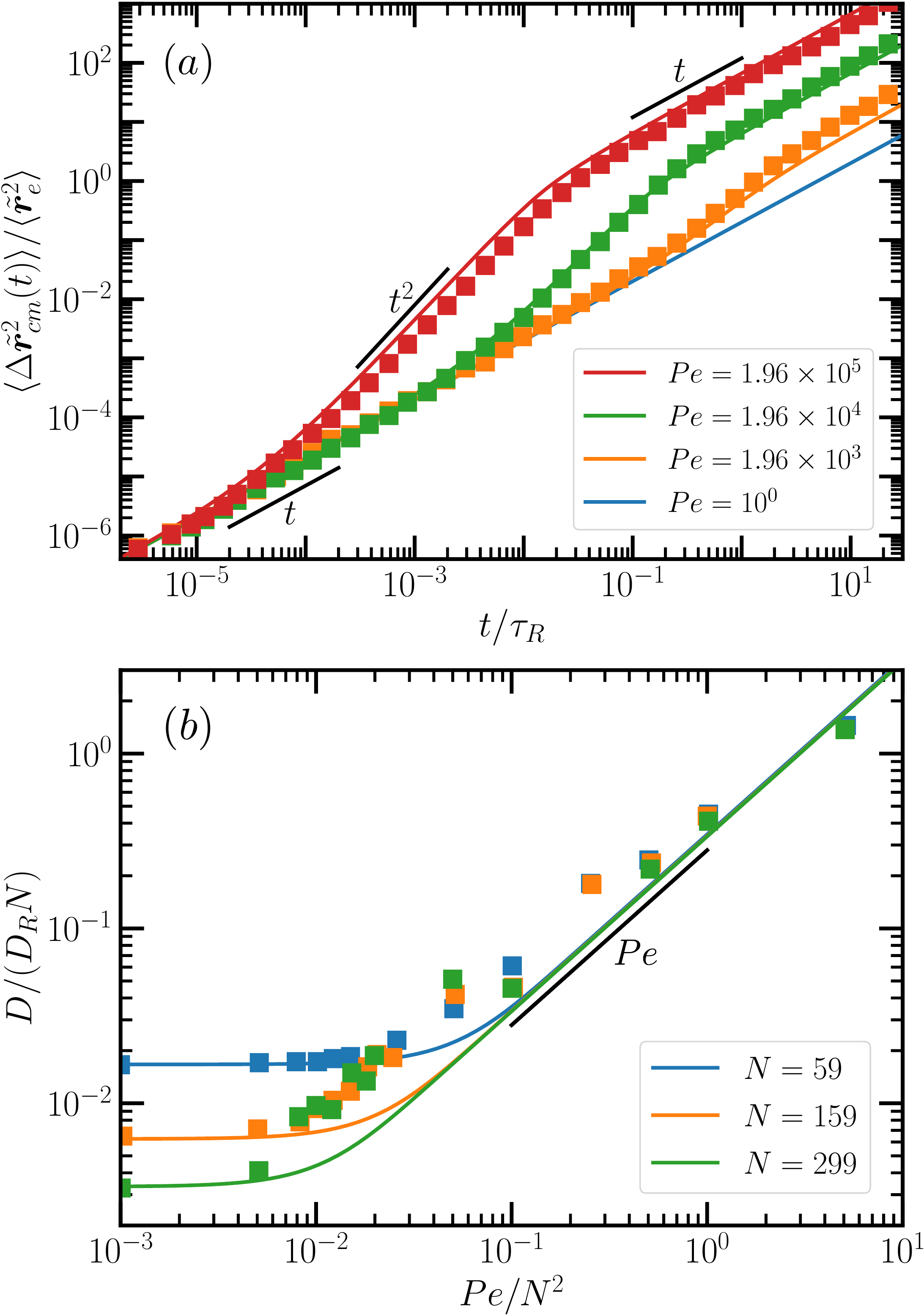}
	\caption{$(a)$ Polymer center-of-mass mean-square displacements as a function of the time $t/\tau_R$ for discrete APLPs of length $L/l=N=99$ and various Péclet numbers $Pe$ as indicated in the legend. The blue solid line for $Pe=1$ represents the CM-MSD of a passive polymer. $(b)$ Long-time polymer center-of-mass diffusion coefficients, $D$, normalized by $D_R N=k_BT/\tilde \gamma$ as a function of the activity  $Pe/N^2$. The symbols (squares) are simulation results for polymers with $(a)$ $100$, and $(b)$ $50$, $160$, and $300$ beads, taken from Ref.~[\onlinecite{bianco_globulelike_2018}], and the lines are calculated via $(a)$ Eq.~\eqref{eq:MSD_cm_dis_long_time_approx} and $(b)$ Eq.~\eqref{align:msd_cm_dis} for the same number of beads. The black solid lines indicate power laws for time $t$ and the Pélect number $Pe$, respectively.}
	\label{fig:diff_coeff_comparison}
\end{figure}

Figure~\ref{fig:diff_coeff_comparison} presents a comparison between simulation results of Ref.~[\onlinecite{bianco_globulelike_2018}] and our analytical expression \eqref{eq:MSD_cm_dis_long_time_approx} for the polymer CM-MSD and long-time diffusion coefficients. It is important to note that the diffusion coefficient of Eq.~\eqref{eq:MSD_cm_con_long_time_approx} is almost identical with Eq.~\eqref{eq:MSD_cm_dis_long_time_approx} of the discrete polymer over the considered range of P\'eclet numbers. The theoretical results capture the active force dependence obtained in simulations very well, both for the CM-MSD in Fig.~\ref{fig:diff_coeff_comparison}$(a)$ as well as the diffusion coefficients in Fig.~\ref{fig:diff_coeff_comparison}$(b)$. In our approach, the polymer conformations are independent of $Pe$, hence it does not show the shrinkage of active polymers as observed in simulations for $Pe/N^2>1$,\cite{bianco_globulelike_2018,anan:18} with a corresponding reduction in the CM-MSD and $D$. In general, the theoretical curves in Fig.~\ref{fig:diff_coeff_comparison}$(b)$ are somewhat shifted toward larger $Pe$, and multiplication with a factor of approximately $1.5$ yields better agreement, specifically for $N=59$. This quantitative difference might originate from the different models, e.g., no self-avoidance in our model, and the different implementation of the active force. 

In addition, the crossover time, $\tau_c = (\gamma /k_BT)L^3/Pe$, from the active ballistic to the active diffusive time regime, agrees with the scaling relation found by the simulations of   Ref.~[\onlinecite{bianco_globulelike_2018}].

\begin{figure}[t]
	\centering
	\includegraphics[width=\columnwidth]{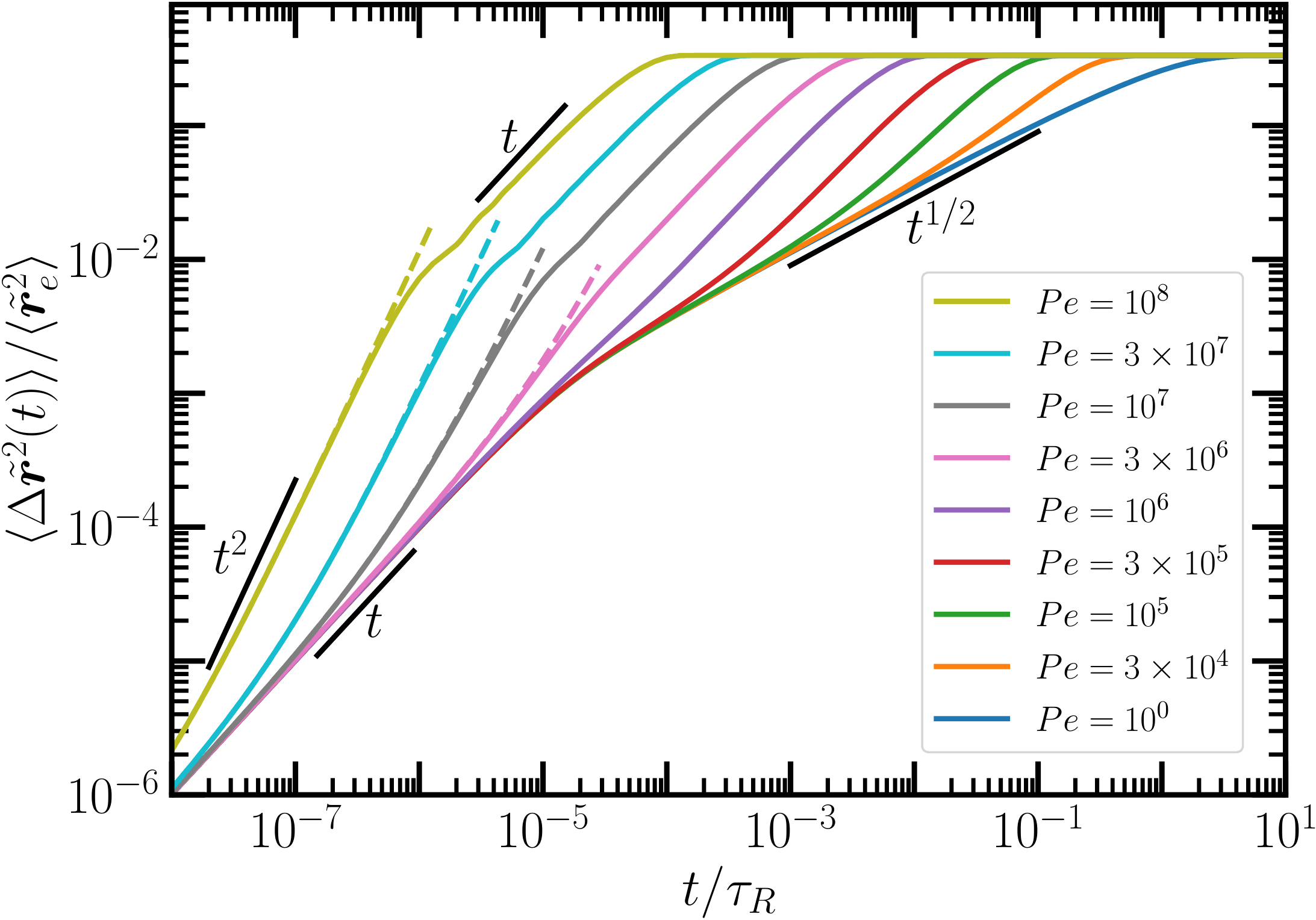}
	\caption{Normalized bead mean-square displacements in the center-of-mass reference frame $\langle \Delta \tilde{\bm{r}}^2(t) \rangle$ as a function of the time $t/\tau_R$ for a discrete polymer of length $L/l=N=500$ and various Péclet numbers $Pe$ as indicated in the legend. The colored dashed lines represent the second-order approximation of Eq. (S36). The black lines indicate power laws.}
	\label{fig:MSD_seg}
\end{figure}

\subsection{Mean-Square Displacement in the Center-of-Mass Reference Frame}

Figure~\ref{fig:MSD_seg} presents the contour-averaged mean-square displacement of the beads of the discrete polymer model in the center-of-mass reference frame,
\begin{align}
\langle \Delta \tilde {\bm r}^2(t)  \rangle = \frac{1}{N+1}\sum_{j=0}^N \lla \left(\Delta \tilde {\bm r}_j(t) - \Delta \tilde{\bm r}_j(0) \right)^2 \rra , 
\end{align}
with $\Delta \tilde{\bm{r}}_j(t) = \tilde{\bm r}_j(t) - \tilde{\bm r}_{cm}(t)$. Explicitly, it reads
\begin{align} \nonumber \label{eq:msd_seg_dis}
    \langle \Delta \tilde{\bm{r}}^2(t)\rangle = & \ \frac{12 k_BT}{(N+1)}   \sum_{m,n=1}^{N} \frac{\bm{b}_m^{\dagger} \cdot \bm{b}_n^{\dagger}}{\txi_m + \txi_n}  \left(1 - e^{-\tilde \xi_m t/ \tilde \gamma}  \right) \\ & \times
    \Big[\bm{b}_m \cdot \bm{b}_n - \frac{1}{N+1} \sum_{i,j=0}^{N} b_m^{(i)} b_n^{(j)} \Big]
\end{align}
in terms of the eigenvectors, Eqs.~\eqref{align:eigen_dis_b} and \eqref{align:eigen_dis_b_dag}.

In the limit $Pe \to 0$, the various time regimes well-known for flexible polymers are obtained, with a linear increase of $\langle \Delta \tilde {\bm r}^2(t) \rangle$ with increasing time for $t/\tilde \tau_R \ll 1/N^2$, $\langle \Delta \tilde {\bm r}^2(t) \rangle \sim \sqrt{t/\tilde \tau_R}$ in the interval $1/N^2 < t/\tilde \tau_R \ll 1$, and the plateau value $\langle \Delta \tilde {\bm r}^2(t) \rangle = 2 \langle \tilde{\bm r}_g^2 \rangle = \langle \tilde{\bm{r}}_e^2 \rangle / 3$ for $t/\tilde \tau_R \gg 1$ (cf. Eqs.~\eqref{align:msq_end_dis} and \eqref{align:msq_end_con}).

With increasing activity, gradually a linear time regime appears, and $\langle \Delta \tilde {\bm r}^2(t) \rangle$ growths more strongly than $t^{1/2}$ of the passive polymer. The crossover time, $t_c$, to the linear time regime depends on the P\'eclet number. As long as $Pe/(6 N^2) <1$, Eq.~(S30) of the continuous polymer model yields $t_c/\tau_R \approx (6 \pi N/Pe)^2$, consistent with Figure~\ref{fig:MSD_seg}. For $Pe/(6 N^2)>1$, the eigenvalues $\txi_m$ are complex. The linear time regime is determined by the difference of the $\cos(\omega_m t)$ and $\sin(\omega_m t)$ terms, which appear from the exponential $e^{i \omega_m t}$ of the imaginary part of the eigenvalue $\txi_m$. From the condition $\omega_1 t_c \approx 1$ for the largest frequency $\omega_1$, the crossover time  $t_c/\tilde \tau_R \approx 3 \pi^2/Pe$ is found, consistent with Figure~\ref{fig:MSD_seg}. For times $t/\tilde \tau_R \ll 3 \pi^2/Pe$, an active quadratic time regime is present. The dashed lines in Fig.~\ref{fig:MSD_seg} represent Eq.~\eqref{eq:msd_seg_dis} with the exponential function expanded up to second order in $\txi_m t/\tilde \gamma$. Here, the linear terms of the real and imaginary part of $\txi_m$ and their product are most important and determine the time dependence. 
Remarkably, the MSD is real, as expected for a physical quantity, despite imaginary eigenvalues, as shown in Sec.~S-II of the SM.
For times $t/\tilde \tau_R \gg \pi^2 N/Pe$, the plateau value $Nl^2/3$ is assumed, where the crossover time $t_c/\tilde \tau_R = \pi^2N/Pe$  follows from the condition $Nl^2/3=6 D t_c$, with $D$ in Eq.~\eqref{eq:MSD_cm_dis_long_time_approx}. 

As our numerical calculations show, the two terms in Eq.~(S30) of the SM cancel each other to some extent. This possess a major challenge in the evaluation of the sums and hampers the confirmation and interpretation of the observed linear time dependence. However, the $Pe$ dependence of the characteristic crossover time  $t_c/\tilde \tau_R = \pi^2N/Pe$ to the plateau regime is remarkable. This relation does not follow from the longest relaxation time, $(1 + (Pe/(6 \pi pL))^2)t/ \tau_R \ll 1$, which is significantly smaller than $t_c/\tilde \tau_R$ and exhibits a stronger dependence on $Pe$,  but is rather determined by the frequency $\omega_1$ in case of $Pe/(6 N^2) >1$.

\subsection{Total Bead Mean-Square Displacement}

\begin{figure}[t]
	\centering
	\includegraphics[width=\columnwidth]{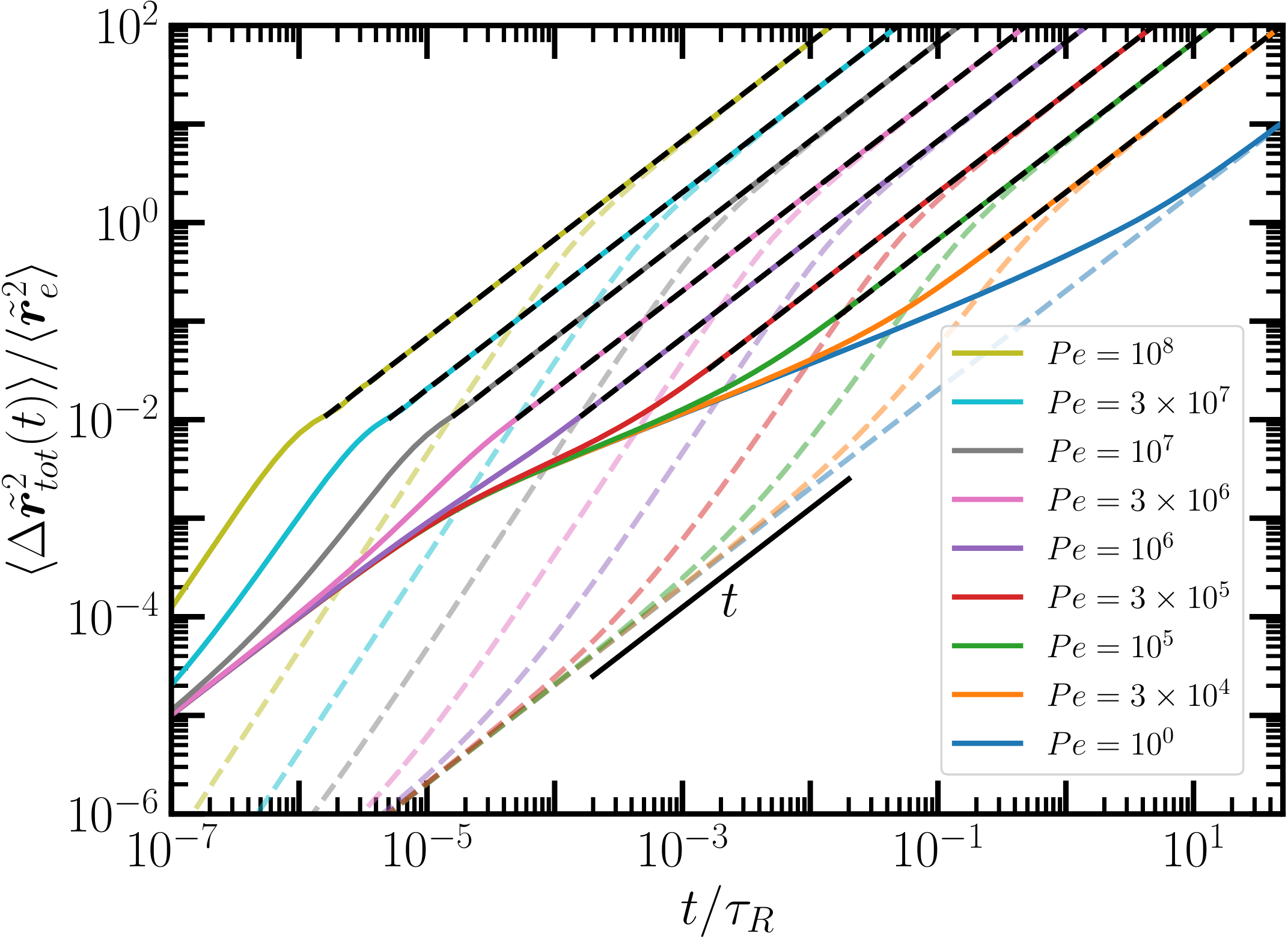}
	\caption{Normalized total bead mean-square displacements $\langle \Delta \bm{r}_{tot}^2(t) \rangle$ (solid lines) as a function of time $t/\tau_R$ for a discrete polymer of length $L/l=N=500$ and various Péclet numbers $Pe$ as indicated in the legend. The dashed lines present the corresponding center-of-mass MSDs, $\langle \Delta \tilde{\bm{r}}_{cm}^2(t) \rangle$. The black dashed lines indicate the active long-time diffusion with the diffusion coefficient of continuous polymers, Eq.~\eqref{eq:MSD_cm_con_long_time_approx}. The black solid line indicates a power law.}
	\label{fig:MSD_tot}
\end{figure}

The total MSD of the beads, Eqs.~\eqref{eq:msd} and \eqref{eq:msd_sum},
\begin{align} \nonumber \label{eq:msd_tot_dis}
	& \lla \Delta \tilde{\bm{r}}^2_{tot}(t) \rra  =   \frac{2N}{\pi^2} \frac{\coth(d(N+1))}{\coth(d)} \frac{t}{\tilde \tau_R}  \\ \nonumber & 
        +
        \frac{12 k_BT}{(N+1)^2N l^2 } \sum_{m=1}^{N} \frac{\bm{b}_m^{\dagger} \cdot \bm{b}_0^{\dagger}}{\txi_m} \sum_{i,j=0}^{N} b_m^{(i)} b_0^{(j)} \Big[1 - e^{- \tilde \xi_m t/ \tilde{\gamma}} \Big] 
        \\ & + \frac{12 k_BT}{(N+1)}   \sum_{m,n=1}^{N}  \frac{\bm{b}_m^{\dagger} \cdot \bm{b}_n^{\dagger}}{\txi_m + \txi_n}   
    \bm{b}_m \cdot \bm{b}_n  \left(1 - e^{-\tilde \xi_m t/ \tilde \gamma}  \right).
\end{align}
(see SM,  Eq.~(S33)) is displayed in Figure~\ref{fig:MSD_tot}. Evidently, $\langle \Delta \tilde {\bm r}^2_{tot}(t) \rangle$ is dominated by the bead MSD in the center-of-mass reference frame up to $\langle \Delta \tilde {\bm r}^2_{tot}(t) \rangle \approx \langle \tilde {\bm r}^2_{e} \rangle$. Strikingly, the linear time regime of $\langle \Delta \tilde {\bm r}^2(t) \rangle$ for $t/\tilde \tau_R \gg (6 \pi N/Pe)^2$ and $Pe/(6N^2) < 1$,  and $t/\tilde \tau_R \gg 3 \pi^2 /Pe$ and $Pe/(6N^2)>1$ joins smoothly with the respective regime in the CM-MSD, although $\langle \Delta \tilde {\bm r}^2 (t) \rangle$ assumes the plateau value $2 \langle \bm {\tilde r}_g^2 \rangle =  \langle \bm {\tilde r}_e^2 \rangle/3$, and the CM-MSD has not yet reached the long-time asymptotic value. As shown in Fig.~S1 of the SM, the contribution of the sums over modes $m, n \neq 0$ in Eq.~\eqref{eq:msd_tot_dis} to $\langle \Delta \tilde {\bm r}^2_{tot}(t) \rangle$ becomes smaller with increasing $Pe$ for $t/\tilde \tau_R \gg (6 \pi N/Pe)^2$ and $t/\tilde \tau_R \gg 3 \pi^2 /Pe$, respectively, and the contribution of the sums over modes with $m, n \neq 0$ assumes a time-independent value much smaller than $\langle \tilde {\bm r}_g^2 \rangle$. Thus, the total MSD is dominated by the term linear in time in Eq.~\eqref{eq:msd_tot_dis}.  
Contributions from the internal dynamics cancel in Eq.~\eqref{eq:msd_tot_dis} and the share from the mode $m=0$ prevails. At shorter times $t/\tilde \tau_R \ll (6 \pi N/Pe)^2$ for $Pe/(6 N^2) <1$, the passive polymer total MSD is assumed. Here, the sums over modes $m, n \neq 0$ determine the dynamics. In the case $Pe/(6 N^2)>1$ and $t/\tilde \tau_R \ll 3 \pi^2 /Pe$, cancellation of the linear terms in the sums over modes $m, n \neq 0$ yield a ballistic time regime, as for the MSD in the center-of-mass reference frame (Fig.~\ref{fig:MSD_seg}).

The MSD reflects a complex interdependence of the dynamics of the active beads. The active motion along the continuously changing polymer contour implies strong correlations in the bead translational motion, specifically on shorter time scales.       

The discussion of the MSD is focused on discrete APLPs. The numerical evaluation of the sums of modes of the continuous polymer model possess major challenges, since it requires summation over a huge number of modes, specifically for large $Pe$, and a high precision to compensate the large value of exponential factors containing $Pe/(6\pi pL)$. However, based on our continuum approximations for the CM-MSD, one can confidently state that the MSD in the center-of-mass reference frame of the continuous polymer model agrees with that of a discrete polymer model as long as $Pe \lesssim 10^6$ ($Pe/(6 N^2) <1$) and $t/\tilde \tau_R \gtrsim 1/N^2$. 

The total MSD of APLPs differs from that of active Brownian polymers, where the MSD is dominated by the center-of-mass motion at large P\'eclet numbers. \cite{eisenstecken_conformational_2016,eisenstecken_internal_2017} Hence, for APLPs the internal dynamics is more important for the overall displacement than for active Brownian polymers. \\

\FloatBarrier

\subsection{End-to-End Vector Correlation Function}

The temporal end-to-end vector correlation functions $\langle \tilde{\bm{r}}_e(t) \cdot \tilde{\bm{r}}_e(0) \rangle$ and $\langle {\bm{r}}_e(t) \cdot {\bm{r}}_e(0) \rangle$, normalized by their equilibrium values, are 
\begin{widetext}
    \begin{align}
    	 C_d(t) = & \ \frac{\lla \tilde{\bm{r}}_e(t) \cdot \tilde{\bm{r}}_e(0) \rra}{\lla \tilde{\bm r}_e^2 \rra}  =  6 k_BT  \sum_{m,n=1}^{N}  \frac{\bm{b}_m^{\dagger} \cdot \bm{b}_n^{\dagger}}{\txi_m+\txi_n} \left[b_m^{(N)} - b_m^{(0)}\right] \left[ b_n^{(N)} - b_n^{(0)} \right] e^{-\tilde \xi_m t/ \tilde \gamma} , \\ 
    C_c(t) = & \	\frac{\langle \bm{r}_e(t) \cdot \bm{r}_e(0) \rangle}{\langle \bm{r}_e^2 \rangle}
    	= \frac{16r_c}{\pi^3} 
    	\sum_{m,n=1}^{\infty}   \ \frac{m^2n^2 (-1 + (-1)^me^{-\pi r_c}) (-1 + (-1)^ne^{-\pi r_c}) (-1+(-1)^{m+n}e^{2\pi r_c})}{[(m-n)^2 + 4 r_c^2] [(m+n)^2 + 4r_c^2] (m^2+r_c^2) (n^2+r_c^2)} e^{-t/\tau_m} ,
    \end{align}
\end{widetext}
where $r_c=Pe/(6 \pi pL)$. 

\begin{figure}[t]
	\centering
    \includegraphics[width=\columnwidth]{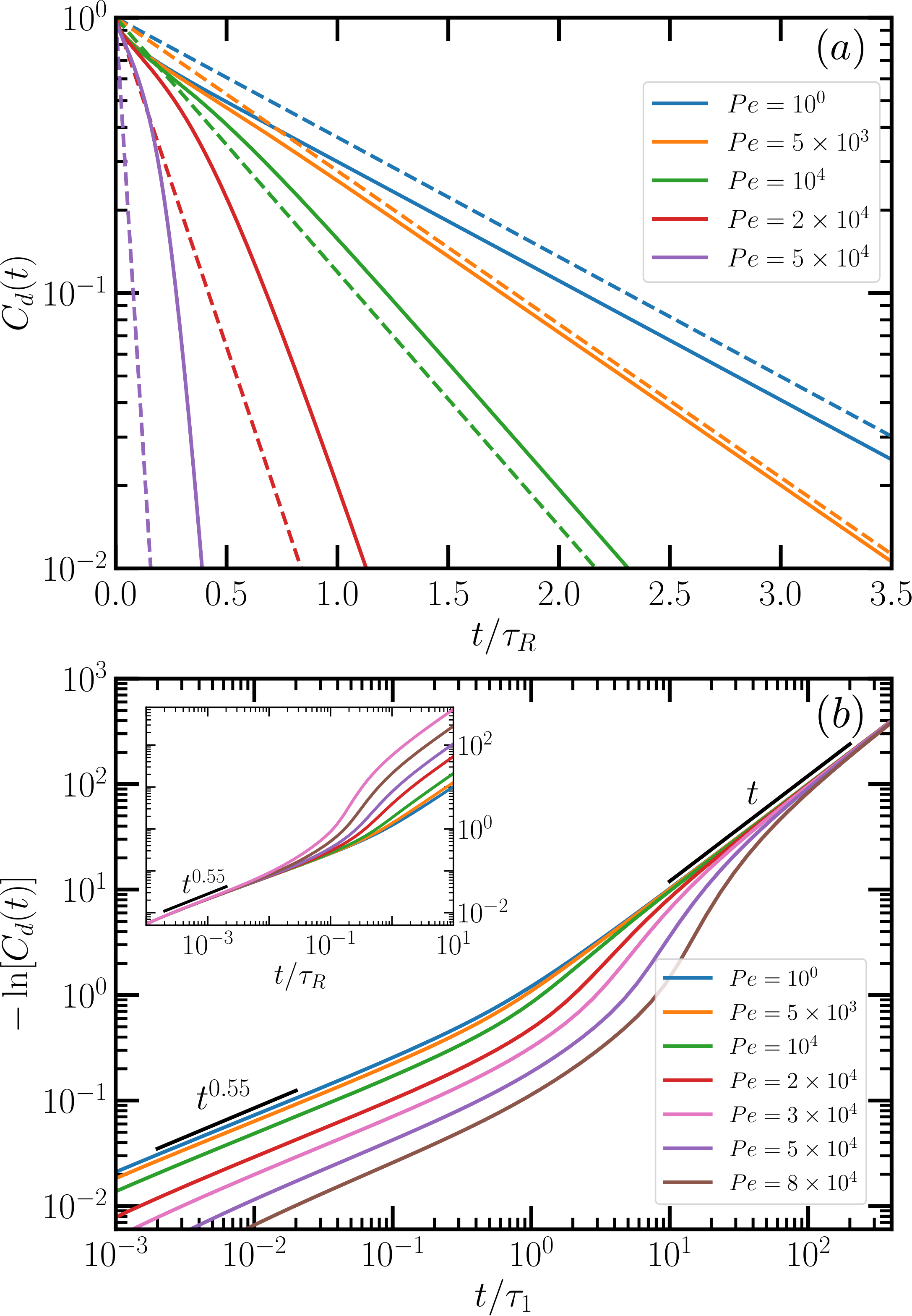}
    \caption{$(a)$ Semi-logarithmic representation of the normalized end-to-end vector correlation function $C_d(t) =\langle \tilde{\bm{r}}_e(t) \cdot \tilde{\bm{r}}_e(0) \rangle$ (solid lines) as a function of the time $t/ \tau_R$ for discrete APLPs of length $L/l=N=500$ and various Péclet numbers as indicated in the legend. The dashed lines represent the correlation functions including only the longest relaxation time $\tau_1$.
    $(b)$ Normalized end-to-end vector correlation function $-\ln\big[C_d(t)\big]$ as a function of $t/\tau_1$, where $\tau_1$ is the longest relaxation time, for various P\'eclet numbers. The inset shows $-\ln\big[C_d(t)\big]$ as a function of $t/\tau_R$. The short solid lines indicate power laws. 
    }
	\label{fig:end_end_corr}
\end{figure}

Figure~\ref{fig:end_end_corr} displays end-to-end vector correlation functions for various P\'eclet numbers. The correlation functions for the various $Pe$ numbers decay approximately exponentially for $t > \tau_R$, and approach the asymptotic behavior $C_d(t) \sim e^{-t/\tau_1}$ in the limit $t \to \infty$ (Fig.~\ref{fig:end_end_corr}$(b)$). In the limit $t \to 0$, all $C_d(t)$ curves approach the passive asymptotic time dependence (inset Fig.~\ref{fig:end_end_corr}$(b)$). The time dependence of the passive polymer crosses over from a linear regime for $t < \tau_R/N^2$ to a compressed exponential $- \ln (C_d(t)) \sim t^{0.55}$, and finally approaches the exponential decay. The correlation functions of APLPs exhibit a $Pe$-dependent crossover from the passive polymer behavior to the asymptotic exponential decay. They decay faster with increasing $Pe$, with a significantly slower decay at short times (Fig.~\ref{fig:end_end_corr}$(a)$). This is clearly visible when $C_d(t)$ is presented as a function of $t/\tau_1$, which shows a slower decay with increasing $Pe$ before the asymptotic exponential decay is assumed. Hence, activity leads to a faster decorrelation of the end-to-end vector in time due to the decreasing relaxation times with increasing activity. However, relative to the longest relaxation time, larger $Pe$ lead to a slower decay for $t/\tau_1 <1$.


\section{Summary and Conclusions} \label{sec:summary}

We have presented analytical results for the conformational and dynamical properties of discrete and continuous flexible Gaussian polymer models, where they are propelled by forces along bonds for the discrete and along the local tangent for the continuous polymer.
The propulsion forces imply non-symmetric/non-Hermitian eigenvalue equations, which are solved by an expansion into a biorthogonal basis set. The polar nature of the polymers leads to a mode coupling in the mode amplitude correlation functions. This is in contrast to active Brownian polymers (ABPOs),\cite{eisenstecken_internal_2017,wink:20} and gives rise to the emergence of distinctively different conformational and dynamical features.

Within the applied models, the polymer conformational properties are independent of the active forces and are identical to those of passive polymers. This is in contrast to ABPOs, which swell with increasing activity,\cite{eisenstecken_conformational_2016,eisenstecken_internal_2017,wink:20} and some computer simulations of tangentially driven active polymers, which reveal polymer shrinkage with increasing activity. \cite{isel:15,bianco_globulelike_2018,anan:18} However, in the latter cases, either a different propulsion force is considered, and/or excluded-volume interactions are additionally taken into account. This restricts a direct comparison of the various results. In any case, our results can be compared with simulations in three dimensions for not too large active forces.      

The non-symmetric matrix in the eigenvalue problem of the discrete APLPs model yields for $Pe/(6 N^2) <1$ real eigenvalues, whereas complex eigenvalues appear for $Pe/(6 N^2)>1$, with a single relaxation time and activity-dependent frequencies. These features of the active Gaussian model might appear also in simulations of slightly different models, since the respective P\'eclet numbers are easily reached. 

The coupling of modes leads to a maximum in the mode-amplitude autocorrelation function --- for a continuous polymer at the mode number $m_a = \mathrm{Integer}\big[Pe/(6N^2)\big]$. This requires, for the continuous polymer model, summation over an increasing number of modes with increasing $Pe$ to achieve convergence, and poses a major computational difficulty for large activities.

The polymer center-of-mass mean-square displacement, $\langle \Delta \bm r_{cm}^2(t) \rangle$, exhibits an active ballistic time regime for $\langle \Delta \bm r_{cm}^2(t) \rangle/\langle \bm r_e^2 \rangle <1$, followed by a diffusive regime with an activity-dependent diffusion coefficient, in agreement with simulations.\cite{isel:15,bianco_globulelike_2018} The effective velocity in the ballistic regime increase linearly with the P\'eclet number. Similarly, the effective diffusion coefficient increases linearly with $Pe$ and becomes independent of polymer length for $Pe \gg 1$, in agreement with results of computer simulations.\cite{bianco_globulelike_2018} The bead mean-square displacement in the center-of-mass reference frame also shows a ballistic time regime for $t/\tilde \tau_R \ll 3 \pi^2/Pe$ and $Pe/(6 N^2) \gg 1$. The contribution of the center-of-mass motion to the total bead mean-square displacement is negligible for $\langle \Delta \bm r_{tot}^2(t) \rangle/\langle \bm r_e^2 \rangle <1$, i.e., it is dominated by the active internal polymer dynamics. For $Pe \gg1$, the diffusive dynamics is quantitatively described by the diffusion coefficient of Eq.~\eqref{eq:MSD_cm_con_long_time_approx} even on length scales much smaller than $\langle \bm r_e^2 \rangle$.        

The dynamics of APLPs differs qualitatively from that of active Brownian polymers, where the center-of-mass motion dominates the overall polymer dynamics at large activities.\cite{eisenstecken_internal_2017} 
This reflects a complex interdependence of the dynamics of the active beads in APLPs. The active motion along the continuously change polymer contour implies strong correlations in the bead translational motion, specifically on shorter time scales. 

The active dynamics of APLPs exhibits similarities to the tank-treating motion observed for active polar ring polymers (APRPs),\cite{philipps_ring_2022} which reveal a  motion along the polymer contour. This is particularly pronounced for stiff rings. 
Thereby, the polar nature of the linear active polymer, with a nonzero overall active force, plays an important role and leads to different time regimes compared to flexible APRPs. In particular, the long-time MSD of APRPs is independent of activity and solely determined by thermal fluctuations. 

In Ref.~[\onlinecite{bianco_globulelike_2018}], the influence of the active force on the polymer center-of-mass motion is phenomenological described by representing it as a colored noise random process. This is similar to the active process in active Brownian polymers, whose beads or sites are indeed exposed to colored noise. \cite{eisenstecken_conformational_2016,wink:20}
In contrast, in our approach the dynamics of the mode-amplitudes (Eq.~\eqref{align:discrete_EOM_AMF}) is governed by white noise thermal fluctuations, and the mode correlation functions decay exponentially  (Eq.~\eqref{eq:msq_mode_dis}) as for a passive polymer. The complexity of the dynamical behavior results from the tight coupling of the various modes. The difference in the noise --- colored versus white noise --- is reflected, e.g., in the dependence of the ballistic motion and active long-time diffusion on the active force, which exhibit a distinct dependence on the P\'eclet number, namely, linear for APLPs and quadratic for ABPOs.          

Our analytical studies shed light onto the unique dynamical properties of APLPs, and provide theoretical insight into yet unrevealed dependencies on activity. This will be helpful in the interpretation of experimental findings as well as the design of functional active soft matter systems.       

\section*{SUPPLEMENTARY MATERIAL}
The supplementary material provides a derivation of the eigenvalues and eigenvectors of  discrete APLPs, as well as the proof that position vectors $\bm \tilde {\bm r}_j(t)$ are real. In addition, various definitions of conformational properties, displacements and correlation functions are given, and derivations of approximations for the short time center-of-mass dynamics are presented. 

\section*{AUTHOR DECLARATIONS}

\section*{Conflict of Interest}

The authors have no conflicts of interest to declare.

\section*{Data Availability Statement}

The data that support the findings of this study are available from the corresponding author upon reasonable request.



\begin{thebibliography}{100}%
\makeatletter
\providecommand \@ifxundefined [1]{%
 \@ifx{#1\undefined}
}%
\providecommand \@ifnum [1]{%
 \ifnum #1\expandafter \@firstoftwo
 \else \expandafter \@secondoftwo
 \fi
}%
\providecommand \@ifx [1]{%
 \ifx #1\expandafter \@firstoftwo
 \else \expandafter \@secondoftwo
 \fi
}%
\providecommand \natexlab [1]{#1}%
\providecommand \enquote  [1]{``#1''}%
\providecommand \bibnamefont  [1]{#1}%
\providecommand \bibfnamefont [1]{#1}%
\providecommand \citenamefont [1]{#1}%
\providecommand \href@noop [0]{\@secondoftwo}%
\providecommand \href [0]{\begingroup \@sanitize@url \@href}%
\providecommand \@href[1]{\@@startlink{#1}\@@href}%
\providecommand \@@href[1]{\endgroup#1\@@endlink}%
\providecommand \@sanitize@url [0]{\catcode `\\12\catcode `\$12\catcode
  `\&12\catcode `\#12\catcode `\^12\catcode `\_12\catcode `\%12\relax}%
\providecommand \@@startlink[1]{}%
\providecommand \@@endlink[0]{}%
\providecommand \url  [0]{\begingroup\@sanitize@url \@url }%
\providecommand \@url [1]{\endgroup\@href {#1}{\urlprefix }}%
\providecommand \urlprefix  [0]{URL }%
\providecommand \Eprint [0]{\href }%
\providecommand \doibase [0]{https://doi.org/}%
\providecommand \selectlanguage [0]{\@gobble}%
\providecommand \bibinfo  [0]{\@secondoftwo}%
\providecommand \bibfield  [0]{\@secondoftwo}%
\providecommand \translation [1]{[#1]}%
\providecommand \BibitemOpen [0]{}%
\providecommand \bibitemStop [0]{}%
\providecommand \bibitemNoStop [0]{.\EOS\space}%
\providecommand \EOS [0]{\spacefactor3000\relax}%
\providecommand \BibitemShut  [1]{\csname bibitem#1\endcsname}%
\let\auto@bib@innerbib\@empty
\bibitem [{\citenamefont {Demirel}(2010)}]{demi:10}%
  \BibitemOpen
  \bibfield  {author} {\bibinfo {author} {\bibfnamefont {Y.}~\bibnamefont
  {Demirel}},\ }\bibfield  {title} {\enquote {\bibinfo {title} {Nonequilibrium
  thermodynamics modeling of coupled biochemical cycles in living cells},}\
  }\href {https://doi.org/https://doi.org/10.1016/j.jnnfm.2010.02.006}
  {\bibfield  {journal} {\bibinfo  {journal} {J. Non-Newtonian Fluid Mech.}\
  }\textbf {\bibinfo {volume} {165}},\ \bibinfo {pages} {953} (\bibinfo {year}
  {2010})}\BibitemShut {NoStop}%
\bibitem [{\citenamefont {Fang}\ \emph {et~al.}(2019)\citenamefont {Fang},
  \citenamefont {Kruse}, \citenamefont {Lu},\ and\ \citenamefont
  {Wang}}]{fang:19}%
  \BibitemOpen
  \bibfield  {author} {\bibinfo {author} {\bibfnamefont {X.}~\bibnamefont
  {Fang}}, \bibinfo {author} {\bibfnamefont {K.}~\bibnamefont {Kruse}},
  \bibinfo {author} {\bibfnamefont {T.}~\bibnamefont {Lu}},\ and\ \bibinfo
  {author} {\bibfnamefont {J.}~\bibnamefont {Wang}},\ }\bibfield  {title}
  {\enquote {\bibinfo {title} {Nonequilibrium physics in biology},}\ }\href
  {https://doi.org/10.1103/RevModPhys.91.045004} {\bibfield  {journal}
  {\bibinfo  {journal} {Rev. Mod. Phys.}\ }\textbf {\bibinfo {volume} {91}},\
  \bibinfo {pages} {045004} (\bibinfo {year} {2019})}\BibitemShut {NoStop}%
\bibitem [{\citenamefont {Winkler}\ and\ \citenamefont
  {Gompper}(2020)}]{wink:20}%
  \BibitemOpen
  \bibfield  {author} {\bibinfo {author} {\bibfnamefont {R.~G.}\ \bibnamefont
  {Winkler}}\ and\ \bibinfo {author} {\bibfnamefont {G.}~\bibnamefont
  {Gompper}},\ }\bibfield  {title} {\enquote {\bibinfo {title} {The physics of
  active polymers and filaments},}\ }\href {https://doi.org/10.1063/5.0011466}
  {\bibfield  {journal} {\bibinfo  {journal} {J. Chem. Phys.}\ }\textbf
  {\bibinfo {volume} {153}},\ \bibinfo {pages} {040901} (\bibinfo {year}
  {2020})}\BibitemShut {NoStop}%
\bibitem [{\citenamefont {Kapral}(2013)}]{kapr:13}%
  \BibitemOpen
  \bibfield  {author} {\bibinfo {author} {\bibfnamefont {R.}~\bibnamefont
  {Kapral}},\ }\bibfield  {title} {\enquote {\bibinfo {title} {Perspective:
  Nanomotors without moving parts that propel themselves in solution},}\ }\href
  {https://doi.org/10.1063/1.4773981} {\bibfield  {journal} {\bibinfo
  {journal} {J. Chem. Phys.}\ }\textbf {\bibinfo {volume} {138}},\ \bibinfo
  {pages} {020901} (\bibinfo {year} {2013})}\BibitemShut {NoStop}%
\bibitem [{\citenamefont {Lau}\ \emph {et~al.}(2003)\citenamefont {Lau},
  \citenamefont {Hoffman}, \citenamefont {Davies}, \citenamefont {Crocker},\
  and\ \citenamefont {Lubensky}}]{lau:03}%
  \BibitemOpen
  \bibfield  {author} {\bibinfo {author} {\bibfnamefont {A.~W.~C.}\
  \bibnamefont {Lau}}, \bibinfo {author} {\bibfnamefont {B.~D.}\ \bibnamefont
  {Hoffman}}, \bibinfo {author} {\bibfnamefont {A.}~\bibnamefont {Davies}},
  \bibinfo {author} {\bibfnamefont {J.~C.}\ \bibnamefont {Crocker}},\ and\
  \bibinfo {author} {\bibfnamefont {T.~C.}\ \bibnamefont {Lubensky}},\
  }\bibfield  {title} {\enquote {\bibinfo {title} {Microrheology, stress
  fluctuations, and active behavior of living cells},}\ }\href
  {https://doi.org/10.1103/PhysRevLett.91.198101} {\bibfield  {journal}
  {\bibinfo  {journal} {Phys. Rev. Lett.}\ }\textbf {\bibinfo {volume} {91}},\
  \bibinfo {pages} {198101} (\bibinfo {year} {2003})}\BibitemShut {NoStop}%
\bibitem [{\citenamefont {Brangwynne}\ \emph {et~al.}(2008)\citenamefont
  {Brangwynne}, \citenamefont {Koenderink}, \citenamefont {MacKintosh},\ and\
  \citenamefont {Weitz}}]{bran:08}%
  \BibitemOpen
  \bibfield  {author} {\bibinfo {author} {\bibfnamefont {C.~P.}\ \bibnamefont
  {Brangwynne}}, \bibinfo {author} {\bibfnamefont {G.~H.}\ \bibnamefont
  {Koenderink}}, \bibinfo {author} {\bibfnamefont {F.~C.}\ \bibnamefont
  {MacKintosh}},\ and\ \bibinfo {author} {\bibfnamefont {D.~A.}\ \bibnamefont
  {Weitz}},\ }\bibfield  {title} {\enquote {\bibinfo {title} {Cytoplasmic
  diffusion: molecular motors mix it up},}\ }\href
  {https://doi.org/10.1083/jcb.200806149} {\bibfield  {journal} {\bibinfo
  {journal} {J. Cell. Biol.}\ }\textbf {\bibinfo {volume} {183}},\ \bibinfo
  {pages} {583} (\bibinfo {year} {2008})}\BibitemShut {NoStop}%
\bibitem [{\citenamefont {Robert}\ \emph {et~al.}(2010)\citenamefont {Robert},
  \citenamefont {Nguyen}, \citenamefont {Gallet},\ and\ \citenamefont
  {Wilhelm}}]{robe:10}%
  \BibitemOpen
  \bibfield  {author} {\bibinfo {author} {\bibfnamefont {D.}~\bibnamefont
  {Robert}}, \bibinfo {author} {\bibfnamefont {T.-H.}\ \bibnamefont {Nguyen}},
  \bibinfo {author} {\bibfnamefont {F.}~\bibnamefont {Gallet}},\ and\ \bibinfo
  {author} {\bibfnamefont {C.}~\bibnamefont {Wilhelm}},\ }\bibfield  {title}
  {\enquote {\bibinfo {title} {In vivo determination of fluctuating forces
  during endosome trafficking using a combination of active and passive
  microrheology},}\ }\href {https://doi.org/10.1371/journal.pone.0010046}
  {\bibfield  {journal} {\bibinfo  {journal} {PLOS ONE}\ }\textbf {\bibinfo
  {volume} {5}},\ \bibinfo {pages} {e10046} (\bibinfo {year}
  {2010})}\BibitemShut {NoStop}%
\bibitem [{\citenamefont {Fakhri}\ \emph {et~al.}(2014)\citenamefont {Fakhri},
  \citenamefont {Wessel}, \citenamefont {Willms}, \citenamefont {Pasquali},
  \citenamefont {Klopfenstein}, \citenamefont {MacKintosh},\ and\ \citenamefont
  {Schmidt}}]{fakh:14}%
  \BibitemOpen
  \bibfield  {author} {\bibinfo {author} {\bibfnamefont {N.}~\bibnamefont
  {Fakhri}}, \bibinfo {author} {\bibfnamefont {A.~D.}\ \bibnamefont {Wessel}},
  \bibinfo {author} {\bibfnamefont {C.}~\bibnamefont {Willms}}, \bibinfo
  {author} {\bibfnamefont {M.}~\bibnamefont {Pasquali}}, \bibinfo {author}
  {\bibfnamefont {D.~R.}\ \bibnamefont {Klopfenstein}}, \bibinfo {author}
  {\bibfnamefont {F.~C.}\ \bibnamefont {MacKintosh}},\ and\ \bibinfo {author}
  {\bibfnamefont {C.~F.}\ \bibnamefont {Schmidt}},\ }\bibfield  {title}
  {\enquote {\bibinfo {title} {High-resolution mapping of intracellular
  fluctuations using carbon nanotubes},}\ }\href
  {https://doi.org/10.1126/science.1250170} {\bibfield  {journal} {\bibinfo
  {journal} {Science}\ }\textbf {\bibinfo {volume} {344}},\ \bibinfo {pages}
  {1031} (\bibinfo {year} {2014})}\BibitemShut {NoStop}%
\bibitem [{\citenamefont {Guo}\ \emph {et~al.}(2014)\citenamefont {Guo},
  \citenamefont {Ehrlicher}, \citenamefont {Jensen}, \citenamefont {Renz},
  \citenamefont {Moore}, \citenamefont {Goldman}, \citenamefont
  {Lippincott-Schwartz}, \citenamefont {Mackintosh},\ and\ \citenamefont
  {Weitz}}]{guo:14}%
  \BibitemOpen
  \bibfield  {author} {\bibinfo {author} {\bibfnamefont {M.}~\bibnamefont
  {Guo}}, \bibinfo {author} {\bibfnamefont {A.~J.}\ \bibnamefont {Ehrlicher}},
  \bibinfo {author} {\bibfnamefont {M.~H.}\ \bibnamefont {Jensen}}, \bibinfo
  {author} {\bibfnamefont {M.}~\bibnamefont {Renz}}, \bibinfo {author}
  {\bibfnamefont {J.~R.}\ \bibnamefont {Moore}}, \bibinfo {author}
  {\bibfnamefont {R.~D.}\ \bibnamefont {Goldman}}, \bibinfo {author}
  {\bibfnamefont {J.}~\bibnamefont {Lippincott-Schwartz}}, \bibinfo {author}
  {\bibfnamefont {F.~C.}\ \bibnamefont {Mackintosh}},\ and\ \bibinfo {author}
  {\bibfnamefont {D.~A.}\ \bibnamefont {Weitz}},\ }\bibfield  {title} {\enquote
  {\bibinfo {title} {Probing the stochastic, motor-driven properties of the
  cytoplasm using force spectrum microscopy},}\ }\href
  {https://doi.org/https://doi.org/10.1016/j.cell.2014.06.051} {\bibfield
  {journal} {\bibinfo  {journal} {Cell}\ }\textbf {\bibinfo {volume} {158}},\
  \bibinfo {pages} {822} (\bibinfo {year} {2014})}\BibitemShut {NoStop}%
\bibitem [{\citenamefont {Parry}\ \emph {et~al.}(2014)\citenamefont {Parry},
  \citenamefont {Surovtsev}, \citenamefont {Cabeen}, \citenamefont {O'Hern},
  \citenamefont {Dufresne},\ and\ \citenamefont {Jacobs-Wagner}}]{parr:14}%
  \BibitemOpen
  \bibfield  {author} {\bibinfo {author} {\bibfnamefont {B.~R.}\ \bibnamefont
  {Parry}}, \bibinfo {author} {\bibfnamefont {I.~V.}\ \bibnamefont
  {Surovtsev}}, \bibinfo {author} {\bibfnamefont {M.~T.}\ \bibnamefont
  {Cabeen}}, \bibinfo {author} {\bibfnamefont {C.~S.}\ \bibnamefont {O'Hern}},
  \bibinfo {author} {\bibfnamefont {E.~R.}\ \bibnamefont {Dufresne}},\ and\
  \bibinfo {author} {\bibfnamefont {C.}~\bibnamefont {Jacobs-Wagner}},\
  }\bibfield  {title} {\enquote {\bibinfo {title} {The bacterial cytoplasm has
  glass-like properties and is fluidized by metabolic activity},}\ }\href
  {https://doi.org/https://doi.org/10.1016/j.cell.2013.11.028} {\bibfield
  {journal} {\bibinfo  {journal} {Cell}\ }\textbf {\bibinfo {volume} {156}},\
  \bibinfo {pages} {183} (\bibinfo {year} {2014})}\BibitemShut {NoStop}%
\bibitem [{\citenamefont {Golestanian}(2015)}]{gole:15}%
  \BibitemOpen
  \bibfield  {author} {\bibinfo {author} {\bibfnamefont {R.}~\bibnamefont
  {Golestanian}},\ }\bibfield  {title} {\enquote {\bibinfo {title} {Enhanced
  diffusion of enzymes that catalyze exothermic reactions},}\ }\href
  {https://doi.org/10.1103/PhysRevLett.115.108102} {\bibfield  {journal}
  {\bibinfo  {journal} {Phys. Rev. Lett.}\ }\textbf {\bibinfo {volume} {115}},\
  \bibinfo {pages} {108102} (\bibinfo {year} {2015})}\BibitemShut {NoStop}%
\bibitem [{\citenamefont {Mikhailov}\ and\ \citenamefont
  {Kapral}(2015)}]{mikh:15}%
  \BibitemOpen
  \bibfield  {author} {\bibinfo {author} {\bibfnamefont {A.~S.}\ \bibnamefont
  {Mikhailov}}\ and\ \bibinfo {author} {\bibfnamefont {R.}~\bibnamefont
  {Kapral}},\ }\bibfield  {title} {\enquote {\bibinfo {title} {Hydrodynamic
  collective effects of active protein machines in solution and lipid
  bilayers},}\ }\href {https://doi.org/10.1073/pnas.1506825112} {\bibfield
  {journal} {\bibinfo  {journal} {Proc. Natl. Acad. Sci. USA}\ }\textbf
  {\bibinfo {volume} {112}},\ \bibinfo {pages} {E3639} (\bibinfo {year}
  {2015})}\BibitemShut {NoStop}%
\bibitem [{\citenamefont {Weber}, \citenamefont {Spakowitz},\ and\
  \citenamefont {Theriot}(2012)}]{webe:12}%
  \BibitemOpen
  \bibfield  {author} {\bibinfo {author} {\bibfnamefont {S.~C.}\ \bibnamefont
  {Weber}}, \bibinfo {author} {\bibfnamefont {A.~J.}\ \bibnamefont
  {Spakowitz}},\ and\ \bibinfo {author} {\bibfnamefont {J.~A.}\ \bibnamefont
  {Theriot}},\ }\bibfield  {title} {\enquote {\bibinfo {title} {Nonthermal
  {ATP}-dependent fluctuations contribute to the in vivo motion of chromosomal
  loci},}\ }\href {https://doi.org/10.1073/pnas.1119505109} {\bibfield
  {journal} {\bibinfo  {journal} {Proc. Natl. Acad. Sci. USA}\ }\textbf
  {\bibinfo {volume} {109}},\ \bibinfo {pages} {7338} (\bibinfo {year}
  {2012})}\BibitemShut {NoStop}%
\bibitem [{\citenamefont {Kapral}\ and\ \citenamefont
  {Mikhailov}(2016)}]{kapr:16}%
  \BibitemOpen
  \bibfield  {author} {\bibinfo {author} {\bibfnamefont {R.}~\bibnamefont
  {Kapral}}\ and\ \bibinfo {author} {\bibfnamefont {A.~S.}\ \bibnamefont
  {Mikhailov}},\ }\bibfield  {title} {\enquote {\bibinfo {title} {Stirring a
  fluid at low {R}eynolds numbers: Hydrodynamic collective effects of active
  proteins in biological cells},}\ }\href
  {https://doi.org/10.1016/j.physd.2015.10.024} {\bibfield  {journal} {\bibinfo
   {journal} {Physica D}\ }\textbf {\bibinfo {volume} {318-319}},\ \bibinfo
  {pages} {100} (\bibinfo {year} {2016})}\BibitemShut {NoStop}%
\bibitem [{\citenamefont {Gnesotto}\ \emph {et~al.}(2018)\citenamefont
  {Gnesotto}, \citenamefont {Mura}, \citenamefont {Gladrow},\ and\
  \citenamefont {Broedersz}}]{gnes:18}%
  \BibitemOpen
  \bibfield  {author} {\bibinfo {author} {\bibfnamefont {F.~S.}\ \bibnamefont
  {Gnesotto}}, \bibinfo {author} {\bibfnamefont {F.}~\bibnamefont {Mura}},
  \bibinfo {author} {\bibfnamefont {J.}~\bibnamefont {Gladrow}},\ and\ \bibinfo
  {author} {\bibfnamefont {C.~P.}\ \bibnamefont {Broedersz}},\ }\bibfield
  {title} {\enquote {\bibinfo {title} {Broken detailed balance and
  non-equilibrium dynamics in living systems: a review},}\ }\href
  {https://doi.org/10.1088/1361-6633/aab3ed} {\bibfield  {journal} {\bibinfo
  {journal} {Rep. Prog. Phys.}\ }\textbf {\bibinfo {volume} {81}},\ \bibinfo
  {pages} {066601} (\bibinfo {year} {2018})}\BibitemShut {NoStop}%
\bibitem [{\citenamefont {Wu}\ \emph {et~al.}(2019)\citenamefont {Wu},
  \citenamefont {Japaridze}, \citenamefont {Zheng}, \citenamefont {Wiktor},
  \citenamefont {Kerssemakers},\ and\ \citenamefont {Dekker}}]{wu:19}%
  \BibitemOpen
  \bibfield  {author} {\bibinfo {author} {\bibfnamefont {F.}~\bibnamefont
  {Wu}}, \bibinfo {author} {\bibfnamefont {A.}~\bibnamefont {Japaridze}},
  \bibinfo {author} {\bibfnamefont {X.}~\bibnamefont {Zheng}}, \bibinfo
  {author} {\bibfnamefont {J.}~\bibnamefont {Wiktor}}, \bibinfo {author}
  {\bibfnamefont {J.~W.~J.}\ \bibnamefont {Kerssemakers}},\ and\ \bibinfo
  {author} {\bibfnamefont {C.}~\bibnamefont {Dekker}},\ }\bibfield  {title}
  {\enquote {\bibinfo {title} {Direct imaging of the circular chromosome in a
  live bacterium},}\ }\href {https://doi.org/10.1038/s41467-019-10221-0}
  {\bibfield  {journal} {\bibinfo  {journal} {Nat. Commun.}\ }\textbf {\bibinfo
  {volume} {10}},\ \bibinfo {pages} {2194} (\bibinfo {year}
  {2019})}\BibitemShut {NoStop}%
\bibitem [{\citenamefont {MacKintosh}\ and\ \citenamefont
  {Levine}(2008)}]{mack:08}%
  \BibitemOpen
  \bibfield  {author} {\bibinfo {author} {\bibfnamefont {F.~C.}\ \bibnamefont
  {MacKintosh}}\ and\ \bibinfo {author} {\bibfnamefont {A.~J.}\ \bibnamefont
  {Levine}},\ }\bibfield  {title} {\enquote {\bibinfo {title} {Nonequilibrium
  mechanics and dynamics of motor-activated gels},}\ }\href
  {https://doi.org/10.1103/PhysRevLett.100.018104} {\bibfield  {journal}
  {\bibinfo  {journal} {Phys. Rev. Lett.}\ }\textbf {\bibinfo {volume} {100}},\
  \bibinfo {pages} {018104} (\bibinfo {year} {2008})}\BibitemShut {NoStop}%
\bibitem [{\citenamefont {Lu}\ \emph {et~al.}(2016)\citenamefont {Lu},
  \citenamefont {Winding}, \citenamefont {Lakonishok}, \citenamefont
  {Wildonger},\ and\ \citenamefont {Gelfand}}]{lu:16}%
  \BibitemOpen
  \bibfield  {author} {\bibinfo {author} {\bibfnamefont {W.}~\bibnamefont
  {Lu}}, \bibinfo {author} {\bibfnamefont {M.}~\bibnamefont {Winding}},
  \bibinfo {author} {\bibfnamefont {M.}~\bibnamefont {Lakonishok}}, \bibinfo
  {author} {\bibfnamefont {J.}~\bibnamefont {Wildonger}},\ and\ \bibinfo
  {author} {\bibfnamefont {V.~I.}\ \bibnamefont {Gelfand}},\ }\bibfield
  {title} {\enquote {\bibinfo {title} {Microtubule--microtubule sliding by
  kinesin-1 is essential for normal cytoplasmic streaming in
  \emph{{D}rosophila} oocytes},}\ }\href
  {https://doi.org/10.1073/pnas.1522424113} {\bibfield  {journal} {\bibinfo
  {journal} {Proc. Natl. Acad. Sci. USA}\ }\textbf {\bibinfo {volume} {113}},\
  \bibinfo {pages} {E4995} (\bibinfo {year} {2016})}\BibitemShut {NoStop}%
\bibitem [{\citenamefont {Ravichandran}\ \emph {et~al.}(2017)\citenamefont
  {Ravichandran}, \citenamefont {Vliegenthart}, \citenamefont {Saggiorato},
  \citenamefont {Auth},\ and\ \citenamefont {Gompper}}]{ravi:17}%
  \BibitemOpen
  \bibfield  {author} {\bibinfo {author} {\bibfnamefont {A.}~\bibnamefont
  {Ravichandran}}, \bibinfo {author} {\bibfnamefont {G.~A.}\ \bibnamefont
  {Vliegenthart}}, \bibinfo {author} {\bibfnamefont {G.}~\bibnamefont
  {Saggiorato}}, \bibinfo {author} {\bibfnamefont {T.}~\bibnamefont {Auth}},\
  and\ \bibinfo {author} {\bibfnamefont {G.}~\bibnamefont {Gompper}},\
  }\bibfield  {title} {\enquote {\bibinfo {title} {Enhanced dynamics of
  confined cytoskeletal filaments driven by asymmetric motors},}\ }\href
  {https://doi.org/https://doi.org/10.1016/j.bpj.2017.07.016} {\bibfield
  {journal} {\bibinfo  {journal} {Biophys. J.}\ }\textbf {\bibinfo {volume}
  {113}},\ \bibinfo {pages} {1121} (\bibinfo {year} {2017})}\BibitemShut
  {NoStop}%
\bibitem [{\citenamefont {Guthold}\ \emph {et~al.}(1999)\citenamefont
  {Guthold}, \citenamefont {Zhu}, \citenamefont {Rivetti}, \citenamefont
  {Yang}, \citenamefont {Thomson}, \citenamefont {Kasas}, \citenamefont
  {Hansma}, \citenamefont {Smith}, \citenamefont {Hansma},\ and\ \citenamefont
  {Bustamante}}]{guth:99}%
  \BibitemOpen
  \bibfield  {author} {\bibinfo {author} {\bibfnamefont {M.}~\bibnamefont
  {Guthold}}, \bibinfo {author} {\bibfnamefont {X.}~\bibnamefont {Zhu}},
  \bibinfo {author} {\bibfnamefont {C.}~\bibnamefont {Rivetti}}, \bibinfo
  {author} {\bibfnamefont {G.}~\bibnamefont {Yang}}, \bibinfo {author}
  {\bibfnamefont {N.~H.}\ \bibnamefont {Thomson}}, \bibinfo {author}
  {\bibfnamefont {S.}~\bibnamefont {Kasas}}, \bibinfo {author} {\bibfnamefont
  {H.~G.}\ \bibnamefont {Hansma}}, \bibinfo {author} {\bibfnamefont
  {B.}~\bibnamefont {Smith}}, \bibinfo {author} {\bibfnamefont {P.~K.}\
  \bibnamefont {Hansma}},\ and\ \bibinfo {author} {\bibfnamefont
  {C.}~\bibnamefont {Bustamante}},\ }\bibfield  {title} {\enquote {\bibinfo
  {title} {Direct observation of one-dimensional diffusion and transcription by
  \emph{{E}scherichia coli} {RNA} polymerase},}\ }\href
  {https://doi.org/https://doi.org/10.1016/S0006-3495(99)77067-0} {\bibfield
  {journal} {\bibinfo  {journal} {Biophys. J.}\ }\textbf {\bibinfo {volume}
  {77}},\ \bibinfo {pages} {2284} (\bibinfo {year} {1999})}\BibitemShut
  {NoStop}%
\bibitem [{\citenamefont {Mejia}, \citenamefont {Nudler},\ and\ \citenamefont
  {Bustamante}(2015)}]{meji:15}%
  \BibitemOpen
  \bibfield  {author} {\bibinfo {author} {\bibfnamefont {Y.~X.}\ \bibnamefont
  {Mejia}}, \bibinfo {author} {\bibfnamefont {E.}~\bibnamefont {Nudler}},\ and\
  \bibinfo {author} {\bibfnamefont {C.}~\bibnamefont {Bustamante}},\ }\bibfield
   {title} {\enquote {\bibinfo {title} {Trigger loop folding determines
  transcription rate of \emph{{E}scherichia coli's} {RNA} polymerase},}\ }\href
  {https://doi.org/10.1073/pnas.1421067112} {\bibfield  {journal} {\bibinfo
  {journal} {Proc. Natl. Acad. Sci. USA}\ }\textbf {\bibinfo {volume} {112}},\
  \bibinfo {pages} {743} (\bibinfo {year} {2015})}\BibitemShut {NoStop}%
\bibitem [{\citenamefont {Belitsky}\ and\ \citenamefont
  {Sch{\"u}tz}(2019)}]{beli:19}%
  \BibitemOpen
  \bibfield  {author} {\bibinfo {author} {\bibfnamefont {V.}~\bibnamefont
  {Belitsky}}\ and\ \bibinfo {author} {\bibfnamefont {G.~M.}\ \bibnamefont
  {Sch{\"u}tz}},\ }\bibfield  {title} {\enquote {\bibinfo {title} {Stationary
  {RNA} polymerase fluctuations during transcription elongation},}\ }\href
  {https://doi.org/10.1103/PhysRevE.99.012405} {\bibfield  {journal} {\bibinfo
  {journal} {Phys. Rev. E}\ }\textbf {\bibinfo {volume} {99}},\ \bibinfo
  {pages} {012405} (\bibinfo {year} {2019})}\BibitemShut {NoStop}%
\bibitem [{\citenamefont {Javer}\ \emph {et~al.}(2013)\citenamefont {Javer},
  \citenamefont {Long}, \citenamefont {Nugent}, \citenamefont {Grisi},
  \citenamefont {Siriwatwetchakul}, \citenamefont {Dorfman}, \citenamefont
  {Cicuta},\ and\ \citenamefont {Cosentino~Lagomarsino}}]{jave:13}%
  \BibitemOpen
  \bibfield  {author} {\bibinfo {author} {\bibfnamefont {A.}~\bibnamefont
  {Javer}}, \bibinfo {author} {\bibfnamefont {Z.}~\bibnamefont {Long}},
  \bibinfo {author} {\bibfnamefont {E.}~\bibnamefont {Nugent}}, \bibinfo
  {author} {\bibfnamefont {M.}~\bibnamefont {Grisi}}, \bibinfo {author}
  {\bibfnamefont {K.}~\bibnamefont {Siriwatwetchakul}}, \bibinfo {author}
  {\bibfnamefont {K.~D.}\ \bibnamefont {Dorfman}}, \bibinfo {author}
  {\bibfnamefont {P.}~\bibnamefont {Cicuta}},\ and\ \bibinfo {author}
  {\bibfnamefont {M.}~\bibnamefont {Cosentino~Lagomarsino}},\ }\bibfield
  {title} {\enquote {\bibinfo {title} {Short-time movement of {{\em {E} coli}}
  chromosomal loci depends on coordinate and subcellular localization},}\
  }\href {https://doi.org/10.1038/ncomms3003} {\bibfield  {journal} {\bibinfo
  {journal} {Nat. Commun.}\ }\textbf {\bibinfo {volume} {4}},\ \bibinfo {pages}
  {3003} (\bibinfo {year} {2013})}\BibitemShut {NoStop}%
\bibitem [{\citenamefont {Zidovska}, \citenamefont {Weitz},\ and\ \citenamefont
  {Mitchison}(2013)}]{zido:13}%
  \BibitemOpen
  \bibfield  {author} {\bibinfo {author} {\bibfnamefont {A.}~\bibnamefont
  {Zidovska}}, \bibinfo {author} {\bibfnamefont {D.~A.}\ \bibnamefont
  {Weitz}},\ and\ \bibinfo {author} {\bibfnamefont {T.~J.}\ \bibnamefont
  {Mitchison}},\ }\bibfield  {title} {\enquote {\bibinfo {title} {Micron-scale
  coherence in interphase chromatin dynamics},}\ }\href
  {http://www.pnas.org/content/110/39/15555.abstract} {\bibfield  {journal}
  {\bibinfo  {journal} {Proc. Natl. Acad. Sci. USA}\ }\textbf {\bibinfo
  {volume} {110}},\ \bibinfo {pages} {15555} (\bibinfo {year}
  {2013})}\BibitemShut {NoStop}%
\bibitem [{\citenamefont {Lieberman-Aiden}\ \emph {et~al.}(2009)\citenamefont
  {Lieberman-Aiden}, \citenamefont {van Berkum}, \citenamefont {Williams},
  \citenamefont {Imakaev}, \citenamefont {Ragoczy}, \citenamefont {Telling},
  \citenamefont {Amit}, \citenamefont {Lajoie}, \citenamefont {Sabo},
  \citenamefont {Dorschner}, \citenamefont {Sandstrom}, \citenamefont
  {Bernstein}, \citenamefont {Bender}, \citenamefont {Groudine}, \citenamefont
  {Gnirke}, \citenamefont {Stamatoyannopoulos}, \citenamefont {Mirny},
  \citenamefont {Lander},\ and\ \citenamefont {Dekker}}]{lieb:09}%
  \BibitemOpen
  \bibfield  {author} {\bibinfo {author} {\bibfnamefont {E.}~\bibnamefont
  {Lieberman-Aiden}}, \bibinfo {author} {\bibfnamefont {N.~L.}\ \bibnamefont
  {van Berkum}}, \bibinfo {author} {\bibfnamefont {L.}~\bibnamefont
  {Williams}}, \bibinfo {author} {\bibfnamefont {M.}~\bibnamefont {Imakaev}},
  \bibinfo {author} {\bibfnamefont {T.}~\bibnamefont {Ragoczy}}, \bibinfo
  {author} {\bibfnamefont {A.}~\bibnamefont {Telling}}, \bibinfo {author}
  {\bibfnamefont {I.}~\bibnamefont {Amit}}, \bibinfo {author} {\bibfnamefont
  {B.~R.}\ \bibnamefont {Lajoie}}, \bibinfo {author} {\bibfnamefont {P.~J.}\
  \bibnamefont {Sabo}}, \bibinfo {author} {\bibfnamefont {M.~O.}\ \bibnamefont
  {Dorschner}}, \bibinfo {author} {\bibfnamefont {R.}~\bibnamefont
  {Sandstrom}}, \bibinfo {author} {\bibfnamefont {B.}~\bibnamefont
  {Bernstein}}, \bibinfo {author} {\bibfnamefont {M.~A.}\ \bibnamefont
  {Bender}}, \bibinfo {author} {\bibfnamefont {M.}~\bibnamefont {Groudine}},
  \bibinfo {author} {\bibfnamefont {A.}~\bibnamefont {Gnirke}}, \bibinfo
  {author} {\bibfnamefont {J.}~\bibnamefont {Stamatoyannopoulos}}, \bibinfo
  {author} {\bibfnamefont {L.~A.}\ \bibnamefont {Mirny}}, \bibinfo {author}
  {\bibfnamefont {E.~S.}\ \bibnamefont {Lander}},\ and\ \bibinfo {author}
  {\bibfnamefont {J.}~\bibnamefont {Dekker}},\ }\bibfield  {title} {\enquote
  {\bibinfo {title} {Comprehensive mapping of long-range interactions reveals
  folding principles of the human genome},}\ }\href
  {https://doi.org/10.1126/science.1181369} {\bibfield  {journal} {\bibinfo
  {journal} {Science}\ }\textbf {\bibinfo {volume} {326}},\ \bibinfo {pages}
  {289} (\bibinfo {year} {2009})}\BibitemShut {NoStop}%
\bibitem [{\citenamefont {Cremer}\ \emph {et~al.}(2015)\citenamefont {Cremer},
  \citenamefont {Cremer}, \citenamefont {H{\"u}bner}, \citenamefont
  {Strickfaden}, \citenamefont {Smeets}, \citenamefont {Popken}, \citenamefont
  {Sterr}, \citenamefont {Markaki}, \citenamefont {Rippe},\ and\ \citenamefont
  {Cremer}}]{crem:15}%
  \BibitemOpen
  \bibfield  {author} {\bibinfo {author} {\bibfnamefont {T.}~\bibnamefont
  {Cremer}}, \bibinfo {author} {\bibfnamefont {M.}~\bibnamefont {Cremer}},
  \bibinfo {author} {\bibfnamefont {B.}~\bibnamefont {H{\"u}bner}}, \bibinfo
  {author} {\bibfnamefont {H.}~\bibnamefont {Strickfaden}}, \bibinfo {author}
  {\bibfnamefont {D.}~\bibnamefont {Smeets}}, \bibinfo {author} {\bibfnamefont
  {J.}~\bibnamefont {Popken}}, \bibinfo {author} {\bibfnamefont
  {M.}~\bibnamefont {Sterr}}, \bibinfo {author} {\bibfnamefont
  {Y.}~\bibnamefont {Markaki}}, \bibinfo {author} {\bibfnamefont
  {K.}~\bibnamefont {Rippe}},\ and\ \bibinfo {author} {\bibfnamefont
  {C.}~\bibnamefont {Cremer}},\ }\bibfield  {title} {\enquote {\bibinfo {title}
  {The 4d nucleome: Evidence for a dynamic nuclear landscape based on
  co-aligned active and inactive nuclear compartments},}\ }\href
  {https://doi.org/10.1016/j.febslet.2015.05.037} {\bibfield  {journal}
  {\bibinfo  {journal} {FEBS Lett.}\ }\textbf {\bibinfo {volume} {589}},\
  \bibinfo {pages} {2931} (\bibinfo {year} {2015})}\BibitemShut {NoStop}%
\bibitem [{\citenamefont {Solovei}, \citenamefont {Thanisch},\ and\
  \citenamefont {Feodorova}(2016)}]{solo:16}%
  \BibitemOpen
  \bibfield  {author} {\bibinfo {author} {\bibfnamefont {I.}~\bibnamefont
  {Solovei}}, \bibinfo {author} {\bibfnamefont {K.}~\bibnamefont {Thanisch}},\
  and\ \bibinfo {author} {\bibfnamefont {Y.}~\bibnamefont {Feodorova}},\
  }\bibfield  {title} {\enquote {\bibinfo {title} {How to rule the nucleus:
  divide et impera},}\ }\bibfield  {booktitle} {\emph {\bibinfo {booktitle}
  {Cell nucleus}},\ }\href
  {https://doi.org/https://doi.org/10.1016/j.ceb.2016.02.014} {\bibfield
  {journal} {\bibinfo  {journal} {Curr. Opin. Cell Biol.}\ }\textbf {\bibinfo
  {volume} {40}},\ \bibinfo {pages} {47} (\bibinfo {year} {2016})}\BibitemShut
  {NoStop}%
\bibitem [{\citenamefont {Saintillan}, \citenamefont {Shelley},\ and\
  \citenamefont {Zidovska}(2018)}]{sain:18.1}%
  \BibitemOpen
  \bibfield  {author} {\bibinfo {author} {\bibfnamefont {D.}~\bibnamefont
  {Saintillan}}, \bibinfo {author} {\bibfnamefont {M.~J.}\ \bibnamefont
  {Shelley}},\ and\ \bibinfo {author} {\bibfnamefont {A.}~\bibnamefont
  {Zidovska}},\ }\bibfield  {title} {\enquote {\bibinfo {title} {Extensile
  motor activity drives coherent motions in a model of interphase chromatin},}\
  }\href {https://doi.org/10.1073/pnas.1807073115} {\bibfield  {journal}
  {\bibinfo  {journal} {Proc. Natl. Acad. Sci. USA}\ }\textbf {\bibinfo
  {volume} {115}},\ \bibinfo {pages} {11442} (\bibinfo {year}
  {2018})}\BibitemShut {NoStop}%
\bibitem [{\citenamefont {Ganai}, \citenamefont {Sengupta},\ and\ \citenamefont
  {Menon}(2014)}]{gana:14}%
  \BibitemOpen
  \bibfield  {author} {\bibinfo {author} {\bibfnamefont {N.}~\bibnamefont
  {Ganai}}, \bibinfo {author} {\bibfnamefont {S.}~\bibnamefont {Sengupta}},\
  and\ \bibinfo {author} {\bibfnamefont {G.~I.}\ \bibnamefont {Menon}},\
  }\bibfield  {title} {\enquote {\bibinfo {title} {Chromosome positioning from
  activity-based segregation},}\ }\href {https://doi.org/10.1093/nar/gkt1417}
  {\bibfield  {journal} {\bibinfo  {journal} {Nucleic Acids Res.}\ }\textbf
  {\bibinfo {volume} {42}},\ \bibinfo {pages} {4145} (\bibinfo {year}
  {2014})}\BibitemShut {NoStop}%
\bibitem [{\citenamefont {Smrek}\ and\ \citenamefont {Kremer}(2017)}]{smre:17}%
  \BibitemOpen
  \bibfield  {author} {\bibinfo {author} {\bibfnamefont {J.}~\bibnamefont
  {Smrek}}\ and\ \bibinfo {author} {\bibfnamefont {K.}~\bibnamefont {Kremer}},\
  }\bibfield  {title} {\enquote {\bibinfo {title} {Small activity differences
  drive phase separation in active-passive polymer mixtures},}\ }\href
  {https://doi.org/10.1103/PhysRevLett.118.098002} {\bibfield  {journal}
  {\bibinfo  {journal} {Phys. Rev. Lett.}\ }\textbf {\bibinfo {volume} {118}},\
  \bibinfo {pages} {098002} (\bibinfo {year} {2017})}\BibitemShut {NoStop}%
\bibitem [{\citenamefont {Kawamura}\ \emph {et~al.}(2008)\citenamefont
  {Kawamura}, \citenamefont {Kakugo}, \citenamefont {Shikinaka}, \citenamefont
  {Osada},\ and\ \citenamefont {Gong}}]{kawa:08}%
  \BibitemOpen
  \bibfield  {author} {\bibinfo {author} {\bibfnamefont {R.}~\bibnamefont
  {Kawamura}}, \bibinfo {author} {\bibfnamefont {A.}~\bibnamefont {Kakugo}},
  \bibinfo {author} {\bibfnamefont {K.}~\bibnamefont {Shikinaka}}, \bibinfo
  {author} {\bibfnamefont {Y.}~\bibnamefont {Osada}},\ and\ \bibinfo {author}
  {\bibfnamefont {J.~P.}\ \bibnamefont {Gong}},\ }\bibfield  {title} {\enquote
  {\bibinfo {title} {Ring-{{Shaped Assembly}} of {{Microtubules Shows
  Preferential Counterclockwise Motion}}},}\ }\href
  {https://doi.org/10.1021/bm800639w} {\bibfield  {journal} {\bibinfo
  {journal} {Biomacromolecules}\ }\textbf {\bibinfo {volume} {9}},\ \bibinfo
  {pages} {2277} (\bibinfo {year} {2008})}\BibitemShut {NoStop}%
\bibitem [{\citenamefont {Liu}, \citenamefont {Powers},\ and\ \citenamefont
  {Breuer}(2011)}]{liu:11}%
  \BibitemOpen
  \bibfield  {author} {\bibinfo {author} {\bibfnamefont {B.}~\bibnamefont
  {Liu}}, \bibinfo {author} {\bibfnamefont {T.~R.}\ \bibnamefont {Powers}},\
  and\ \bibinfo {author} {\bibfnamefont {K.~S.}\ \bibnamefont {Breuer}},\
  }\bibfield  {title} {\enquote {\bibinfo {title} {Force-free swimming of a
  model helical flagellum in viscoelastic fluids},}\ }\href
  {https://doi.org/10.1073/pnas.1113082108} {\bibfield  {journal} {\bibinfo
  {journal} {Proc. Natl. Acad. Sci. USA}\ }\textbf {\bibinfo {volume} {108}},\
  \bibinfo {pages} {19516} (\bibinfo {year} {2011})}\BibitemShut {NoStop}%
\bibitem [{\citenamefont {Keya}, \citenamefont {Kabir},\ and\ \citenamefont
  {Kakugo}(2020{\natexlab{a}})}]{keya:20}%
  \BibitemOpen
  \bibfield  {author} {\bibinfo {author} {\bibfnamefont {J.~J.}\ \bibnamefont
  {Keya}}, \bibinfo {author} {\bibfnamefont {A.~M.~R.}\ \bibnamefont {Kabir}},\
  and\ \bibinfo {author} {\bibfnamefont {A.}~\bibnamefont {Kakugo}},\
  }\bibfield  {title} {\enquote {\bibinfo {title} {Synchronous operation of
  biomolecular engines},}\ }\href {https://doi.org/10.1007/s12551-020-00651-2}
  {\bibfield  {journal} {\bibinfo  {journal} {Biophys. Rev.}\ }\textbf
  {\bibinfo {volume} {12}},\ \bibinfo {pages} {40} (\bibinfo {year}
  {2020}{\natexlab{a}})}\BibitemShut {NoStop}%
\bibitem [{\citenamefont {Butt}\ \emph {et~al.}(2010)\citenamefont {Butt},
  \citenamefont {Mufti}, \citenamefont {Humayun}, \citenamefont {Rosenthal},
  \citenamefont {Khan}, \citenamefont {Khan},\ and\ \citenamefont
  {Molloy}}]{butt:10}%
  \BibitemOpen
  \bibfield  {author} {\bibinfo {author} {\bibfnamefont {T.}~\bibnamefont
  {Butt}}, \bibinfo {author} {\bibfnamefont {T.}~\bibnamefont {Mufti}},
  \bibinfo {author} {\bibfnamefont {A.}~\bibnamefont {Humayun}}, \bibinfo
  {author} {\bibfnamefont {P.~B.}\ \bibnamefont {Rosenthal}}, \bibinfo {author}
  {\bibfnamefont {S.}~\bibnamefont {Khan}}, \bibinfo {author} {\bibfnamefont
  {S.}~\bibnamefont {Khan}},\ and\ \bibinfo {author} {\bibfnamefont {J.~E.}\
  \bibnamefont {Molloy}},\ }\bibfield  {title} {\enquote {\bibinfo {title}
  {Myosin motors drive long range alignment of actin filaments},}\ }\href
  {https://doi.org/https://doi.org/10.1074/jbc.M109.044792} {\bibfield
  {journal} {\bibinfo  {journal} {J. Biol. Chem.}\ }\textbf {\bibinfo {volume}
  {285}},\ \bibinfo {pages} {4964} (\bibinfo {year} {2010})}\BibitemShut
  {NoStop}%
\bibitem [{\citenamefont {Schaller}\ \emph {et~al.}(2010)\citenamefont
  {Schaller}, \citenamefont {Weber}, \citenamefont {Semmrich}, \citenamefont
  {Frey},\ and\ \citenamefont {Bausch}}]{scha:10}%
  \BibitemOpen
  \bibfield  {author} {\bibinfo {author} {\bibfnamefont {V.}~\bibnamefont
  {Schaller}}, \bibinfo {author} {\bibfnamefont {C.}~\bibnamefont {Weber}},
  \bibinfo {author} {\bibfnamefont {C.}~\bibnamefont {Semmrich}}, \bibinfo
  {author} {\bibfnamefont {E.}~\bibnamefont {Frey}},\ and\ \bibinfo {author}
  {\bibfnamefont {A.~R.}\ \bibnamefont {Bausch}},\ }\bibfield  {title}
  {\enquote {\bibinfo {title} {Polar patterns of driven filaments},}\ }\href
  {https://doi.org/10.1038/nature09312} {\bibfield  {journal} {\bibinfo
  {journal} {Nature}\ }\textbf {\bibinfo {volume} {467}},\ \bibinfo {pages}
  {73} (\bibinfo {year} {2010})}\BibitemShut {NoStop}%
\bibitem [{\citenamefont {Schaller}\ \emph {et~al.}(2011)\citenamefont
  {Schaller}, \citenamefont {Weber}, \citenamefont {Frey},\ and\ \citenamefont
  {Bausch}}]{scha:11}%
  \BibitemOpen
  \bibfield  {author} {\bibinfo {author} {\bibfnamefont {V.}~\bibnamefont
  {Schaller}}, \bibinfo {author} {\bibfnamefont {C.}~\bibnamefont {Weber}},
  \bibinfo {author} {\bibfnamefont {E.}~\bibnamefont {Frey}},\ and\ \bibinfo
  {author} {\bibfnamefont {A.~R.}\ \bibnamefont {Bausch}},\ }\bibfield  {title}
  {\enquote {\bibinfo {title} {Polar pattern formation: hydrodynamic coupling
  of driven filaments},}\ }\href {https://doi.org/10.1039/C0SM01063D}
  {\bibfield  {journal} {\bibinfo  {journal} {Soft Matter}\ }\textbf {\bibinfo
  {volume} {7}},\ \bibinfo {pages} {3213} (\bibinfo {year} {2011})}\BibitemShut
  {NoStop}%
\bibitem [{\citenamefont {Doostmohammadi}\ \emph {et~al.}(2018)\citenamefont
  {Doostmohammadi}, \citenamefont {Ign{\'e}s-Mullol}, \citenamefont {Yeomans},\
  and\ \citenamefont {Sagu{\'e}s}}]{doos:18}%
  \BibitemOpen
  \bibfield  {author} {\bibinfo {author} {\bibfnamefont {A.}~\bibnamefont
  {Doostmohammadi}}, \bibinfo {author} {\bibfnamefont {J.}~\bibnamefont
  {Ign{\'e}s-Mullol}}, \bibinfo {author} {\bibfnamefont {J.~M.}\ \bibnamefont
  {Yeomans}},\ and\ \bibinfo {author} {\bibfnamefont {F.}~\bibnamefont
  {Sagu{\'e}s}},\ }\bibfield  {title} {\enquote {\bibinfo {title} {Active
  nematics},}\ }\href {https://doi.org/10.1038/s41467-018-05666-8} {\bibfield
  {journal} {\bibinfo  {journal} {Nat. Commun.}\ }\textbf {\bibinfo {volume}
  {9}},\ \bibinfo {pages} {3246} (\bibinfo {year} {2018})}\BibitemShut
  {NoStop}%
\bibitem [{\citenamefont {Sanchez}\ \emph {et~al.}(2012)\citenamefont
  {Sanchez}, \citenamefont {Chen}, \citenamefont {DeCamp}, \citenamefont
  {Heymann},\ and\ \citenamefont {Dogic}}]{sanc:12}%
  \BibitemOpen
  \bibfield  {author} {\bibinfo {author} {\bibfnamefont {T.}~\bibnamefont
  {Sanchez}}, \bibinfo {author} {\bibfnamefont {D.~T.~N.}\ \bibnamefont
  {Chen}}, \bibinfo {author} {\bibfnamefont {S.~J.}\ \bibnamefont {DeCamp}},
  \bibinfo {author} {\bibfnamefont {M.}~\bibnamefont {Heymann}},\ and\ \bibinfo
  {author} {\bibfnamefont {Z.}~\bibnamefont {Dogic}},\ }\bibfield  {title}
  {\enquote {\bibinfo {title} {Spontaneous motion in hierarchically assembled
  active matter},}\ }\href {https://doi.org/10.1038/nature11591} {\bibfield
  {journal} {\bibinfo  {journal} {Nature}\ }\textbf {\bibinfo {volume} {491}},\
  \bibinfo {pages} {431} (\bibinfo {year} {2012})}\BibitemShut {NoStop}%
\bibitem [{\citenamefont {Needleman}\ and\ \citenamefont
  {Dogic}(2017)}]{need:17}%
  \BibitemOpen
  \bibfield  {author} {\bibinfo {author} {\bibfnamefont {D.}~\bibnamefont
  {Needleman}}\ and\ \bibinfo {author} {\bibfnamefont {Z.}~\bibnamefont
  {Dogic}},\ }\bibfield  {title} {\enquote {\bibinfo {title} {Active matter at
  the interface between materials science and cell biology},}\ }\href
  {https://doi.org/10.1038/natrevmats.2017.48} {\bibfield  {journal} {\bibinfo
  {journal} {Nat. Rev. Mater.}\ }\textbf {\bibinfo {volume} {2}},\ \bibinfo
  {pages} {17048} (\bibinfo {year} {2017})}\BibitemShut {NoStop}%
\bibitem [{\citenamefont {Vliegenthart}\ \emph {et~al.}(2020)\citenamefont
  {Vliegenthart}, \citenamefont {Ravichandran}, \citenamefont {Ripoll},
  \citenamefont {Auth},\ and\ \citenamefont {Gompper}}]{vliegenthart_2019}%
  \BibitemOpen
  \bibfield  {author} {\bibinfo {author} {\bibfnamefont {G.}~\bibnamefont
  {Vliegenthart}}, \bibinfo {author} {\bibfnamefont {A.}~\bibnamefont
  {Ravichandran}}, \bibinfo {author} {\bibfnamefont {M.}~\bibnamefont
  {Ripoll}}, \bibinfo {author} {\bibfnamefont {T.}~\bibnamefont {Auth}},\ and\
  \bibinfo {author} {\bibfnamefont {G.}~\bibnamefont {Gompper}},\ }\bibfield
  {title} {\enquote {\bibinfo {title} {Filamentous active matter: Band
  formation, bending, buckling, and defects},}\ }\href
  {https://doi.org/10.1126/sciadv.aaw9975} {\bibfield  {journal} {\bibinfo
  {journal} {Sci. Adv.}\ }\textbf {\bibinfo {volume} {6}},\ \bibinfo {pages}
  {eaaw9957} (\bibinfo {year} {2020})}\BibitemShut {NoStop}%
\bibitem [{\citenamefont {Alert}, \citenamefont {Joanny},\ and\ \citenamefont
  {Casademunt}(2020)}]{aler:20}%
  \BibitemOpen
  \bibfield  {author} {\bibinfo {author} {\bibfnamefont {R.}~\bibnamefont
  {Alert}}, \bibinfo {author} {\bibfnamefont {J.-F.}\ \bibnamefont {Joanny}},\
  and\ \bibinfo {author} {\bibfnamefont {J.}~\bibnamefont {Casademunt}},\
  }\bibfield  {title} {\enquote {\bibinfo {title} {Universal scaling of active
  nematic turbulence},}\ }\href {https://doi.org/10.1038/s41567-020-0854-4}
  {\bibfield  {journal} {\bibinfo  {journal} {Nat. Phys.}\ }\textbf {\bibinfo
  {volume} {16}},\ \bibinfo {pages} {682} (\bibinfo {year} {2020})}\BibitemShut
  {NoStop}%
\bibitem [{\citenamefont {Mart{\'\i}nez-Prat}\ \emph
  {et~al.}(2021)\citenamefont {Mart{\'\i}nez-Prat}, \citenamefont {Alert},
  \citenamefont {Meng}, \citenamefont {Ign{\'e}s-Mullol}, \citenamefont
  {Joanny}, \citenamefont {Casademunt}, \citenamefont {Golestanian},\ and\
  \citenamefont {Sagu{\'e}s}}]{mart:21}%
  \BibitemOpen
  \bibfield  {author} {\bibinfo {author} {\bibfnamefont {B.}~\bibnamefont
  {Mart{\'\i}nez-Prat}}, \bibinfo {author} {\bibfnamefont {R.}~\bibnamefont
  {Alert}}, \bibinfo {author} {\bibfnamefont {F.}~\bibnamefont {Meng}},
  \bibinfo {author} {\bibfnamefont {J.}~\bibnamefont {Ign{\'e}s-Mullol}},
  \bibinfo {author} {\bibfnamefont {J.-F.}\ \bibnamefont {Joanny}}, \bibinfo
  {author} {\bibfnamefont {J.}~\bibnamefont {Casademunt}}, \bibinfo {author}
  {\bibfnamefont {R.}~\bibnamefont {Golestanian}},\ and\ \bibinfo {author}
  {\bibfnamefont {F.}~\bibnamefont {Sagu{\'e}s}},\ }\bibfield  {title}
  {\enquote {\bibinfo {title} {Scaling regimes of active turbulence with
  external dissipation},}\ }\href {https://doi.org/10.1103/PhysRevX.11.031065}
  {\bibfield  {journal} {\bibinfo  {journal} {Physical Review X}\ }\textbf
  {\bibinfo {volume} {11}},\ \bibinfo {pages} {031065} (\bibinfo {year}
  {2021})}\BibitemShut {NoStop}%
\bibitem [{\citenamefont {Sasaki}\ \emph {et~al.}(2014)\citenamefont {Sasaki},
  \citenamefont {Takikawa}, \citenamefont {Jampani}, \citenamefont {Hoshikawa},
  \citenamefont {Seto}, \citenamefont {Bahr}, \citenamefont {Herminghaus},
  \citenamefont {Hidaka},\ and\ \citenamefont {Orihara}}]{sasa:14}%
  \BibitemOpen
  \bibfield  {author} {\bibinfo {author} {\bibfnamefont {Y.}~\bibnamefont
  {Sasaki}}, \bibinfo {author} {\bibfnamefont {Y.}~\bibnamefont {Takikawa}},
  \bibinfo {author} {\bibfnamefont {V.~S.~R.}\ \bibnamefont {Jampani}},
  \bibinfo {author} {\bibfnamefont {H.}~\bibnamefont {Hoshikawa}}, \bibinfo
  {author} {\bibfnamefont {T.}~\bibnamefont {Seto}}, \bibinfo {author}
  {\bibfnamefont {C.}~\bibnamefont {Bahr}}, \bibinfo {author} {\bibfnamefont
  {S.}~\bibnamefont {Herminghaus}}, \bibinfo {author} {\bibfnamefont
  {Y.}~\bibnamefont {Hidaka}},\ and\ \bibinfo {author} {\bibfnamefont
  {H.}~\bibnamefont {Orihara}},\ }\bibfield  {title} {\enquote {\bibinfo
  {title} {Colloidal caterpillars for cargo transportation},}\ }\href
  {https://doi.org/10.1039/C4SM01354A} {\bibfield  {journal} {\bibinfo
  {journal} {Soft Matter}\ }\textbf {\bibinfo {volume} {10}},\ \bibinfo {pages}
  {8813} (\bibinfo {year} {2014})}\BibitemShut {NoStop}%
\bibitem [{\citenamefont {Martinez-Pedrero}\ \emph {et~al.}(2015)\citenamefont
  {Martinez-Pedrero}, \citenamefont {Ortiz-Ambriz}, \citenamefont
  {Pagonabarraga},\ and\ \citenamefont {Tierno}}]{mart:15}%
  \BibitemOpen
  \bibfield  {author} {\bibinfo {author} {\bibfnamefont {F.}~\bibnamefont
  {Martinez-Pedrero}}, \bibinfo {author} {\bibfnamefont {A.}~\bibnamefont
  {Ortiz-Ambriz}}, \bibinfo {author} {\bibfnamefont {I.}~\bibnamefont
  {Pagonabarraga}},\ and\ \bibinfo {author} {\bibfnamefont {P.}~\bibnamefont
  {Tierno}},\ }\bibfield  {title} {\enquote {\bibinfo {title} {Colloidal
  microworms propelling via a cooperative hydrodynamic conveyor belt},}\ }\href
  {https://doi.org/10.1103/PhysRevLett.115.138301} {\bibfield  {journal}
  {\bibinfo  {journal} {Phys. Rev. Lett.}\ }\textbf {\bibinfo {volume} {115}},\
  \bibinfo {pages} {138301} (\bibinfo {year} {2015})}\BibitemShut {NoStop}%
\bibitem [{\citenamefont {Yan}\ \emph {et~al.}(2016)\citenamefont {Yan},
  \citenamefont {Han}, \citenamefont {Zhang}, \citenamefont {Xu}, \citenamefont
  {Luijten},\ and\ \citenamefont {Granick}}]{yan:16}%
  \BibitemOpen
  \bibfield  {author} {\bibinfo {author} {\bibfnamefont {J.}~\bibnamefont
  {Yan}}, \bibinfo {author} {\bibfnamefont {M.}~\bibnamefont {Han}}, \bibinfo
  {author} {\bibfnamefont {J.}~\bibnamefont {Zhang}}, \bibinfo {author}
  {\bibfnamefont {C.}~\bibnamefont {Xu}}, \bibinfo {author} {\bibfnamefont
  {E.}~\bibnamefont {Luijten}},\ and\ \bibinfo {author} {\bibfnamefont
  {S.}~\bibnamefont {Granick}},\ }\bibfield  {title} {\enquote {\bibinfo
  {title} {Reconfiguring active particles by electrostatic imbalance},}\ }\href
  {https://doi.org/10.1038/nmat4696} {\bibfield  {journal} {\bibinfo  {journal}
  {Nat. Mater.}\ }\textbf {\bibinfo {volume} {15}},\ \bibinfo {pages} {1095}
  (\bibinfo {year} {2016})}\BibitemShut {NoStop}%
\bibitem [{\citenamefont {Di~Leonardo}(2016)}]{dile:16}%
  \BibitemOpen
  \bibfield  {author} {\bibinfo {author} {\bibfnamefont {R.}~\bibnamefont
  {Di~Leonardo}},\ }\bibfield  {title} {\enquote {\bibinfo {title} {Active
  colloids: Controlled collective motions},}\ }\href
  {https://doi.org/10.1038/nmat4761} {\bibfield  {journal} {\bibinfo  {journal}
  {Nat. Mater.}\ }\textbf {\bibinfo {volume} {15}},\ \bibinfo {pages} {1057}
  (\bibinfo {year} {2016})}\BibitemShut {NoStop}%
\bibitem [{\citenamefont {Zhang}, \citenamefont {Yan},\ and\ \citenamefont
  {Granick}(2016)}]{zhan:16}%
  \BibitemOpen
  \bibfield  {author} {\bibinfo {author} {\bibfnamefont {J.}~\bibnamefont
  {Zhang}}, \bibinfo {author} {\bibfnamefont {J.}~\bibnamefont {Yan}},\ and\
  \bibinfo {author} {\bibfnamefont {S.}~\bibnamefont {Granick}},\ }\bibfield
  {title} {\enquote {\bibinfo {title} {Directed self-assembly pathways of
  active colloidal clusters},}\ }\href {https://doi.org/10.1002/anie.201509978}
  {\bibfield  {journal} {\bibinfo  {journal} {Angew. Chem. Int. Ed.}\ }\textbf
  {\bibinfo {volume} {55}},\ \bibinfo {pages} {5166} (\bibinfo {year}
  {2016})}\BibitemShut {NoStop}%
\bibitem [{\citenamefont {Zhang}\ and\ \citenamefont
  {Granick}(2016)}]{zhan:16.1}%
  \BibitemOpen
  \bibfield  {author} {\bibinfo {author} {\bibfnamefont {J.}~\bibnamefont
  {Zhang}}\ and\ \bibinfo {author} {\bibfnamefont {S.}~\bibnamefont
  {Granick}},\ }\bibfield  {title} {\enquote {\bibinfo {title} {Natural
  selection in the colloid world: active chiral spirals},}\ }\href
  {https://doi.org/10.1039/C6FD00077K} {\bibfield  {journal} {\bibinfo
  {journal} {Faraday Discuss.}\ }\textbf {\bibinfo {volume} {191}},\ \bibinfo
  {pages} {35} (\bibinfo {year} {2016})}\BibitemShut {NoStop}%
\bibitem [{\citenamefont {Vutukuri}\ \emph {et~al.}(2017)\citenamefont
  {Vutukuri}, \citenamefont {Bet}, \citenamefont {van Roij}, \citenamefont
  {Dijkstra},\ and\ \citenamefont {Huck}}]{vutu:17}%
  \BibitemOpen
  \bibfield  {author} {\bibinfo {author} {\bibfnamefont {H.~R.}\ \bibnamefont
  {Vutukuri}}, \bibinfo {author} {\bibfnamefont {B.}~\bibnamefont {Bet}},
  \bibinfo {author} {\bibfnamefont {R.}~\bibnamefont {van Roij}}, \bibinfo
  {author} {\bibfnamefont {M.}~\bibnamefont {Dijkstra}},\ and\ \bibinfo
  {author} {\bibfnamefont {W.~T.~S.}\ \bibnamefont {Huck}},\ }\bibfield
  {title} {\enquote {\bibinfo {title} {Rational design and dynamics of
  self-propelled colloidal bead chains: from rotators to flagella},}\ }\href
  {https://doi.org/10.1038/s41598-017-16731-5} {\bibfield  {journal} {\bibinfo
  {journal} {Sci. Rep.}\ }\textbf {\bibinfo {volume} {7}},\ \bibinfo {pages}
  {16758} (\bibinfo {year} {2017})}\BibitemShut {NoStop}%
\bibitem [{\citenamefont {Kokot}\ \emph {et~al.}(2017)\citenamefont {Kokot},
  \citenamefont {Das}, \citenamefont {Winkler}, \citenamefont {Gompper},
  \citenamefont {Aranson},\ and\ \citenamefont {Snezhko}}]{koko:17}%
  \BibitemOpen
  \bibfield  {author} {\bibinfo {author} {\bibfnamefont {G.}~\bibnamefont
  {Kokot}}, \bibinfo {author} {\bibfnamefont {S.}~\bibnamefont {Das}}, \bibinfo
  {author} {\bibfnamefont {R.~G.}\ \bibnamefont {Winkler}}, \bibinfo {author}
  {\bibfnamefont {G.}~\bibnamefont {Gompper}}, \bibinfo {author} {\bibfnamefont
  {I.~S.}\ \bibnamefont {Aranson}},\ and\ \bibinfo {author} {\bibfnamefont
  {A.}~\bibnamefont {Snezhko}},\ }\bibfield  {title} {\enquote {\bibinfo
  {title} {Active turbulence in a gas of self-assembled spinners},}\ }\href
  {https://doi.org/10.1073/pnas.1710188114} {\bibfield  {journal} {\bibinfo
  {journal} {Proc. Natl. Acad. Sci. USA}\ }\textbf {\bibinfo {volume} {114}},\
  \bibinfo {pages} {12870} (\bibinfo {year} {2017})}\BibitemShut {NoStop}%
\bibitem [{\citenamefont {Biswas}\ \emph {et~al.}(2017)\citenamefont {Biswas},
  \citenamefont {Manna}, \citenamefont {Laskar}, \citenamefont {Kumar},
  \citenamefont {Adhikari},\ and\ \citenamefont {Kumaraswamy}}]{bisw:17}%
  \BibitemOpen
  \bibfield  {author} {\bibinfo {author} {\bibfnamefont {B.}~\bibnamefont
  {Biswas}}, \bibinfo {author} {\bibfnamefont {R.~K.}\ \bibnamefont {Manna}},
  \bibinfo {author} {\bibfnamefont {A.}~\bibnamefont {Laskar}}, \bibinfo
  {author} {\bibfnamefont {P.~B.~S.}\ \bibnamefont {Kumar}}, \bibinfo {author}
  {\bibfnamefont {R.}~\bibnamefont {Adhikari}},\ and\ \bibinfo {author}
  {\bibfnamefont {G.}~\bibnamefont {Kumaraswamy}},\ }\bibfield  {title}
  {\enquote {\bibinfo {title} {Linking catalyst-coated isotropic colloids into
  ``active''flexible chains enhances their diffusivity},}\ }\href
  {https://doi.org/10.1021/acsnano.7b04265} {\bibfield  {journal} {\bibinfo
  {journal} {ACS Nano}\ }\textbf {\bibinfo {volume} {11}},\ \bibinfo {pages}
  {10025} (\bibinfo {year} {2017})}\BibitemShut {NoStop}%
\bibitem [{\citenamefont {Nishiguchi}\ \emph {et~al.}(2018)\citenamefont
  {Nishiguchi}, \citenamefont {Iwasawa}, \citenamefont {Jiang},\ and\
  \citenamefont {Sano}}]{nish:18}%
  \BibitemOpen
  \bibfield  {author} {\bibinfo {author} {\bibfnamefont {D.}~\bibnamefont
  {Nishiguchi}}, \bibinfo {author} {\bibfnamefont {J.}~\bibnamefont {Iwasawa}},
  \bibinfo {author} {\bibfnamefont {H.-R.}\ \bibnamefont {Jiang}},\ and\
  \bibinfo {author} {\bibfnamefont {M.}~\bibnamefont {Sano}},\ }\bibfield
  {title} {\enquote {\bibinfo {title} {Flagellar dynamics of chains of active
  {J}anus particles fueled by an {AC} electric field},}\ }\href
  {https://doi.org/10.1088/1367-2630/aa9b48} {\bibfield  {journal} {\bibinfo
  {journal} {New J. Phys.}\ }\textbf {\bibinfo {volume} {20}},\ \bibinfo
  {pages} {015002} (\bibinfo {year} {2018})}\BibitemShut {NoStop}%
\bibitem [{\citenamefont {L{\"o}wen}(2018)}]{loew:18}%
  \BibitemOpen
  \bibfield  {author} {\bibinfo {author} {\bibfnamefont {H.}~\bibnamefont
  {L{\"o}wen}},\ }\bibfield  {title} {\enquote {\bibinfo {title} {Active
  colloidal molecules},}\ }\href {https://doi.org/10.1209/0295-5075/121/58001}
  {\bibfield  {journal} {\bibinfo  {journal} {EPL}\ }\textbf {\bibinfo {volume}
  {121}},\ \bibinfo {pages} {58001} (\bibinfo {year} {2018})}\BibitemShut
  {NoStop}%
\bibitem [{\citenamefont {Shafiei~Aporvari}\ \emph {et~al.}(2020)\citenamefont
  {Shafiei~Aporvari}, \citenamefont {Utkur}, \citenamefont {Saritas},
  \citenamefont {Volpe},\ and\ \citenamefont {Stenhammar}}]{shaf:20}%
  \BibitemOpen
  \bibfield  {author} {\bibinfo {author} {\bibfnamefont {M.}~\bibnamefont
  {Shafiei~Aporvari}}, \bibinfo {author} {\bibfnamefont {M.}~\bibnamefont
  {Utkur}}, \bibinfo {author} {\bibfnamefont {E.~U.}\ \bibnamefont {Saritas}},
  \bibinfo {author} {\bibfnamefont {G.}~\bibnamefont {Volpe}},\ and\ \bibinfo
  {author} {\bibfnamefont {J.}~\bibnamefont {Stenhammar}},\ }\bibfield  {title}
  {\enquote {\bibinfo {title} {Anisotropic dynamics of a self-assembled
  colloidal chain in an active bath},}\ }\href
  {https://doi.org/10.1039/D0SM00318B} {\bibfield  {journal} {\bibinfo
  {journal} {Soft Matter}\ }\textbf {\bibinfo {volume} {16}},\ \bibinfo {pages}
  {5609} (\bibinfo {year} {2020})}\BibitemShut {NoStop}%
\bibitem [{\citenamefont {Howse}\ \emph
  {et~al.}(2007{\natexlab{a}})\citenamefont {Howse}, \citenamefont {Jones},
  \citenamefont {Ryan}, \citenamefont {Gough}, \citenamefont {Vafabakhsh},\
  and\ \citenamefont {Golestanian}}]{hows:07}%
  \BibitemOpen
  \bibfield  {author} {\bibinfo {author} {\bibfnamefont {J.~R.}\ \bibnamefont
  {Howse}}, \bibinfo {author} {\bibfnamefont {R.~A.~L.}\ \bibnamefont {Jones}},
  \bibinfo {author} {\bibfnamefont {A.~J.}\ \bibnamefont {Ryan}}, \bibinfo
  {author} {\bibfnamefont {T.}~\bibnamefont {Gough}}, \bibinfo {author}
  {\bibfnamefont {R.}~\bibnamefont {Vafabakhsh}},\ and\ \bibinfo {author}
  {\bibfnamefont {R.}~\bibnamefont {Golestanian}},\ }\bibfield  {title}
  {\enquote {\bibinfo {title} {Self-motile colloidal particles: From directed
  propulsion to random walk},}\ }\href
  {https://doi.org/10.1103/PhysRevLett.99.048102} {\bibfield  {journal}
  {\bibinfo  {journal} {Phys. Rev. Lett.}\ }\textbf {\bibinfo {volume} {99}},\
  \bibinfo {pages} {048102} (\bibinfo {year} {2007}{\natexlab{a}})}\BibitemShut
  {NoStop}%
\bibitem [{\citenamefont {Jiang}, \citenamefont {Yoshinaga},\ and\
  \citenamefont {Sano}(2010)}]{jian:10}%
  \BibitemOpen
  \bibfield  {author} {\bibinfo {author} {\bibfnamefont {H.-R.}\ \bibnamefont
  {Jiang}}, \bibinfo {author} {\bibfnamefont {N.}~\bibnamefont {Yoshinaga}},\
  and\ \bibinfo {author} {\bibfnamefont {M.}~\bibnamefont {Sano}},\ }\bibfield
  {title} {\enquote {\bibinfo {title} {Active motion of a janus particle by
  self-thermophoresis in a defocused laser beam},}\ }\href
  {https://doi.org/10.1103/PhysRevLett.105.268302} {\bibfield  {journal}
  {\bibinfo  {journal} {Phys. Rev. Lett.}\ }\textbf {\bibinfo {volume} {105}},\
  \bibinfo {pages} {268302} (\bibinfo {year} {2010})}\BibitemShut {NoStop}%
\bibitem [{\citenamefont {Valadares}\ \emph {et~al.}(2010)\citenamefont
  {Valadares}, \citenamefont {Tao}, \citenamefont {Zacharia}, \citenamefont
  {Kitaev}, \citenamefont {Galembeck}, \citenamefont {Kapral},\ and\
  \citenamefont {Ozin}}]{vala:10}%
  \BibitemOpen
  \bibfield  {author} {\bibinfo {author} {\bibfnamefont {L.~F.}\ \bibnamefont
  {Valadares}}, \bibinfo {author} {\bibfnamefont {Y.-G.}\ \bibnamefont {Tao}},
  \bibinfo {author} {\bibfnamefont {N.~S.}\ \bibnamefont {Zacharia}}, \bibinfo
  {author} {\bibfnamefont {V.}~\bibnamefont {Kitaev}}, \bibinfo {author}
  {\bibfnamefont {F.}~\bibnamefont {Galembeck}}, \bibinfo {author}
  {\bibfnamefont {R.}~\bibnamefont {Kapral}},\ and\ \bibinfo {author}
  {\bibfnamefont {G.~A.}\ \bibnamefont {Ozin}},\ }\bibfield  {title} {\enquote
  {\bibinfo {title} {Catalytic nanomotors: Self-propelled sphere dimers},}\
  }\href {https://doi.org/10.1002/smll.200901976} {\bibfield  {journal}
  {\bibinfo  {journal} {Small}\ }\textbf {\bibinfo {volume} {6}},\ \bibinfo
  {pages} {565--572} (\bibinfo {year} {2010})}\BibitemShut {NoStop}%
\bibitem [{\citenamefont {W{\"u}rger}(2010)}]{wurg:10}%
  \BibitemOpen
  \bibfield  {author} {\bibinfo {author} {\bibfnamefont {A.}~\bibnamefont
  {W{\"u}rger}},\ }\bibfield  {title} {\enquote {\bibinfo {title} {Thermal
  non-equilibrium transport in colloids},}\ }\href
  {https://doi.org/10.1088/0034-4885/73/12/126601} {\bibfield  {journal}
  {\bibinfo  {journal} {Rep. Prog. Phys.}\ }\textbf {\bibinfo {volume} {73}},\
  \bibinfo {pages} {126601} (\bibinfo {year} {2010})}\BibitemShut {NoStop}%
\bibitem [{\citenamefont {Volpe}\ \emph {et~al.}(2011)\citenamefont {Volpe},
  \citenamefont {Buttinoni}, \citenamefont {Vogt}, \citenamefont
  {K{\"u}mmerer},\ and\ \citenamefont {Bechinger}}]{volp:11}%
  \BibitemOpen
  \bibfield  {author} {\bibinfo {author} {\bibfnamefont {G.}~\bibnamefont
  {Volpe}}, \bibinfo {author} {\bibfnamefont {I.}~\bibnamefont {Buttinoni}},
  \bibinfo {author} {\bibfnamefont {D.}~\bibnamefont {Vogt}}, \bibinfo {author}
  {\bibfnamefont {H.~J.}\ \bibnamefont {K{\"u}mmerer}},\ and\ \bibinfo {author}
  {\bibfnamefont {C.}~\bibnamefont {Bechinger}},\ }\bibfield  {title} {\enquote
  {\bibinfo {title} {Microswimmers in patterned environments},}\ }\href
  {https://doi.org/10.1039/C1SM05960B} {\bibfield  {journal} {\bibinfo
  {journal} {Soft Matter}\ }\textbf {\bibinfo {volume} {7}},\ \bibinfo {pages}
  {8810} (\bibinfo {year} {2011})}\BibitemShut {NoStop}%
\bibitem [{\citenamefont {Thutupalli}, \citenamefont {Seemann},\ and\
  \citenamefont {Herminghaus}(2011)}]{thut:11}%
  \BibitemOpen
  \bibfield  {author} {\bibinfo {author} {\bibfnamefont {S.}~\bibnamefont
  {Thutupalli}}, \bibinfo {author} {\bibfnamefont {R.}~\bibnamefont
  {Seemann}},\ and\ \bibinfo {author} {\bibfnamefont {S.}~\bibnamefont
  {Herminghaus}},\ }\bibfield  {title} {\enquote {\bibinfo {title} {Swarming
  behavior of simple model squirmers},}\ }\href
  {https://doi.org/10.1088/1367-2630/13/7/073021} {\bibfield  {journal}
  {\bibinfo  {journal} {New J. Phys.}\ }\textbf {\bibinfo {volume} {13}},\
  \bibinfo {pages} {073021} (\bibinfo {year} {2011})}\BibitemShut {NoStop}%
\bibitem [{\citenamefont {Buttinoni}\ \emph {et~al.}(2013)\citenamefont
  {Buttinoni}, \citenamefont {Bialk{\'e}}, \citenamefont {K{\"u}mmel},
  \citenamefont {L{\"o}wen}, \citenamefont {Bechinger},\ and\ \citenamefont
  {Speck}}]{butt:13}%
  \BibitemOpen
  \bibfield  {author} {\bibinfo {author} {\bibfnamefont {I.}~\bibnamefont
  {Buttinoni}}, \bibinfo {author} {\bibfnamefont {J.}~\bibnamefont
  {Bialk{\'e}}}, \bibinfo {author} {\bibfnamefont {F.}~\bibnamefont
  {K{\"u}mmel}}, \bibinfo {author} {\bibfnamefont {H.}~\bibnamefont
  {L{\"o}wen}}, \bibinfo {author} {\bibfnamefont {C.}~\bibnamefont
  {Bechinger}},\ and\ \bibinfo {author} {\bibfnamefont {T.}~\bibnamefont
  {Speck}},\ }\bibfield  {title} {\enquote {\bibinfo {title} {Dynamical
  clustering and phase separation in suspensions of self-propelled colloidal
  particles},}\ }\href {https://doi.org/10.1103/PhysRevLett.110.238301}
  {\bibfield  {journal} {\bibinfo  {journal} {Phys. Rev. Lett.}\ }\textbf
  {\bibinfo {volume} {110}},\ \bibinfo {pages} {238301} (\bibinfo {year}
  {2013})}\BibitemShut {NoStop}%
\bibitem [{\citenamefont {ten Hagen}\ \emph {et~al.}(2014)\citenamefont {ten
  Hagen}, \citenamefont {K{\"u}mmel}, \citenamefont {Wittkowski}, \citenamefont
  {Takagi}, \citenamefont {L{\"o}wen},\ and\ \citenamefont
  {Bechinger}}]{hage:14}%
  \BibitemOpen
  \bibfield  {author} {\bibinfo {author} {\bibfnamefont {B.}~\bibnamefont {ten
  Hagen}}, \bibinfo {author} {\bibfnamefont {F.}~\bibnamefont {K{\"u}mmel}},
  \bibinfo {author} {\bibfnamefont {R.}~\bibnamefont {Wittkowski}}, \bibinfo
  {author} {\bibfnamefont {D.}~\bibnamefont {Takagi}}, \bibinfo {author}
  {\bibfnamefont {H.}~\bibnamefont {L{\"o}wen}},\ and\ \bibinfo {author}
  {\bibfnamefont {C.}~\bibnamefont {Bechinger}},\ }\bibfield  {title} {\enquote
  {\bibinfo {title} {Gravitaxis of asymmetric self-propelled colloidal
  particles},}\ }\href {https://doi.org/10.1038/ncomms5829} {\bibfield
  {journal} {\bibinfo  {journal} {Nat. Commun.}\ }\textbf {\bibinfo {volume}
  {5}},\ \bibinfo {pages} {4829} (\bibinfo {year} {2014})}\BibitemShut
  {NoStop}%
\bibitem [{\citenamefont {Bechinger}\ \emph {et~al.}(2016)\citenamefont
  {Bechinger}, \citenamefont {Di~Leonardo}, \citenamefont {L{\"o}wen},
  \citenamefont {Reichhardt}, \citenamefont {Volpe},\ and\ \citenamefont
  {Volpe}}]{bech:16}%
  \BibitemOpen
  \bibfield  {author} {\bibinfo {author} {\bibfnamefont {C.}~\bibnamefont
  {Bechinger}}, \bibinfo {author} {\bibfnamefont {R.}~\bibnamefont
  {Di~Leonardo}}, \bibinfo {author} {\bibfnamefont {H.}~\bibnamefont
  {L{\"o}wen}}, \bibinfo {author} {\bibfnamefont {C.}~\bibnamefont
  {Reichhardt}}, \bibinfo {author} {\bibfnamefont {G.}~\bibnamefont {Volpe}},\
  and\ \bibinfo {author} {\bibfnamefont {G.}~\bibnamefont {Volpe}},\ }\bibfield
   {title} {\enquote {\bibinfo {title} {Active particles in complex and crowded
  environments},}\ }\href {https://doi.org/10.1103/RevModPhys.88.045006}
  {\bibfield  {journal} {\bibinfo  {journal} {Rev. Mod. Phys.}\ }\textbf
  {\bibinfo {volume} {88}},\ \bibinfo {pages} {045006} (\bibinfo {year}
  {2016})}\BibitemShut {NoStop}%
\bibitem [{\citenamefont {Maass}\ \emph {et~al.}(2016)\citenamefont {Maass},
  \citenamefont {Kr{\"u}ger}, \citenamefont {Herminghaus},\ and\ \citenamefont
  {Bahr}}]{maas:16}%
  \BibitemOpen
  \bibfield  {author} {\bibinfo {author} {\bibfnamefont {C.~C.}\ \bibnamefont
  {Maass}}, \bibinfo {author} {\bibfnamefont {C.}~\bibnamefont {Kr{\"u}ger}},
  \bibinfo {author} {\bibfnamefont {S.}~\bibnamefont {Herminghaus}},\ and\
  \bibinfo {author} {\bibfnamefont {C.}~\bibnamefont {Bahr}},\ }\bibfield
  {title} {\enquote {\bibinfo {title} {Swimming droplets},}\ }\href
  {https://doi.org/10.1146/annurev-conmatphys-031115-011517} {\bibfield
  {journal} {\bibinfo  {journal} {Annu. Rev. Cond. Mat. Phys.}\ }\textbf
  {\bibinfo {volume} {7}},\ \bibinfo {pages} {171} (\bibinfo {year}
  {2016})}\BibitemShut {NoStop}%
\bibitem [{\citenamefont {Elgeti}, \citenamefont {Winkler},\ and\ \citenamefont
  {Gompper}(2015)}]{elge:15}%
  \BibitemOpen
  \bibfield  {author} {\bibinfo {author} {\bibfnamefont {J.}~\bibnamefont
  {Elgeti}}, \bibinfo {author} {\bibfnamefont {R.~G.}\ \bibnamefont
  {Winkler}},\ and\ \bibinfo {author} {\bibfnamefont {G.}~\bibnamefont
  {Gompper}},\ }\bibfield  {title} {\enquote {\bibinfo {title} {Physics of
  microswimmers---single particle motion and collective behavior: a review},}\
  }\href {https://doi.org/10.1088/0034-4885/78/5/056601} {\bibfield  {journal}
  {\bibinfo  {journal} {Rep. Prog. Phys.}\ }\textbf {\bibinfo {volume} {78}},\
  \bibinfo {pages} {056601} (\bibinfo {year} {2015})}\BibitemShut {NoStop}%
\bibitem [{\citenamefont {Harder}, \citenamefont {Valeriani},\ and\
  \citenamefont {Cacciuto}(2014)}]{hard:14}%
  \BibitemOpen
  \bibfield  {author} {\bibinfo {author} {\bibfnamefont {J.}~\bibnamefont
  {Harder}}, \bibinfo {author} {\bibfnamefont {C.}~\bibnamefont {Valeriani}},\
  and\ \bibinfo {author} {\bibfnamefont {A.}~\bibnamefont {Cacciuto}},\
  }\bibfield  {title} {\enquote {\bibinfo {title} {Activity-induced collapse
  and reexpansion of rigid polymers},}\ }\href
  {https://doi.org/10.1103/PhysRevE.90.062312} {\bibfield  {journal} {\bibinfo
  {journal} {Phys. Rev. E}\ }\textbf {\bibinfo {volume} {90}},\ \bibinfo
  {pages} {062312} (\bibinfo {year} {2014})}\BibitemShut {NoStop}%
\bibitem [{\citenamefont {Kaiser}\ and\ \citenamefont
  {L{\"o}wen}(2014)}]{kais:14}%
  \BibitemOpen
  \bibfield  {author} {\bibinfo {author} {\bibfnamefont {A.}~\bibnamefont
  {Kaiser}}\ and\ \bibinfo {author} {\bibfnamefont {H.}~\bibnamefont
  {L{\"o}wen}},\ }\bibfield  {title} {\enquote {\bibinfo {title} {Unusual
  swelling of a polymer in a bacterial bath},}\ }\href
  {https://doi.org/http://dx.doi.org/10.1063/1.4891095} {\bibfield  {journal}
  {\bibinfo  {journal} {J. Chem. Phys.}\ }\textbf {\bibinfo {volume} {141}},\
  \bibinfo {eid} {044903} (\bibinfo {year} {2014})}\BibitemShut {NoStop}%
\bibitem [{\citenamefont {Bianco}, \citenamefont {Locatelli},\ and\
  \citenamefont {Malgaretti}(2018)}]{bianco_globulelike_2018}%
  \BibitemOpen
  \bibfield  {author} {\bibinfo {author} {\bibfnamefont {V.}~\bibnamefont
  {Bianco}}, \bibinfo {author} {\bibfnamefont {E.}~\bibnamefont {Locatelli}},\
  and\ \bibinfo {author} {\bibfnamefont {P.}~\bibnamefont {Malgaretti}},\
  }\bibfield  {title} {\enquote {\bibinfo {title} {Globulelike {{Conformation}}
  and {{Enhanced Diffusion}} of {{Active Polymers}}},}\ }\href
  {https://doi.org/10.1103/PhysRevLett.121.217802} {\bibfield  {journal}
  {\bibinfo  {journal} {Phys. Rev. Lett.}\ }\textbf {\bibinfo {volume} {121}},\
  \bibinfo {pages} {217802} (\bibinfo {year} {2018})}\BibitemShut {NoStop}%
\bibitem [{\citenamefont {Anand}\ and\ \citenamefont {Singh}(2020)}]{anan:20}%
  \BibitemOpen
  \bibfield  {author} {\bibinfo {author} {\bibfnamefont {S.~K.}\ \bibnamefont
  {Anand}}\ and\ \bibinfo {author} {\bibfnamefont {S.~P.}\ \bibnamefont
  {Singh}},\ }\bibfield  {title} {\enquote {\bibinfo {title} {Conformation and
  dynamics of a self-avoiding active flexible polymer},}\ }\href
  {https://doi.org/10.1103/PhysRevE.101.030501} {\bibfield  {journal} {\bibinfo
   {journal} {Phys. Rev. E}\ }\textbf {\bibinfo {volume} {101}},\ \bibinfo
  {pages} {030501} (\bibinfo {year} {2020})}\BibitemShut {NoStop}%
\bibitem [{\citenamefont {Das}, \citenamefont {Kennedy},\ and\ \citenamefont
  {Cacciuto}(2021)}]{das:21}%
  \BibitemOpen
  \bibfield  {author} {\bibinfo {author} {\bibfnamefont {S.}~\bibnamefont
  {Das}}, \bibinfo {author} {\bibfnamefont {N.}~\bibnamefont {Kennedy}},\ and\
  \bibinfo {author} {\bibfnamefont {A.}~\bibnamefont {Cacciuto}},\ }\bibfield
  {title} {\enquote {\bibinfo {title} {The coil--globule transition in
  self-avoiding active polymers},}\ }\href {https://doi.org/10.1039/D0SM01526A}
  {\bibfield  {journal} {\bibinfo  {journal} {Soft Matter}\ }\textbf {\bibinfo
  {volume} {17}},\ \bibinfo {pages} {160} (\bibinfo {year} {2021})}\BibitemShut
  {NoStop}%
\bibitem [{\citenamefont {Ghosh}\ and\ \citenamefont {Gov}(2014)}]{ghos:14}%
  \BibitemOpen
  \bibfield  {author} {\bibinfo {author} {\bibfnamefont {A.}~\bibnamefont
  {Ghosh}}\ and\ \bibinfo {author} {\bibfnamefont {N.~S.}\ \bibnamefont
  {Gov}},\ }\bibfield  {title} {\enquote {\bibinfo {title} {Dynamics of active
  semiflexible polymers},}\ }\href {https://doi.org/10.1016/j.bpj.2014.07.034}
  {\bibfield  {journal} {\bibinfo  {journal} {Biophys. J.}\ }\textbf {\bibinfo
  {volume} {107}},\ \bibinfo {pages} {1065} (\bibinfo {year}
  {2014})}\BibitemShut {NoStop}%
\bibitem [{\citenamefont {Shin}\ \emph {et~al.}(2015)\citenamefont {Shin},
  \citenamefont {Cherstvy}, \citenamefont {Kim},\ and\ \citenamefont
  {Metzler}}]{shin:15}%
  \BibitemOpen
  \bibfield  {author} {\bibinfo {author} {\bibfnamefont {J.}~\bibnamefont
  {Shin}}, \bibinfo {author} {\bibfnamefont {A.~G.}\ \bibnamefont {Cherstvy}},
  \bibinfo {author} {\bibfnamefont {W.~K.}\ \bibnamefont {Kim}},\ and\ \bibinfo
  {author} {\bibfnamefont {R.}~\bibnamefont {Metzler}},\ }\bibfield  {title}
  {\enquote {\bibinfo {title} {Facilitation of polymer looping and giant
  polymer diffusivity in crowded solutions of active particles},}\ }\href
  {https://doi.org/10.1088/1367-2630/17/11/113008} {\bibfield  {journal}
  {\bibinfo  {journal} {New J. Phys.}\ }\textbf {\bibinfo {volume} {17}},\
  \bibinfo {pages} {113008} (\bibinfo {year} {2015})}\BibitemShut {NoStop}%
\bibitem [{\citenamefont {Eisenstecken}, \citenamefont {Gompper},\ and\
  \citenamefont {Winkler}(2016)}]{eisenstecken_conformational_2016}%
  \BibitemOpen
  \bibfield  {author} {\bibinfo {author} {\bibfnamefont {T.}~\bibnamefont
  {Eisenstecken}}, \bibinfo {author} {\bibfnamefont {G.}~\bibnamefont
  {Gompper}},\ and\ \bibinfo {author} {\bibfnamefont {R.~G.}\ \bibnamefont
  {Winkler}},\ }\bibfield  {title} {\enquote {\bibinfo {title} {Conformational
  {{Properties}} of {{Active Semiflexible Polymers}}},}\ }\href
  {https://doi.org/10.3390/polym8080304} {\bibfield  {journal} {\bibinfo
  {journal} {Polymers}\ }\textbf {\bibinfo {volume} {8}},\ \bibinfo {pages}
  {304} (\bibinfo {year} {2016})}\BibitemShut {NoStop}%
\bibitem [{\citenamefont {Eisenstecken}, \citenamefont {Gompper},\ and\
  \citenamefont {Winkler}(2017)}]{eisenstecken_internal_2017}%
  \BibitemOpen
  \bibfield  {author} {\bibinfo {author} {\bibfnamefont {T.}~\bibnamefont
  {Eisenstecken}}, \bibinfo {author} {\bibfnamefont {G.}~\bibnamefont
  {Gompper}},\ and\ \bibinfo {author} {\bibfnamefont {R.~G.}\ \bibnamefont
  {Winkler}},\ }\bibfield  {title} {\enquote {\bibinfo {title} {Internal
  dynamics of semiflexible polymers with active noise},}\ }\href
  {https://doi.org/10.1063/1.4981012} {\bibfield  {journal} {\bibinfo
  {journal} {J. Chem. Phys.}\ }\textbf {\bibinfo {volume} {146}},\ \bibinfo
  {pages} {154903} (\bibinfo {year} {2017})}\BibitemShut {NoStop}%
\bibitem [{\citenamefont {Mart{{\'\i}}n-G{{\'o}}mez}, \citenamefont {Gompper},\
  and\ \citenamefont {Winkler}(2018)}]{mart:18.1}%
  \BibitemOpen
  \bibfield  {author} {\bibinfo {author} {\bibfnamefont {A.}~\bibnamefont
  {Mart{{\'\i}}n-G{{\'o}}mez}}, \bibinfo {author} {\bibfnamefont
  {G.}~\bibnamefont {Gompper}},\ and\ \bibinfo {author} {\bibfnamefont {R.~G.}\
  \bibnamefont {Winkler}},\ }\bibfield  {title} {\enquote {\bibinfo {title}
  {Active {B}rownian filamentous polymers under shear flow},}\ }\href
  {https://doi.org/10.3390/polym10080837} {\bibfield  {journal} {\bibinfo
  {journal} {Polymers}\ }\textbf {\bibinfo {volume} {10}},\ \bibinfo {pages}
  {837} (\bibinfo {year} {2018})}\BibitemShut {NoStop}%
\bibitem [{\citenamefont {Mousavi}, \citenamefont {Gompper},\ and\
  \citenamefont {Winkler}(2019)}]{mous:19}%
  \BibitemOpen
  \bibfield  {author} {\bibinfo {author} {\bibfnamefont {S.~M.}\ \bibnamefont
  {Mousavi}}, \bibinfo {author} {\bibfnamefont {G.}~\bibnamefont {Gompper}},\
  and\ \bibinfo {author} {\bibfnamefont {R.~G.}\ \bibnamefont {Winkler}},\
  }\bibfield  {title} {\enquote {\bibinfo {title} {Active {B}rownian ring
  polymers},}\ }\href {https://doi.org/10.1063/1.5082723} {\bibfield  {journal}
  {\bibinfo  {journal} {J. Chem. Phys.}\ }\textbf {\bibinfo {volume} {150}},\
  \bibinfo {pages} {064913} (\bibinfo {year} {2019})}\BibitemShut {NoStop}%
\bibitem [{\citenamefont {Suma}\ \emph {et~al.}(2014)\citenamefont {Suma},
  \citenamefont {Gonnella}, \citenamefont {Marenduzzo},\ and\ \citenamefont
  {Orlandini}}]{suma:14}%
  \BibitemOpen
  \bibfield  {author} {\bibinfo {author} {\bibfnamefont {A.}~\bibnamefont
  {Suma}}, \bibinfo {author} {\bibfnamefont {G.}~\bibnamefont {Gonnella}},
  \bibinfo {author} {\bibfnamefont {D.}~\bibnamefont {Marenduzzo}},\ and\
  \bibinfo {author} {\bibfnamefont {E.}~\bibnamefont {Orlandini}},\ }\bibfield
  {title} {\enquote {\bibinfo {title} {Motility-induced phase separation in an
  active dumbbell fluid},}\ }\href
  {https://doi.org/10.1209/0295-5075/108/56004} {\bibfield  {journal} {\bibinfo
   {journal} {EPL}\ }\textbf {\bibinfo {volume} {108}},\ \bibinfo {pages}
  {56004} (\bibinfo {year} {2014})}\BibitemShut {NoStop}%
\bibitem [{\citenamefont {Siebert}\ \emph {et~al.}(2017)\citenamefont
  {Siebert}, \citenamefont {Letz}, \citenamefont {Speck},\ and\ \citenamefont
  {Virnau}}]{sieb:17}%
  \BibitemOpen
  \bibfield  {author} {\bibinfo {author} {\bibfnamefont {J.~T.}\ \bibnamefont
  {Siebert}}, \bibinfo {author} {\bibfnamefont {J.}~\bibnamefont {Letz}},
  \bibinfo {author} {\bibfnamefont {T.}~\bibnamefont {Speck}},\ and\ \bibinfo
  {author} {\bibfnamefont {P.}~\bibnamefont {Virnau}},\ }\bibfield  {title}
  {\enquote {\bibinfo {title} {Phase behavior of active {B}rownian disks,
  spheres, and dimers},}\ }\href {https://doi.org/10.1039/C6SM02622B}
  {\bibfield  {journal} {\bibinfo  {journal} {Soft Matter}\ }\textbf {\bibinfo
  {volume} {13}},\ \bibinfo {pages} {1020} (\bibinfo {year}
  {2017})}\BibitemShut {NoStop}%
\bibitem [{\citenamefont {Martin-Gomez}\ \emph {et~al.}(2020)\citenamefont
  {Martin-Gomez}, \citenamefont {Eisenstecken}, \citenamefont {Gompper},\ and\
  \citenamefont {Winkler}}]{mart:20}%
  \BibitemOpen
  \bibfield  {author} {\bibinfo {author} {\bibfnamefont {A.}~\bibnamefont
  {Martin-Gomez}}, \bibinfo {author} {\bibfnamefont {T.}~\bibnamefont
  {Eisenstecken}}, \bibinfo {author} {\bibfnamefont {G.}~\bibnamefont
  {Gompper}},\ and\ \bibinfo {author} {\bibfnamefont {R.~G.}\ \bibnamefont
  {Winkler}},\ }\bibfield  {title} {\enquote {\bibinfo {title} {Hydrodynamics
  of polymers in an active bath},}\ }\href
  {https://doi.org/10.1103/PhysRevE.101.052612} {\bibfield  {journal} {\bibinfo
   {journal} {Phys. Rev. E}\ }\textbf {\bibinfo {volume} {101}},\ \bibinfo
  {pages} {052612} (\bibinfo {year} {2020})}\BibitemShut {NoStop}%
\bibitem [{\citenamefont {Mart{{\'\i}}n-G{{\'o}}mez}\ \emph
  {et~al.}(2019)\citenamefont {Mart{{\'\i}}n-G{{\'o}}mez}, \citenamefont
  {Eisenstecken}, \citenamefont {Gompper},\ and\ \citenamefont
  {Winkler}}]{mart:19}%
  \BibitemOpen
  \bibfield  {author} {\bibinfo {author} {\bibfnamefont {A.}~\bibnamefont
  {Mart{{\'\i}}n-G{{\'o}}mez}}, \bibinfo {author} {\bibfnamefont
  {T.}~\bibnamefont {Eisenstecken}}, \bibinfo {author} {\bibfnamefont
  {G.}~\bibnamefont {Gompper}},\ and\ \bibinfo {author} {\bibfnamefont {R.~G.}\
  \bibnamefont {Winkler}},\ }\bibfield  {title} {\enquote {\bibinfo {title}
  {Active {B}rownian filaments with hydrodynamic interactions: conformations
  and dynamics},}\ }\href {https://doi.org/10.1039/C9SM00391F} {\bibfield
  {journal} {\bibinfo  {journal} {Soft Matter}\ }\textbf {\bibinfo {volume}
  {15}},\ \bibinfo {pages} {3957} (\bibinfo {year} {2019})}\BibitemShut
  {NoStop}%
\bibitem [{\citenamefont {Isele-Holder}, \citenamefont {Elgeti},\ and\
  \citenamefont {Gompper}(2015)}]{isel:15}%
  \BibitemOpen
  \bibfield  {author} {\bibinfo {author} {\bibfnamefont {R.~E.}\ \bibnamefont
  {Isele-Holder}}, \bibinfo {author} {\bibfnamefont {J.}~\bibnamefont
  {Elgeti}},\ and\ \bibinfo {author} {\bibfnamefont {G.}~\bibnamefont
  {Gompper}},\ }\bibfield  {title} {\enquote {\bibinfo {title} {Self-propelled
  worm-like filaments: spontaneous spiral formation, structure, and
  dynamics},}\ }\href {https://doi.org/10.1039/C5SM01683E} {\bibfield
  {journal} {\bibinfo  {journal} {Soft Matter}\ }\textbf {\bibinfo {volume}
  {11}},\ \bibinfo {pages} {7181} (\bibinfo {year} {2015})}\BibitemShut
  {NoStop}%
\bibitem [{\citenamefont {Duman}\ \emph {et~al.}(2018)\citenamefont {Duman},
  \citenamefont {Isele-Holder}, \citenamefont {Elgeti},\ and\ \citenamefont
  {Gompper}}]{duma:18}%
  \BibitemOpen
  \bibfield  {author} {\bibinfo {author} {\bibfnamefont {O.}~\bibnamefont
  {Duman}}, \bibinfo {author} {\bibfnamefont {R.~E.}\ \bibnamefont
  {Isele-Holder}}, \bibinfo {author} {\bibfnamefont {J.}~\bibnamefont
  {Elgeti}},\ and\ \bibinfo {author} {\bibfnamefont {G.}~\bibnamefont
  {Gompper}},\ }\bibfield  {title} {\enquote {\bibinfo {title} {Collective
  dynamics of self-propelled semiflexible filaments},}\ }\href
  {https://doi.org/10.1039/C8SM00282G} {\bibfield  {journal} {\bibinfo
  {journal} {Soft Matter}\ }\textbf {\bibinfo {volume} {14}},\ \bibinfo {pages}
  {4483} (\bibinfo {year} {2018})}\BibitemShut {NoStop}%
\bibitem [{\citenamefont {Anand}\ and\ \citenamefont {Singh}(2018)}]{anan:18}%
  \BibitemOpen
  \bibfield  {author} {\bibinfo {author} {\bibfnamefont {S.~K.}\ \bibnamefont
  {Anand}}\ and\ \bibinfo {author} {\bibfnamefont {S.~P.}\ \bibnamefont
  {Singh}},\ }\bibfield  {title} {\enquote {\bibinfo {title} {Structure and
  dynamics of a self-propelled semiflexible filament},}\ }\href
  {https://doi.org/10.1103/PhysRevE.98.042501} {\bibfield  {journal} {\bibinfo
  {journal} {Phys. Rev. E}\ }\textbf {\bibinfo {volume} {98}},\ \bibinfo
  {pages} {042501} (\bibinfo {year} {2018})}\BibitemShut {NoStop}%
\bibitem [{\citenamefont {Prathyusha}, \citenamefont {Henkes},\ and\
  \citenamefont {Sknepnek}(2018)}]{prat:18}%
  \BibitemOpen
  \bibfield  {author} {\bibinfo {author} {\bibfnamefont {K.~R.}\ \bibnamefont
  {Prathyusha}}, \bibinfo {author} {\bibfnamefont {S.}~\bibnamefont {Henkes}},\
  and\ \bibinfo {author} {\bibfnamefont {R.}~\bibnamefont {Sknepnek}},\
  }\bibfield  {title} {\enquote {\bibinfo {title} {Dynamically generated
  patterns in dense suspensions of active filaments},}\ }\href
  {https://doi.org/10.1103/PhysRevE.97.022606} {\bibfield  {journal} {\bibinfo
  {journal} {Phys. Rev. E}\ }\textbf {\bibinfo {volume} {97}},\ \bibinfo
  {pages} {022606} (\bibinfo {year} {2018})}\BibitemShut {NoStop}%
\bibitem [{\citenamefont {Foglino}\ \emph {et~al.}(2019)\citenamefont
  {Foglino}, \citenamefont {Locatelli}, \citenamefont {Brackley}, \citenamefont
  {Michieletto}, \citenamefont {Likos},\ and\ \citenamefont
  {Marenduzzo}}]{fogl:19}%
  \BibitemOpen
  \bibfield  {author} {\bibinfo {author} {\bibfnamefont {M.}~\bibnamefont
  {Foglino}}, \bibinfo {author} {\bibfnamefont {E.}~\bibnamefont {Locatelli}},
  \bibinfo {author} {\bibfnamefont {C.~A.}\ \bibnamefont {Brackley}}, \bibinfo
  {author} {\bibfnamefont {D.}~\bibnamefont {Michieletto}}, \bibinfo {author}
  {\bibfnamefont {C.~N.}\ \bibnamefont {Likos}},\ and\ \bibinfo {author}
  {\bibfnamefont {D.}~\bibnamefont {Marenduzzo}},\ }\bibfield  {title}
  {\enquote {\bibinfo {title} {Non-equilibrium effects of molecular motors on
  polymers},}\ }\href {https://doi.org/10.1039/C9SM00273A} {\bibfield
  {journal} {\bibinfo  {journal} {Soft Matter}\ }\textbf {\bibinfo {volume}
  {15}},\ \bibinfo {pages} {5995} (\bibinfo {year} {2019})}\BibitemShut
  {NoStop}%
\bibitem [{\citenamefont {Moore}\ \emph {et~al.}(2020)\citenamefont {Moore},
  \citenamefont {Thompson}, \citenamefont {Glaser},\ and\ \citenamefont
  {Betterton}}]{moor:20}%
  \BibitemOpen
  \bibfield  {author} {\bibinfo {author} {\bibfnamefont {J.~M.}\ \bibnamefont
  {Moore}}, \bibinfo {author} {\bibfnamefont {T.~N.}\ \bibnamefont {Thompson}},
  \bibinfo {author} {\bibfnamefont {M.~A.}\ \bibnamefont {Glaser}},\ and\
  \bibinfo {author} {\bibfnamefont {M.~D.}\ \bibnamefont {Betterton}},\
  }\bibfield  {title} {\enquote {\bibinfo {title} {Collective motion of driven
  semiflexible filaments tuned by soft repulsion and stiffness},}\ }\href
  {https://doi.org/10.1039/D0SM01036G} {\bibfield  {journal} {\bibinfo
  {journal} {Soft Matter}\ }\textbf {\bibinfo {volume} {16}},\ \bibinfo {pages}
  {9436} (\bibinfo {year} {2020})}\BibitemShut {NoStop}%
\bibitem [{\citenamefont {Peterson}, \citenamefont {Hagan},\ and\ \citenamefont
  {Baskaran}(2020)}]{peterson_statistical_2020}%
  \BibitemOpen
  \bibfield  {author} {\bibinfo {author} {\bibfnamefont {M.~S.~E.}\
  \bibnamefont {Peterson}}, \bibinfo {author} {\bibfnamefont {M.~F.}\
  \bibnamefont {Hagan}},\ and\ \bibinfo {author} {\bibfnamefont
  {A.}~\bibnamefont {Baskaran}},\ }\bibfield  {title} {\enquote {\bibinfo
  {title} {Statistical properties of a tangentially driven active filament},}\
  }\href {https://doi.org/10.1088/1742-5468/ab6097} {\bibfield  {journal}
  {\bibinfo  {journal} {J. Stat. Mech.}\ }\textbf {\bibinfo {volume} {2020}},\
  \bibinfo {pages} {013216} (\bibinfo {year} {2020})}\BibitemShut {NoStop}%
\bibitem [{\citenamefont {Nguyen}\ \emph {et~al.}(2021)\citenamefont {Nguyen},
  \citenamefont {Ozkan-Aydin}, \citenamefont {Tuazon}, \citenamefont {Goldman},
  \citenamefont {Bhamla},\ and\ \citenamefont {Peleg}}]{nguy:21}%
  \BibitemOpen
  \bibfield  {author} {\bibinfo {author} {\bibfnamefont {C.}~\bibnamefont
  {Nguyen}}, \bibinfo {author} {\bibfnamefont {Y.}~\bibnamefont {Ozkan-Aydin}},
  \bibinfo {author} {\bibfnamefont {H.}~\bibnamefont {Tuazon}}, \bibinfo
  {author} {\bibfnamefont {D.~I.}\ \bibnamefont {Goldman}}, \bibinfo {author}
  {\bibfnamefont {M.~S.}\ \bibnamefont {Bhamla}},\ and\ \bibinfo {author}
  {\bibfnamefont {O.}~\bibnamefont {Peleg}},\ }\bibfield  {title} {\enquote
  {\bibinfo {title} {Emergent collective locomotion in an active polymer model
  of entangled worm blobs},}\ }\href {https://doi.org/10.3389/fphy.2021.734499}
  {\bibfield  {journal} {\bibinfo  {journal} {Front. Phys.}\ }\textbf {\bibinfo
  {volume} {9}},\ \bibinfo {pages} {734499} (\bibinfo {year}
  {2021})}\BibitemShut {NoStop}%
\bibitem [{\citenamefont {Locatelli}, \citenamefont {Bianco},\ and\
  \citenamefont {Malgaretti}(2021)}]{loca:21}%
  \BibitemOpen
  \bibfield  {author} {\bibinfo {author} {\bibfnamefont {E.}~\bibnamefont
  {Locatelli}}, \bibinfo {author} {\bibfnamefont {V.}~\bibnamefont {Bianco}},\
  and\ \bibinfo {author} {\bibfnamefont {P.}~\bibnamefont {Malgaretti}},\
  }\bibfield  {title} {\enquote {\bibinfo {title} {Activity-induced collapse
  and arrest of active polymer rings},}\ }\href
  {https://doi.org/10.1103/PhysRevLett.126.097801} {\bibfield  {journal}
  {\bibinfo  {journal} {Phys. Rev. Lett.}\ }\textbf {\bibinfo {volume} {126}},\
  \bibinfo {pages} {097801} (\bibinfo {year} {2021})}\BibitemShut {NoStop}%
\bibitem [{\citenamefont {Qiao}\ and\ \citenamefont {Kapral}(2021)}]{qiao:22}%
  \BibitemOpen
  \bibfield  {author} {\bibinfo {author} {\bibfnamefont {L.}~\bibnamefont
  {Qiao}}\ and\ \bibinfo {author} {\bibfnamefont {R.}~\bibnamefont {Kapral}},\
  }\href {https://doi.org/10.48550/ARXIV.2109.12602} {\enquote {\bibinfo
  {title} {Control of active polymeric filaments by chemically-powered
  nanomotors},}\ } (\bibinfo {year} {2021})\BibitemShut {NoStop}%
\bibitem [{\citenamefont {Philipps}, \citenamefont {Gompper},\ and\
  \citenamefont {Winkler}(2022)}]{philipps_ring_2022}%
  \BibitemOpen
  \bibfield  {author} {\bibinfo {author} {\bibfnamefont {C.~A.}\ \bibnamefont
  {Philipps}}, \bibinfo {author} {\bibfnamefont {G.}~\bibnamefont {Gompper}},\
  and\ \bibinfo {author} {\bibfnamefont {R.~G.}\ \bibnamefont {Winkler}},\
  }\bibfield  {title} {\enquote {\bibinfo {title} {Dynamics of active polar
  ring polymers},}\ }\href {https://doi.org/10.1103/PhysRevE.105.L062501}
  {\bibfield  {journal} {\bibinfo  {journal} {Phys. Rev. E}\ }\textbf {\bibinfo
  {volume} {105}},\ \bibinfo {pages} {L062501} (\bibinfo {year}
  {2022})}\BibitemShut {NoStop}%
\bibitem [{\citenamefont {Keya}, \citenamefont {Kabir},\ and\ \citenamefont
  {Kakugo}(2020{\natexlab{b}})}]{keya_synchronous_2020}%
  \BibitemOpen
  \bibfield  {author} {\bibinfo {author} {\bibfnamefont {J.~J.}\ \bibnamefont
  {Keya}}, \bibinfo {author} {\bibfnamefont {A.~M.~R.}\ \bibnamefont {Kabir}},\
  and\ \bibinfo {author} {\bibfnamefont {A.}~\bibnamefont {Kakugo}},\
  }\bibfield  {title} {\enquote {\bibinfo {title} {Synchronous operation of
  biomolecular engines},}\ }\href {https://doi.org/10.1007/s12551-020-00651-2}
  {\bibfield  {journal} {\bibinfo  {journal} {Biophys Rev}\ }\textbf {\bibinfo
  {volume} {12}},\ \bibinfo {pages} {401} (\bibinfo {year}
  {2020}{\natexlab{b}})}\BibitemShut {NoStop}%
\bibitem [{\citenamefont {Harnau}, \citenamefont {Winkler},\ and\ \citenamefont
  {Reineker}(1995)}]{harnau_dynamic_1995}%
  \BibitemOpen
  \bibfield  {author} {\bibinfo {author} {\bibfnamefont {L.}~\bibnamefont
  {Harnau}}, \bibinfo {author} {\bibfnamefont {R.~G.}\ \bibnamefont
  {Winkler}},\ and\ \bibinfo {author} {\bibfnamefont {P.}~\bibnamefont
  {Reineker}},\ }\bibfield  {title} {\enquote {\bibinfo {title} {Dynamic
  properties of molecular chains with variable stiffness},}\ }\href
  {https://doi.org/10.1063/1.469027} {\bibfield  {journal} {\bibinfo  {journal}
  {J. Chem. Phys.}\ }\textbf {\bibinfo {volume} {102}},\ \bibinfo {pages}
  {7750} (\bibinfo {year} {1995})}\BibitemShut {NoStop}%
\bibitem [{\citenamefont {Doi}\ and\ \citenamefont
  {Edwards}(1986)}]{doi_theory_1986}%
  \BibitemOpen
  \bibfield  {author} {\bibinfo {author} {\bibfnamefont {M.}~\bibnamefont
  {Doi}}\ and\ \bibinfo {author} {\bibfnamefont {S.~F.}\ \bibnamefont
  {Edwards}},\ }\href@noop {} {\emph {\bibinfo {title} {The Theory of Polymer
  Dynamics}}}\ (\bibinfo  {publisher} {{Clarendon Press}},\ \bibinfo {address}
  {{Oxford}},\ \bibinfo {year} {1986})\BibitemShut {NoStop}%
\bibitem [{\citenamefont {Winkler}, \citenamefont {Reineker},\ and\
  \citenamefont {Harnau}(1994)}]{winkler_models_1994}%
  \BibitemOpen
  \bibfield  {author} {\bibinfo {author} {\bibfnamefont {R.~G.}\ \bibnamefont
  {Winkler}}, \bibinfo {author} {\bibfnamefont {P.}~\bibnamefont {Reineker}},\
  and\ \bibinfo {author} {\bibfnamefont {L.}~\bibnamefont {Harnau}},\
  }\bibfield  {title} {\enquote {\bibinfo {title} {Models and equilibrium
  properties of stiff molecular chains},}\ }\href
  {https://doi.org/10.1063/1.468239} {\bibfield  {journal} {\bibinfo  {journal}
  {J. Chem. Phys.}\ }\textbf {\bibinfo {volume} {101}},\ \bibinfo {pages}
  {8119--8129} (\bibinfo {year} {1994})}\BibitemShut {NoStop}%
\bibitem [{\citenamefont {Winkler}(2003)}]{winkler_deformation_2003}%
  \BibitemOpen
  \bibfield  {author} {\bibinfo {author} {\bibfnamefont {R.~G.}\ \bibnamefont
  {Winkler}},\ }\bibfield  {title} {\enquote {\bibinfo {title} {Deformation of
  semiflexible chains},}\ }\href {https://doi.org/10.1063/1.1537247} {\bibfield
   {journal} {\bibinfo  {journal} {J. Chem. Phys.}\ }\textbf {\bibinfo {volume}
  {118}},\ \bibinfo {pages} {2919} (\bibinfo {year} {2003})}\BibitemShut
  {NoStop}%
\bibitem [{\citenamefont {Risken}(1984)}]{risken_fokker-planck_1984}%
  \BibitemOpen
  \bibfield  {author} {\bibinfo {author} {\bibfnamefont {H.}~\bibnamefont
  {Risken}},\ }\href@noop {} {\emph {\bibinfo {title} {Fokker-{{Planck
  Equation}} - {{Methods}} of {{Solution}} and {{Applications}}}}}\ (\bibinfo
  {publisher} {{Springer-Verlag}},\ \bibinfo {address} {{Place of publication
  not identified}},\ \bibinfo {year} {1984})\BibitemShut {NoStop}%
\bibitem [{\citenamefont {Ha}\ and\ \citenamefont
  {Thirumalai}(1995)}]{ha_1995}%
  \BibitemOpen
  \bibfield  {author} {\bibinfo {author} {\bibfnamefont {B.~Y.}\ \bibnamefont
  {Ha}}\ and\ \bibinfo {author} {\bibfnamefont {D.}~\bibnamefont
  {Thirumalai}},\ }\bibfield  {title} {\enquote {\bibinfo {title} {A mean-field
  model for semiflexible chains},}\ }\href {https://doi.org/10.1063/1.470001}
  {\bibfield  {journal} {\bibinfo  {journal} {J. Chem. Phys.}\ }\textbf
  {\bibinfo {volume} {103}},\ \bibinfo {pages} {9408} (\bibinfo {year}
  {1995})}\BibitemShut {NoStop}%
\bibitem [{\citenamefont {Battle}\ \emph {et~al.}(2016)\citenamefont {Battle},
  \citenamefont {Broedersz}, \citenamefont {Fakhri}, \citenamefont {Geyer},
  \citenamefont {Howard}, \citenamefont {Schmidt},\ and\ \citenamefont
  {MacKintosh}}]{batt:16}%
  \BibitemOpen
  \bibfield  {author} {\bibinfo {author} {\bibfnamefont {C.}~\bibnamefont
  {Battle}}, \bibinfo {author} {\bibfnamefont {C.~P.}\ \bibnamefont
  {Broedersz}}, \bibinfo {author} {\bibfnamefont {N.}~\bibnamefont {Fakhri}},
  \bibinfo {author} {\bibfnamefont {V.~F.}\ \bibnamefont {Geyer}}, \bibinfo
  {author} {\bibfnamefont {J.}~\bibnamefont {Howard}}, \bibinfo {author}
  {\bibfnamefont {C.~F.}\ \bibnamefont {Schmidt}},\ and\ \bibinfo {author}
  {\bibfnamefont {F.~C.}\ \bibnamefont {MacKintosh}},\ }\bibfield  {title}
  {\enquote {\bibinfo {title} {Broken detailed balance at mesoscopic scales in
  active biological systems},}\ }\href
  {https://doi.org/10.1126/science.aac8167} {\bibfield  {journal} {\bibinfo
  {journal} {Science}\ }\textbf {\bibinfo {volume} {352}},\ \bibinfo {pages}
  {604} (\bibinfo {year} {2016})}\BibitemShut {NoStop}%
\bibitem [{\citenamefont {Howse}\ \emph
  {et~al.}(2007{\natexlab{b}})\citenamefont {Howse}, \citenamefont {Jones},
  \citenamefont {Ryan}, \citenamefont {Gough}, \citenamefont {Vafabakhsh},\
  and\ \citenamefont {Golestanian}}]{howse_abp_2007}%
  \BibitemOpen
  \bibfield  {author} {\bibinfo {author} {\bibfnamefont {J.~R.}\ \bibnamefont
  {Howse}}, \bibinfo {author} {\bibfnamefont {R.~A.~L.}\ \bibnamefont {Jones}},
  \bibinfo {author} {\bibfnamefont {A.~J.}\ \bibnamefont {Ryan}}, \bibinfo
  {author} {\bibfnamefont {T.}~\bibnamefont {Gough}}, \bibinfo {author}
  {\bibfnamefont {R.}~\bibnamefont {Vafabakhsh}},\ and\ \bibinfo {author}
  {\bibfnamefont {R.}~\bibnamefont {Golestanian}},\ }\bibfield  {title}
  {\enquote {\bibinfo {title} {Self-motile colloidal particles: From directed
  propulsion to random walk},}\ }\href
  {https://doi.org/10.1103/PhysRevLett.99.048102} {\bibfield  {journal}
  {\bibinfo  {journal} {Phys. Rev. Lett.}\ }\textbf {\bibinfo {volume} {99}},\
  \bibinfo {pages} {048102} (\bibinfo {year} {2007}{\natexlab{b}})}\BibitemShut
  {NoStop}%
\end{thebibliography}

%

\end{document}